\documentclass[range]{ar2e-changed}
\usepackage{amsfonts}

\newcommand {\bc}{\begin{center}}
\newcommand {\ec}{\end{center}}\newcommand {\bea}{\begin{eqnarray}}
\newcommand {\eea}{\end{eqnarray}}
\newcommand {\be}{\begin{equation}}
\newcommand {\ee}{\end{equation}}

\def\lsim{\mathrel{\rlap{\lower4pt\hbox{\hskip1pt$\sim$}}
    \raise1pt\hbox{$<$}}}               
\def\gsim{\mathrel{\rlap{\lower4pt\hbox{\hskip1pt$\sim$}}
    \raise1pt\hbox{$>$}}}                
\usepackage{graphicx}

\begin{document}



\jname{Annual Review of Nuclear and Particle Science}
\jyear{2014}
\jvol{64}

\title{Fluid Dynamics and Viscosity in\\ Strongly Correlated
Fluids}

\markboth{Fluid Dynamics and Viscosity}{Fluid Dynamics and Viscosity}

\author{Thomas Sch\"afer
\affiliation{Department of Physics, North Carolina State University,
Raleigh, NC 27695}}


\begin{abstract}
We review the modern view of fluid dynamics as an effective low energy, 
long wavelength theory of many body systems at finite temperature. We 
introduce the concept of a nearly perfect fluid, defined by a ratio 
$\eta/s$ of shear viscosity to entropy density of order $\hbar/k_B$ or 
less. Nearly perfect fluids exhibit hydrodynamic behavior at all distances
down to the microscopic length scale of the fluid. We summarize arguments 
that suggest that there is fundamental limit to fluidity, and we review 
the current experimental situation of measurements of $\eta/s$ 
in strongly coupled quantum fluids. 

\end{abstract}

\maketitle

\newpage
\begin{flushright}
\begin{minipage}{0.45\hsize}
$\pi\alpha\nu\tau\alpha$ $\rho\varepsilon\iota$
({\it everything flows})

\hspace{0.1\hsize} Heraclitus

{\it The mountains flowed before the Lord.}\hspace{1cm} 

\hspace{0.1\hsize} Prophet Deborah, Judges, 5:5
\end{minipage}
\end{flushright}

\section{Fluid Dynamics}
\label{sec_fluids}
\subsection{Fluid Dynamics as an effective theory}
\label{sec_eft}

 Fluid dynamics is often described as a consequence of applying 
Newton's laws to a continuous deformable medium. However, the ideas
underlying fluid dynamics are much more general. Fluid dynamics 
describes classical and quantum liquids, gases, and plasmas. It 
accounts for the low energy properties of magnetic materials, 
liquid crystals, crystalline solids, supersolids, and many other 
systems. Indeed, fluid dynamics is now understood as an effective 
theory for the long-distance, long-time properties of any material 
\cite{Martin:1963,Forster}. The only requirement for the applicability 
of fluid dynamics is that the system relaxes to approximate local 
thermodynamic equilibrium on the time scale of the observation. 
This idea is captured by the two quotations above: In principle 
everything behaves as a fluid, but in some systems observing fluid 
dynamic behavior may require divine patience \cite{Reiner:1964}.

 Fluid dynamics is based on the observation that there are 
two basic time scales associated with the behavior of a many body 
system. The first is a microscopic time scale $\tau_{\it fluid}$ 
that characterizes the rate at which a generic disturbance relaxes. 
In a typical molecular liquid this rate is governed by the collision 
rate between molecules. The second time scale $\tau_{\it diff}$ is 
associated with the relaxation of conserved charges$^1$. Because conserved 
charges cannot relax locally, but rather have to decay by diffusion or 
collective motion, this time increases with the length scale $\lambda$ 
of the disturbance, $\tau_{\it diff}\sim\lambda$. Fluid dynamics is 
based on the separation of scales $\tau_{\it fluid}\ll \tau_{\it diff}$, 
and $\omega_{\it fluid}=\tau_{\it fluid}^{-1}$ can be viewed as the 
breakdown scale of fluid dynamics as an effective theory. 

 In a simple non-relativistic fluid the conserved charges are
the mass density $\rho$, the momentum density $\vec{\pi}$, and
the energy density ${\cal E}$. The momentum density can be used 
to define the fluid velocity, $\vec{u}=\vec{\pi}/\rho$. By Galilean 
invariance the energy density can then be written as the the sum of 
the internal energy density and kinetic energy density, ${\cal E}=
{\cal E}_0+\frac{1}{2}\rho u^2$. The conservation laws are$^2$ 
\bea
\label{hydro1}
\frac{\partial \rho}{\partial t} &=& 
   - \vec{\nabla}\cdot\vec{\pi}   , \\
\label{hydro2}
 \frac{\partial \pi_i}{\partial t} &=&  
   - \nabla_j\Pi_{ij}, \\
\label{hydro3}
 \frac{\partial {\cal E}}{\partial t} &=&
   - \vec{\nabla} \cdot\vec{\jmath}^{\;\epsilon} .  
\eea 
For these equations to close we have to specify constitutive 
relations for the stress tensor $\Pi_{ij}$ and the energy current 
$\vec{\jmath}^{\;\epsilon}$. Since fluid dynamics is an effective long
wavelength theory we expect that the currents can be systematically
expanded in gradients of the hydrodynamic variables $\rho$, $\vec{u}$
and ${\cal E}_0$. In the case of the stress tensor the leading,
no-derivative, terms are completely fixed by rotational symmetry and
Galilean invariance. We have 
\be 
 \Pi_{ij} = \rho u_i u_j + P\delta_{ij}+ \delta \Pi_{ij}\, ,
\ee
where $P=P(\rho,{\cal E}_0)$ is the equation of state and $\delta\Pi_{ij}$ 
contains gradient terms. The approximation $\delta\Pi_{ij}=0$ is 
known as ideal fluid dynamics. Ideal fluid dynamics is time reversal 
invariant and the entropy is conserved. If gradient terms are included
then time reversal invariance is broken and the entropy increases.
We will refer to  $\delta\Pi_{ij}$ as the dissipative stresses. At
first order  in the gradient expansion  $\delta\Pi_{ij}$ can be 
written as $\delta\Pi_{ij}=-\eta\sigma_{ij}-\zeta\delta_{ij}\langle 
\sigma\rangle$ with
\be 
 \sigma_{ij} = \nabla_i u_j +\nabla_j u_i 
  -\frac{2}{3}\delta_{ij}   \langle\sigma\rangle \, ,
\hspace{0.1\hsize}
 \langle\sigma\rangle =\vec{\nabla}\cdot\vec{u}\, . 
\ee
This expression contains two transport coefficients, the shear
viscosity $\eta$ and the bulk viscosity $\zeta$. The energy current 
is given by $\vec{\jmath}^{\;\epsilon} = \vec{u}w+\delta\vec{\jmath}^{\;
\epsilon}$, where $w=P+{\cal E}$ is the enthalpy. At leading order in 
the gradient expansion $\delta\jmath_i^{\;\epsilon}=u_j\delta\Pi_{ij}
-\kappa\nabla_i T$, where $\kappa$ is the thermal conductivity. The 
second law of thermodynamics implies that $\eta,\zeta$ and $\kappa$ 
must be positive. 

 We can now establish the expansion parameter that controls the
fluid dynamic description. We first note that the ideal stress
tensor contains two terms, which are related to the pressure $P$ 
and the inertial stress $\rho u_iu_j$. The relative importance of 
these two terms is governed by the Mach number ${\it Ma}=v/c_s$, 
where $c_s^2=(\partial P)/(\partial \rho)_{\bar s}$ is the speed of 
sound and $\bar{s}=s/n$ is the entropy per particle. Flows with 
${\it Ma}\sim 1$ are termed compressible, and flows with ${\it Ma}
\ll 1$ incompressible. We are most interested in expanding 
systems, which are certainly compressible. 

 The validity of hydrodynamics requires that dissipative terms
are small relative to ideal terms. We will focus on the role 
of shear viscosity, because it is the dominant source of 
dissipation in the systems considered here, and because both
$\zeta$ and $\kappa$ can become zero in physically realizable
limits. In particular, $\zeta$ vanishes in a scale invariant fluid
like the unitary gas, and $\kappa$ vanishes in a relativistic fluid
with zero baryon chemical potential like the pure gluon plasma. 
In the case ${\it Ma}\sim 1$ the expansion parameter is 
\be 
\label{Re}
 {\it Re}^{-1} = \frac{\eta \nabla u}{\rho u^2} = 
  \frac{\eta}{\rho u L}\, , 
\ee
where ${\it Re}$ is the Reynolds number and $L$ is a 
characteristic length scale of the flow. Before continuing we 
briefly comment on incompressible flows. The expansion parameter 
in this case is ${\it Ma}^2/{\it Re}$. Flows with ${\it Ma}\ll 1$ 
and ${\it Re}^{-1}\ll 1$ are nearly ideal, turbulent flows. The 
regime ${\it Ma}^2/{\it Re}\ll 1$ but ${\it Re}^{-1}\gsim 1$ is that 
of very viscous flow. Today interest in very viscous flow is often
related to classical fluids in confined geometries. A typical 
example is the problem of bacterial swimming \cite{Purcell:1977}.

 We note that ${\it Re}^{-1}$ can be written as 
\be 
{\it Re}^{-1}
= \frac{\eta}{\hbar n} \times \frac{\hbar}{muL}\, , 
\ee 
where both factors are dimensionless. The first factor is solely a 
property of the fluid, and the second factor characterizes the flow. 
For a typical classical flow the second factor is much smaller than 
one, and the validity of fluid dynamics places no constraints on 
$\eta/(\hbar n)$. For the types of experiments that are explored 
in Sects.~\ref{sec_nr} and \ref{sec_rel} the second factor is of 
order one, and the applicability of fluid dynamics requires $\eta 
\lsim \hbar n$. We note that in relativistic flows the inertial term 
is $\Pi_{ij}=sTu_iu_j$, and the analogous requirement is $\eta \lsim 
\hbar s/k_B$. We refer to fluids that satisfy this condition 
as nearly perfect fluids 
\cite{Danielewicz:1984ww,Kovtun:2004de,Schafer:2009dj,Adams:2012th},
and show that  nearly perfect fluids exhibit hydrodynamic behavior 
on remarkably short length and time scales, comparable to microscopic 
scales such as the inverse temperature or the inverse Fermi wave vector. 
Throughout this review, we  use units in which $\hbar=k_B=1$. 

 The long wavelength expansion can be extended beyond the first order 
in gradients of the hydrodynamic variables$^{3,4}$. The classical higher 
order equations are known as Burnett and super-Burnett equations 
\cite{Burnett:1935,Garcia:2008}. Explicit forms of second order terms 
based on kinetic theory were derived by Grad in the non-relativistic 
case \cite{Grad:1949}, and by Israel, Stewart and others for relativistic 
fluids \cite{Israel:1979wp}. Historically, these theories have not been 
used very frequently. One reason is that the effects are not very large. 
In the case of the Navier-Stokes equation dissipative terms can 
exponentiate and alter the motion qualitatively, even if at any given 
time gradient corrections are small. A simple example is a collective 
oscillation of a fluid, see Sect.~\ref{sec_flow_nr}. Without viscosity 
the mode cannot decay, but if dissipation is present the motion is 
exponentially damped. Typically, second order terms do not exponentiate, 
and the gain in accuracy from including higher order terms is frequently 
offset by uncertainties in higher order transport coefficients or the 
need for additional boundary conditions. 

 The second reason that higher order theories are infrequently used is
that the classical equations at second order 
are unstable to short-wavelength perturbations. In relativistic fluid 
dynamics problems with acausality and instability already appear
at the Navier-Stokes level. These difficulties are not fundamental: 
Fluid dynamics is an effective theory, and unstable or acausal modes 
occur outside the domain of validity of the theory. It is nevertheless 
desirable to construct schemes that have second or higher order 
accuracy and satisfy causality and stability requirements. A possible
solution is to promote the dissipative currents to hydrodynamic 
variables and postulate a set of relaxation equations for these 
quantities. Consider the dissipative stress tensor and define 
$\pi_{ij}=\delta \Pi_{ij}$. The relaxation equation for
$\pi_{ij}$ is 
\be 
\label{pi_relax}
\tau_R \dot{\pi}_{ij}= -\pi_{ij} -\eta\sigma_{ij} + \ldots \, , 
\ee
where $\ldots$ contains second order terms such as $(\nabla\cdot u)
\sigma_{ij}$ and $\sigma_{ik}\sigma_{kj}$. To second order accuracy
this equation is equivalent to $\delta \Pi_{ij}=-\eta\sigma_{ij}
+\tau_R\eta\dot{\sigma}_{ij}+\ldots$, which is part of the standard
Burnett theory. Physically, equ.~(\ref{pi_relax}) describes the 
relaxation of the dissipative stresses to the Navier-Stokes form.
The resulting equations are stable and causal, and the sensitivity
to higher order gradients can be checked by varying second order
coefficients like $\tau_R$ \cite{Romatschke:2009im}. 

 Equ.~(\ref{pi_relax}) was first proposed by Maxwell as a model for 
very viscous fluids \cite{Landau:elast}. Cattaneo observed that relaxation 
equations can be used to restore causality and studied a relaxation model 
in the context of Fourier's law $\delta\vec{\jmath}^{\;\epsilon}=-\kappa
\vec{\nabla} T$ \cite{Cattaneo:1948,Muller:2006}. Relaxation equations 
were derived from kinetic theory by M\"uller \cite{Muller:1967}, Israel 
and Stewart \cite{Israel:1976}, and others. To achieve the expected scaling 
of second order terms with ${\it Re}^{-2}$ it is is important to include 
a full set of second order terms that respect the symmetries of the 
theory. This problem was addressed for relativistic scale invariant 
fluids by Baier et al.~\cite{Baier:2007ix}, and in the non-relativistic 
case by Chao et al.~\cite{Chao:2011cy}.

 It is well know that the low energy expansion in effective field
theories$^5$ is not a simple power series in the expansion parameter 
$\omega/\Lambda$. Quantum fluctuations lead to non-analytic terms.
In the case of fluid dynamics $\Lambda=\omega_{\it fluid}$ and 
non-analyticities arise due to thermal fluctuations of the hydrodynamic 
variables. As a consequence the dissipative currents $\delta\Pi_{ij}$ 
and $\delta \vec{\jmath}^{\;\epsilon}$ contain not only gradient terms 
but also stochastic contributions. The magnitude of the stochastic 
terms is determined by fluctuation-dissipation theorems. We have
\bea 
\label{G_S_cont}
\left\langle  \Pi_{ij}(t,\vec{x}) \Pi_{kl}(t',\vec{x}') \right\rangle 
  &=& 2\eta T \left(\delta_{ik}\delta_{jl}+\delta_{il}\delta_{jk}
      -\frac{2}{3} \delta_{ij}\delta_{kl}\right)
   \delta(t-t')\delta(\vec{x}-\vec{x}')\, ,\\
\left\langle \jmath_i^{\epsilon}(t,\vec{x}) 
             \jmath_j^{\epsilon}(t',\vec{x}')\right\rangle 
 &=& 2\kappa T^2 \delta_{ij}  \delta(t-t')\delta(\vec{x}-\vec{x}')\, ,
\eea
where $\langle . \rangle$ denotes a thermal average and we have neglected
bulk viscosity. A calculation of the response function in stochastic
fluid dynamics shows that the hydrodynamic expansion contains 
non-analyticities that are smaller than the Navier-Stokes term, 
but larger than second order terms \cite{Kovtun:2011np,Chafin:2012eq}. 
This implies that, strictly speaking, the second order theory is only 
consistent if stochastic terms  are included. Some studies of 
fluctuating fluid dynamics have been performed \cite{Ahikari:2005}, but 
in particle and nuclear physics this problem has only recently attracted 
interest \cite{Murase:2013tma}.

\subsection{Microscopic models of fluids: Kinetic Theory}
\label{sec_kin}

 Within fluid dynamics the equation of state and the transport
coefficients are parameters that have to be extracted from 
experiment. If a more microscopic description of the fluid 
is available then we can compute these parameters in terms of more 
fundamental quantities. The simplest microscopic description of a 
fluid is kinetic theory. Kinetic theory is itself an effective 
theory that describes the long distance behavior of an underlying
classical or quantum many-body system. It is applicable whenever 
there is a range of energies and momenta in which the excitations
of the fluid are long-lived quasi-particles. Kinetic theory 
can be used to relate properties of these quasi-particles, their 
masses, lifetimes, and scattering cross sections, to the equation 
of state and the transport coefficients. Kinetic theory can also 
be used to extend the description of collective effects such as 
sound or macroscopic flow into the regime where fluid dynamics breaks down. 

 The basic object in kinetic theory is the quasi-particle distribution 
function $f_p(\vec{x},t)$. Hydrodynamic variables can be written
as integrals of $f_p$ over $d\Gamma=d^3p/(2\pi)^3$. For example, 
the off-diagonal component of the stress tensor is given by 
\be 
\label{pi_ij_kin}
\Pi_{ij}\left(\vec{x},t\right) = 
 \int d\Gamma_p\, p_iv_j f_p\left(\vec{x},t\right)\, ,
 \hspace{0.5cm} (i\neq j) \, ,  
\ee
where $\vec{v}=\vec{\nabla}_p E_p$ is the quasi-particle velocity. Similar 
expressions exist for other conserved currents$^6$. The equation
of motion for $f_p(\vec{x},t)$ is the Boltzmann equation
\be 
\label{B_eq}
 \left( \frac{\partial}{\partial t} 
  + \vec{v} \cdot \vec{\nabla}_x
  + \vec{F}  \cdot \vec{\nabla}_p \right)
f_p\left(\vec{x},t\right)
 = C[f_p]\, ,
\ee
where $\vec{F}=-\vec{\nabla}_x E_p$ is the force and $C[f_p]$ is the 
collision term. By taking moments of the Boltzmann equation we can derive 
the conservation laws (\ref{hydro1}-\ref{hydro3}). In order to extract 
the constitutive relations we have to assume that the distribution 
function is close the equilibrium distribution $f_p(\vec{x},t)=f_p^0(
\vec{x},t)+\delta f_p(\vec{x},t)$, and that gradients of $f_p(\vec{x},t)$ 
are small. The equilibrium distribution can be expressed in terms of 
the conserved charges or, more conveniently, in terms of the corresponding 
intensive quantities $\mu,T$ and $\vec{u}$. We find
\be
\label{f_p_0}
f_p^{0}(\vec{x},t)=\frac{1}
 {\exp\left[\beta\left( E_p-\vec{u}\cdot\vec{p}-\mu\right)\right]\pm 1}
 \, , 
\ee
where $\beta= 1/T$, the $\pm$ sign corresponds to fermions and bosons, 
respectively, and $\beta,\mu,\vec{u}$ are functions of $\vec{x}$ and $t$.

 To identify the expansion parameter we have to understand
the scales involved in the collision term. If $\delta f_p\ll f^0_p$ 
we can use $C[f^0_p]=0$ to linearize the collision term. The linearized
collision term is a hermitean, negative semi-definite, operator 
that can be expanded in terms of its eigenvalues and eigenvectors$^7$. 
We refer to the inverse eigenvalues as collision times. In order to 
solve the Boltzmann equation we have to invert the collision term.  At long 
times we can therefore approximate the collision term by the longest 
collision time $\tau_0$ and write
\be 
\label{BGK}
 C[f_p^0+\delta f_p]\simeq - \frac{\delta f_p}{\tau_0}\, , 
\ee
where we have used the fact that at late times $\delta f_p$ is 
dominated by its projection on the lowest eigenvector. Equ.~(\ref{BGK}) 
is known as the BGK (Bhatnagar-Gross-Krook) or relaxation time approximation 
\cite{Bhatnagar:1954}. We can define a mean free path by $l_{\it mfp}=\tau_0 
\bar{v}$ where $\bar{v}=\langle v^2\rangle^{1/2}$. The expansion parameter 
for the gradient expansion is given by the Knudsen number
\be
{\it Kn}= \frac{l_{\it mfp}}{L}\, 
\ee
where $L\sim \nabla^{-1}$ as in equ.~(\ref{Re}). The systematic 
determination of the constitutive equation via an expansion in 
${\it Kn}$ is called the Chapman-Enskog expansion \cite{Chapman:1970}. 
We find, for example, 
\be 
\label{eta_mfp}
\eta = \frac{1}{3} nl_{\it mfp}\bar{p} \, , 
\ee
and $\tau_R=\tau_0=\eta/P$ \cite{Chapman:1970,Chao:2011cy}. In order 
to estimate the Reynolds number we can use ${\it Ma}=u/c_s\sim 1$. In 
kinetic theory we find $c_s^2=\frac{5}{9}\langle v^2\rangle$ and 
${\it Kn}\sim {\it Re}^{-1}$. The Knudsen expansion is equivalent to 
the Reynolds number expansion in fluid dynamics$^8$. 

 Fluid dynamics corresponds to the long time behavior of kinetic
theory. It is also interesting to examine the short time behavior. 
Consider the response of the fluid to an external shear strain 
$h_{xy}$ with frequency $\omega$ and wave number $k$. The solution 
of the Boltzmann equation is of the form 
\be
\label{del_f_w}
 \delta f_p(\omega,k) = \frac{1}{2T}\frac{-i\omega p_x v_y}
     {-i\omega+i\vec{v}\cdot\vec{k}+\tau_0^{-1}}\,
       f_p^0\, h_{xy}\, . 
\ee
This result can be used to compute the spectral function of correlators 
of conserved currents. For $k=0$ the term $(-i\omega+\tau_0^{-1})$ 
in the denominator of equ.~(\ref{del_f_w}) leads to a Lorentzian shape 
of the spectral function, which is a signature of the presence of 
quasi-particles. The spectral function also provides information about 
the breakdown of kinetic theory for large $\omega$ and $k$. There is 
no intrinsic scale in the Boltzmann equation other than the collision 
time $\tau_0$ which sets the scale for the hydrodynamic expansion. The 
high energy scale is set by matching the Boltzmann equation to the 
equation of motion for a non-equilibrium Green function in quantum field 
theory \cite{Baym:1962}. Instead of matching these equations explicitly, 
we can compare the kinetic spectral functions in equ.~(\ref{del_f_w}) 
to the spectral functions in quantum field theory, see Sect.~\ref{sec_uni}. 
The result shows that the breakdown scale is $\omega_{\it micro}\sim T$. 
This scale should be compared to the hydrodynamic scale $\omega_{\it fluid} 
\sim \tau_0^{-1} \sim P/\eta$. For a typical fluid these scales are 
well separated, but for a nearly perfect fluid the two scales are 
comparable. At least parametrically, in a nearly perfect fluid there 
is no room for kinetic theory, that means there is no regime in 
which kinetic theory is more accurate than fluid dynamics. 

 The collision term is determined by the quasi-particle cross section
$\sigma$, and a rough estimate of the mean free path is given by 
$l_{\it mfp}=1/(n\sigma)$. Using equ.~(\ref{eta_mfp}) we find $\eta 
\sim \bar{p}/\sigma$. This result has two interesting consequences:

\begin{enumerate}
\item The viscosity of a dilute gas is independent of its density. 
The physical reason for this behavior is that viscosity is determined
by the rate of momentum diffusion. The number of particles is 
proportional to $n$, but the mean free path scales as $1/n$. As a 
result, the diffusion rate is constant. Maxwell was so surprised by 
this result that he tested it by measuring the damping rate of a 
torsion pendulum in a sealed container as a function of the air 
pressure \cite{Maxwell:1986,Brush:1986}. He confirmed that $\eta$ 
is not a function of $P$ at fixed $T$. Of course, if the air is 
very dilute then $l_{\it mfp}>L$ and the hydrodynamic description 
breaks down. In this limit, known as the Knudsen regime, damping is 
proportional to pressure.

\item The result $\eta\sim 1/\sigma$ also implies that viscosity
of a weakly coupled gas is very large. This is counter-intuitive
because we think of viscosity as friction between fluid layers. 
Consider a fluid sheared between two parallel plates in the $xz$
plane. The force per unit area is 
\be 
 \frac{F}{A} = \eta\nabla_y u_x\, . 
\ee
We naively expect this force to grow with the strength of the interaction. 
Our intuition is shaped by very viscous fluids, for which viscosity 
is indeed determined by force chains and solid friction. This expectation 
is not entirely inappropriate, because the word viscosity is derived
from the Latin word for mistletoe, viscum album.

\end{enumerate}

\subsection{Matching and Kubo relations}
\label{sec_kubo}

 In the case of kinetic theory we can derive the equations of
fluid dynamics from the underlying microscopic theory. In more 
complicated cases, for example if the short distance description 
is a strongly coupled field theory, this may not be possible. 
In that case we can rely on the fact that fluid dynamics is 
a general long distance effective theory, and compute the 
transport coefficients based on the idea of matching. Matching 
expresses the requirement that in the regime of validity of 
the effective theory, correlation functions must agree with 
correlators in the microscopic theory. Consider the retarded
correlation function of the stress tensor
\be
\label{G_R}
G_R^{xyxy}(\omega,{\bf k}) = 
     -i \int dt \int d^3 x\, e^{i\omega t - i{\vec{k} \cdot \vec{x}}} 
   \Theta(t) \langle [\Pi^{xy}(t,\vec{x}), \Pi^{xy}(0,0)]\rangle \, . 
\ee
In linear response theory this function controls the stress 
induced by an external strain. In fluid dynamics $\Pi_{xy}\simeq
\rho u_xu_y$ and we can compute the correlation function from 
linearized hydrodynamics and fluctuation relations. We find$^9$
\be 
\label{Kubo}
 G_R^{xyxy}(\omega,k) = P -i\eta \omega +\tau_R\eta \omega^2
  - \frac{\kappa_R}{2} k^2 + O(\omega^3,\omega k^2)\, , 
\ee
where $\tau_R$ is the relaxation time defined in equ.~(\ref{pi_relax})
and $\kappa_R$ is another second order transport coefficient
\cite{Chafin:2012eq}. Equ.~(\ref{Kubo}) implies the Kubo
relation
\be 
\label{Kubo_eta}
\eta = -\lim_{\omega\to 0}\lim_{k\to 0}
 \, \frac{d}{d\omega}\,{\rm Im}\,G^{xyxy}_R(\omega,\vec{k})\, .
\ee
This equation can be applied to field theory, on the basis of 
equ.~(\ref{G_R}) and the microscopic definition of the stress tensor. 
This method is used to compute transport coefficients on the lattice, 
in both relativistic and non-relativistic field theories 
\cite{Karsch:1986cq,Meyer:2007ic,Sakai:2007cm,Aarts:2007wj,Meyer:2011gj,Wlazlowski:2013owa}. 
The difficulty with using Kubo's formula is that imaginary time 
Monte Carlo simulations do not provide direct access to correlation 
functions for real frequencies. Measuring the shear viscosity 
requires analytic continuation of imaginary time data, which leads 
to uncertainties that are difficult to quantify. We note that some 
transport coefficients, like the parameter $\kappa_R$ in equ.~(\ref{Kubo}) 
can be measured directly from imaginary time data. 

 Equation (\ref{Kubo}) confirms that the expansion parameter 
of the hydrodynamic expansion is $\omega/\omega_{\it fluid}$ with 
$\omega_{\it fluid}\simeq P/\eta\simeq \tau_R^{-1}$. Note that 
fluctuations introduce non-analytic$^{10}$ terms at order $\omega^{3/2}$
\cite{Kovtun:2011np,Chafin:2012eq}. This is a breakdown of the 
gradient expansion, but not a breakdown of hydrodynamics. For 
example, at second order in the low energy expansion the 
$\omega^{3/2}$ term is completely determined by $\eta$ and $P$, 
and the relaxation time $\tau_R$ can be extracted by matching
$G_R(\omega)$ to the low energy expansion in fluid dynamics.

\subsection{Microscopic models of fluids: Holography}
\label{sec_ads_cft}

  Kinetic theory provides explicit theoretical realizations of weakly 
coupled fluids. Holographic dualities and the AdS/CFT correspondence 
have led to controlled realizations of strongly coupled fluids. The 
basic idea originated from the study of black holes. It had been 
known for some time that black holes have entropy, and that the
process of a perturbed black hole settling down to a stationary
configuration bears some resemblance to dissipative relaxation 
in fluids. Indeed, it was shown that one can assign a shear viscosity 
and electric conductivity to the ``stretched horizon'', an imaginary 
surface that hovers just above the event horizon \cite{Thorne:1986}.

 These ideas were made precise in the context of the AdS/CFT 
correspondence \cite{Maldacena:1997re}, see the reviews
\cite{Son:2007vk,Gubser:2009md,CasalderreySolana:2011us,DeWolfe:2013cua}. 
In the simplest case one considers a Schwarzschild black hole embedded 
in five dimensional Anti-de-Sitter (AdS$_5$) space. The full spacetime 
has additional compact dimensions, which are required by string theory 
but play no role in our discussion. Black holes in AdS$_5$ do not 
evaporate and the black hole is in thermal equilibrium. This means 
that the rate of Hawking radiation balances the amount of energy 
falling back into the black hole. Based on its causal structure we 
can view AdS$_5$ as having a ``boundary''which is four-dimensional 
Minkowski space. Matter on the boundary is in thermal equilibrium
with the black hole spacetime.

The AdS/CFT correspondence asserts that the boundary is described
by an ordinary quantum field theory, and that the correlation 
functions of this field theory have a dual description in terms
of boundary correlation functions of a gravitational theory in
the bulk. The correspondence is simplest if the boundary theory 
is strongly coupled and contains a large number $N$ of degrees
of freedom. In this case the bulk theory is simply classical 
Einstein gravity. The partition function of the boundary quantum 
field theory (QFT) is 
\be 
 Z_{\it QFT}[J_i]=\exp\left(-S\left[\left.\phi_i\right|_{\partial{\it M}}
= J_i\right]\right)\, , 
\ee
where $J_i$ is a set of sources in the field theory, $S$ is the 
gravitational action, $\phi_i$ is a dual set of fields in
the gravitational theory, and $\partial{\it M}$ is the boundary 
of $AdS_5$. The fields $\phi_i$ satisfy classical equations of 
motions subject to boundary conditions on $\partial{\it M}$.

The original construction involves a black hole in AdS$_5$ and is dual 
to a relativistic fluid governed by a generalization of QCD known as 
${\cal N}=4$ super Yang-Mills theory. This theory is considered in 
the limit of a large number of colors $N_c$. The gravitational theory 
is Einstein gravity with additional matter fields that are not 
relevant here. The AdS$_5$ black hole metric is  
\be
\label{bh_son}
ds^2 = \frac{(\pi T R)^2}{u}  \left(-f(u) dt^2 + d\vec{x}^2 \right) + 
  \frac{R^2}{4 u^2 f(u)} du^2\, ,
\ee
where $\vec{x},t$ are Minkowski space coordinates, and $u$ is a 
``radial'' coordinate where $u=1$ is the location of the black hole 
horizon and $u=0$ is the boundary. $T$ is the temperature, $R$ is the 
AdS radius, and $f(u)=1-u^2$. In the boundary theory the metric couples
to the stress tensor $\Pi_{\mu\nu}$. Correlation functions of the 
stress tensor can be found by linearizing the bulk action around
the AdS$_5$ solution, $g_{AB}=g_{AB}^0+\delta g_{AB}$, where $A,B=1,
\ldots,5$. Small oscillations of the off-diagonal strain $\delta
g_x^y$ are particularly simple. We consider harmonic dependence 
on the Minkowski coordinates $\delta g_x^y=\phi_k(u)e^{ikx-
i\omega t}$. Fluctuations are governed by the wave equation
\be
\label{lingrav}
\phi_k''(u) - \frac{1+u^2}{uf(u)} \phi_k'(u) 
      + \frac{\omega^2 -k^2f(u)}{(2\pi T)^2 u f(u)^2}
  \phi_k(u) = 0\, .
\ee
This differential equation has two linearly independent solutions.
The retarded correlation function corresponds to picking a solution 
that is purely infalling at the horizon \cite{Son:2007vk}. For small 
(or very large) $\omega,k$ this solution can be found analytically
\cite{Policastro:2002se,Teaney:2006nc}. $G_R(\omega,k)$ is computed 
by inserting the solution into the Einstein-Hilbert action, and then 
computing the variation with respect to the boundary value of 
$\delta g_x^y$. The result$^{11}$ is of the form given in equ.~(\ref{Kubo}) 
with \cite{Policastro:2001yc,Baier:2007ix}
\be 
P=\frac{sT}{4}, \hspace{0.5cm}
\eta= \frac{s}{4\pi}, \hspace{0.5cm}
\tau_R = \frac{2-\log(2)}{2\pi T}\, . 
\ee
Note that in the case of a relativistic fluid $\eta$ is naturally
expressed in units of the entropy density $s$, not the density
$n$. This is because a relativistic fluid need not have a conserved
particle number. As a rough comparison we can use the fact that 
for a weakly interacting relativistic gas $s/n=3.6$. We observe
that the AdS/CFT correspondence describes a very good fluid. In 
particular, $\eta/s<1$ and $\tau_R\sim T^{-1}$. This is a remarkable
result because the AdS/CFT correspondence has provided the first 
reliable theoretical description of a nearly perfect fluid. 

 There are many aspects of the strongly coupled fluid that can 
be studied using AdS/CFT:

\begin{enumerate}
\item The spectral function$^{12}$ $\eta(\omega)=-\frac{1}{\omega}{\it Im}
G_R(\omega)$ does not show evidence for quasi-particles
\cite{Teaney:2006nc,Kovtun:2006pf}. Instead of a Lorentzian of 
width $1/\tau_R$ one finds a smooth function that interpolates 
between the hydrodynamic limit $\eta(0)=\eta$ and the high 
frequency limit $\eta(\omega)\sim \omega^3$. Because of non-renormalization
theorems, the $\omega\to\infty$ limit is given by the correlation
function in free field theory.

\item The relaxation time can written as $\tau_R=c\,\eta/P$ with $c=
(2-\log(2))/2\simeq 0.65$. This value can be compared to the Israel-Stewart 
result $\tau_R=1.5\,\eta/P$. We observe that the relaxation time is very 
short, but in units of $\eta/P$ it is only a factor of 2.3 smaller than 
kinetic theory would predict. The AdS/CFT correspondence has also 
been used to compute other second order transport coefficients 
\cite{Baier:2007ix}. 

\item The validity of the hydrodynamic expansion is controlled
by the location of the poles of $G_R(\omega)$ in the complex 
$\omega$ plane. The hydrodynamic pole of the shear correlator 
is located at $\omega \simeq iD_\eta k^2$, where $D_\eta=\eta/(sT)$
is the momentum diffusion constant. Non-hydrodynamic poles correspond
to so-called quasi-normal modes of the linearized Einstein equations.
These quasi-normal modes come in complex conjugate pairs and are 
located at a minimum distance of order $T$ from the real axis
\cite{Starinets:2002br}. This observation confirms that the expansion 
parameter in a nearly perfect fluid is $\omega/T$.

\item Using the AdS/CFT correspondence one can study the approach
to equilibrium in great detail. For initial conditions that lead
to Bjorken flows the approach to hydrodynamics is very rapid. After 
the quasi-normal modes are damped, on time scales on the order of 
$(\tau T)\lsim 1$, the Navier-Stokes description is very accurate, 
even though non-equilibrium contributions to the pressure can be 
large \cite{Chesler:2010bi,Heller:2011ju}. This phenomenon is sometimes 
referred to as rapid ``hydrodynamization''. 

\item Heller et al.~studied the large order behavior of the hydrodynamic 
expansion for a Bjorken-like flow. They found that the gradient expansion 
is an asymptotic series, and that the radius of convergence is zero 
\cite{Heller:2013fn}. The coefficients of high order terms, and the 
leading singularity in the Borel plane, are governed by the lowest 
quasi-normal mode. We note that this phenomenon is unrelated to the 
non-analytic terms in the expansion mentioned above. The calculation is 
performed in the large $N_c$ limit of the field theory, so  
non-analytic terms in the gradient expansion are suppressed 
\cite{Kovtun:2003vj}. Heller et al.~speculate that the large order
behavior is analogous to the factorial divergence of large orders of 
perturbation theory in quantum field theory. 

\end{enumerate}

 Finally, we note that one can directly derive the equation of 
fluid dynamics by promoting the parameters that label the near
horizon metric to hydrodynamic variables \cite{Bhattacharyya:2008jc}.
Solving the resulting Einstein equations order-by-order in gradients
provides an alternative derivation of the second order transport
coefficients discussed above. This method provides a general 
connection between solutions of the Einstein equation and the 
Navier-Stokes equation, referred to as the fluid-gravity 
correspondence \cite{Rangamani:2009xk}.

\subsection{Viscosity bounds}
\label{sec_visc_bnd}

 The AdS/CFT correspondence provides an explicit, albeit somewhat
theoretical,  example of a nearly perfect fluid, leading to two 
questions: Can nearly perfect fluids be realized in the laboratory, 
and is there a fundamental limit to fluidity? We address the 
first question in Sects.~\ref{sec_nr} and \ref{sec_rel} below. There
are several arguments that the answer to the second question is 
affirmative. We summarize these arguments here: 

{\it Uncertainty relation} \cite{Danielewicz:1984ww}: Kinetic theory
predicts that $\eta= \frac{1}{3}nl_{\it mpf}\bar{p}$, and that low 
viscosity corresponds to a short mean free path. However, the uncertainty 
relation suggests that the product of the mean free path and the mean momentum 
cannot become arbitrarily small. Using $l_{\it mpf}\bar{p}\gsim 1$ implies 
$\eta/n\gsim 1/3$. This argument was originally presented in the context 
of relativistic fluids. In these systems the inverse Reynolds number is 
given by $\eta/(s\tau T)$. Using the entropy per particle of a weakly 
interacting relativistic Bose gas, $s/n=3.6$, we get $\eta/s\gsim 0.09$. 

 There are several issues with this argument. First, it is based on the 
application of kinetic theory in a regime where there are no well-defined
quasi-particles and the theory is not applicable. Second, there is
no obvious reason that the entropy per particle cannot be much larger 
than the free-gas value$^{13}$ \cite{Cohen:2007qr}. Finally, a bound on 
transport coefficients related to the uncertainty relation was first 
proposed by Mott in connection with electric conductivity \cite{Mott:1972}. 
A minimal conductivity implies that the metal-insulator transition 
must be continuous. However, this prediction is known to be false. 
Continuous metal-insulator transitions have been observed 
\cite{Paalanen:1982}, and the physical mechanism of these transitions
can be understood in terms of Anderson localization. 

{\it Holographic dualities} \cite{Kovtun:2004de}: The value $\eta/s=
1/(4\pi)$ is obtained in the strong coupling limit of a large class 
of holographic theories. These theories are characterized by the fact 
that the dual gravitational description involves the Einstein-Hilbert 
action \cite{Buchel:2003tz,Son:2007vk,Iqbal:2008by}. Kovtun, Son, and 
Starinets (KSS) conjectured that the strong coupling result is 
an absolute lower bound for the ratio $\eta/s$ in all fluids, 
\be 
 \frac{\eta}{s} \geq \frac{1}{4\pi}
\ee
This idea is a significant step forward compared to the argument
based on the uncertainty relation. The value $1/(4\pi)$ is the 
result of a reliable calculation. Holographic dualities explain 
why the relevant quantity is $\eta/s$, and they account for the 
difference between momentum and charge diffusion. The diffusion
constant goes to zero in the strong coupling limit
\cite{Herzog:2006gh,CasalderreySolana:2006rq,Gubser:2006bz}, whereas 
the ratio $\eta/s$ remains finite. 

 However, holographic theories exist
that provide counterexamples to the KSS conjecture 
\cite{Kats:2007mq,Cremonini:2011iq}. Finite coupling corrections 
increase the ratio $\eta/s$, but there are cases in which calculable 
finite $N_c$ corrections lower $\eta/s$. In terms of the dual 
description these theories correspond to gravitational theories that 
contain a certain higher derivative correction to the Einstein-Hilbert 
action known as the Gauss-Bonnet term \cite{Brigante:2007nu}. Although 
this result rules out the KSS conjecture, there are compelling arguments 
for a weaker version of the viscosity bound. Given that the violation of 
the KSS bound can be related to the Gauss-Bonnet term one has to
study constraints on the Gauss-Bonnet coefficient $\lambda_{\it GB}$.
It was found that large values of $\lambda_{\it GB}$ lead to 
violations of causality. For the class of theories that are 
known to violate the KSS bound causality implies the slightly
weaker bound  $\eta/s\geq \frac{16}{25}\frac{1}{4\pi}$
\cite{Buchel:2009tt}. It seems likely that this is not the 
final word from holographic dualities. Generalizations of 
Gauss-Bonnet gravity, so-called Lovelock theories, have been 
studied \cite{Camanho:2010ru}, and lower values of $\eta/s$ may be 
possible.  

{\it Fluctuations} \cite{Kovtun:2011np,Chafin:2012eq}: Shear
viscosity is related to momentum diffusion, and $\eta/s=0$ would
imply that mean free path for momentum transport is zero. However, 
in fluid dynamics momentum can also be carried by collective modes
such as sound and shear waves. Indeed, if the viscosity is small 
this process becomes more efficient because the damping rate of 
sound and shear modes is small. This observation suggests that the 
physical viscosity of the fluid cannot be zero. 

 This argument can be made more precise using the low energy 
expansion of hydrodynamic correlation functions. Fluctuations not 
only contribute to non-analytic terms in $G_R(\omega)$, but they 
also correct the polynomial terms that determine the transport 
coefficients. The retarded shear stress correlator in a relativistic 
fluid is of the form $G_R(\omega)=P+\delta P+i\omega (\eta+\delta 
\eta)+\ldots$ where $\delta P$ is a correction to the pressure and 
\be 
\label{eta_eff}
 \eta+\delta\eta = \eta + \frac{17}{120\pi^2}
  \frac{\Lambda_K D_\eta s^2 T^3}{\eta^2}\, . 
\ee
is the physical viscosity. Here, $\Lambda_K$ is the breakdown
momentum of the hydrodynamic description and $D_\eta=\eta/(sT)$
is the momentum diffusion constant. The gradient expansion 
requires $\Lambda_K D_\eta\lsim 1$. We observe that $\delta\eta
\sim 1/\eta^2$, so the physical viscosity cannot become 
arbitrarily small. The bound for $\eta/s$ depends on the 
equation of state. For a quark gluon plasma $\eta/s\gsim 
0.1$ \cite{Kovtun:2011np}, and in a non-relativistic Fermi 
gas $\eta/s\gsim 0.2$ \cite{Chafin:2012eq}.

 The bound is interesting, because it sheds some light on
what is special about shear viscosity. The stress tensor 
is quadratic in the fluid velocity and has a leading order, 
non-linear coupling to shear waves. Other currents do 
not have non-linear mode couplings at leading order. The
bound is not universal, but it is complementary to the 
holographic bounds in the sense that it only operates
at finite $N$, whereas the holographic bounds are rigorous
at infinite $N$. 

 It is difficult to summarize the situation regarding the 
proposed viscosity bounds. There is strong evidence that viscosity
is different from other transport coefficients. We can find systems
for which bulk viscosity, conductivity, or diffusion constants vanish, 
but there are physical effects, the universality of the graviton 
coupling in holographic theories, and the universality of the stress 
tensor in stochastic fluid dynamics, that make it difficult to find 
scenarios in which the shear viscosity vanishes. The precise value of 
the bound is not known, but empirically the value 
$\eta/s=1/(4\pi)$ found in simple holographic theories is a good 
approximation for the viscosity of the best quantum fluids that 
can be studied in the laboratory as discussed further below. 

\section{Non-relativistic Fluids}
\label{sec_nr}
\subsection{The unitary Fermi gas}
\label{sec_uni}

 In the following two sections we describe theoretical and 
experimental results regarding the transport properties of 
the two best fluids that have been studied in the laboratory
\cite{Schafer:2009dj}. These two fluids are ultracold atomic
Fermi gases magnetically tuned to a Feshbach resonance, and
the quark gluon plasma produced in relativistic heavy ion 
collisions at the relativistic heavy ion collider (RHIC) in 
Brookhaven, New York, and the large hadron collider (LHC) at 
CERN in Geneva, Switzerland. 

 Ultracold Fermi gases are composed of atoms with half-integer 
total spin. Experiments focus on alkali atoms such as $^6$Li.
These atoms can be confined in all-optical or magneto-optical 
traps. We  concentrate on systems in which two hyperfine
states are macroscopically occupied. Because the density 
and temperature are very low details of the atomic interaction
and the atomic structure are not resolved, and the two hyperfine
states can be described as the two components of a point-like
non-relativistic spin 1/2 fermion. The fermions are governed 
by the effective Lagrangian
\be 
\label{l_4f}
{\cal L} = \psi^\dagger \left( i\partial_0 +
 \frac{{\vec\nabla}^2}{2m} \right) \psi 
 - \frac{C_0}{2} \left(\psi^\dagger \psi\right)^2 \, . 
\ee
The coupling constants $C_0$ is related to the $s$-wave scattering 
length $a$. At low temperature and density neither higher partial 
waves nor range corrections are important. The two-body $s$-wave 
scattering matrix is
\be 
\label{M_nr}
{\cal M} =  \frac{4\pi}{m} \frac{1}{1/a+iq}\, ,
\ee
where $q$ is the relative momentum. The precise relation between
$C_0$ and $a$ depends on the regularization scheme. In dimensional
regularization $C_0=4\pi a/m$. In the limit of weak coupling this
result follows from the Born approximation. 

 Of particular interest is the ``unitarity'' limit $a\to\infty$.
In this limit the system has no dimensionful parameters and the theory
is scale invariant \cite{Castin:2011}. The scattering amplitude behaves 
as $1/(iq)$, which saturates the $s$-wave unitarity bound. The two-body 
wave function scales as $1/r$ and the many body system is strongly 
correlated even if the density is low. Experimentally, the unitarity 
limit can be studied using magnetically tuned Feshbach resonances 
\cite{Bloch:2007,Giorgini:2008}.

 We note that even at unitarity the dilute Fermi gas has well-defined
quasi-particles if the temperature is large. The average scattering 
amplitude scales as $\sigma\sim \langle q^{-2}\rangle \sim 
\lambda^2_{\it dB}$, where $\lambda_{\it dB}\sim (mT)^{-1/2}$ is the 
thermal wave length. In the high temperature limit the average cross 
section is small, and the collisional width of a fermion quasi-particle 
is $\Gamma\sim zT$ \cite{Schaefer:2013oba}, where $z=(n\lambda^3)/2\ll 1$ 
is the fugacity. In this regime the shear viscosity can be computed 
using kinetic theory$^{14}$. The result is \cite{Massignan:2004,Bruun:2005}
\be
\label{eta_aa}
\eta = \frac{15}{32\sqrt{\pi}}(m T)^{3/2}.
\ee 
As expected, the viscosity is independent of density and increases
with temperature. The ratio $\eta/n$ scales as $1/z$ and is 
parametrically large. We also find $\eta/s\sim1/(z\log(1/z))$.

 In the regime $z\gsim 1$ the unitary gas is strongly coupled. 
At $z\sim 12$ the system undergoes a phase transition to a superfluid 
\cite{Ku:2011}. In the superfluid phase the $U(1)$ symmetry of the 
effective Lagrangian equ.~(\ref{l_4f}) is spontaneously broken, and at 
low temperature there is a well defined bosonic quasi-particle related 
to the $U(1)$ Goldstone mode. Momentum diffusion due to Goldstone 
modes can be studied using kinetic theory, and we find $\eta\sim T^{-5}$
\cite{Rupak:2007vp}. Combined with equ.~(\ref{eta_aa}) this result 
indicates that the viscosity has a minimum in the vicinity of the 
critical temperature. In this regime there are no reliable calculations
of transport properties, but T-matrix calculations suggest that 
$\eta/n$ reaches a value of about $0.5$ \cite{Enss:2010qh}. We note 
that at $T_c$ the entropy per particle is very close to one. Lower 
values of the shear viscosity, $\eta/s\simeq 0.2$, have been found 
in quantum Monte Carlo calculations \cite{Wlazlowski:2013owa}. 

 In kinetic theory the viscosity spectral function has a Lorentzian
line shape with width $\tau_R^{-1}=P/\eta$ \cite{Braby:2010tk}. In 
the strongly coupled regime the shape of the spectral function is
not known, but one can determine the asymptotic behavior$^{15}$ for 
$\omega\to \infty$ as well as the frequency sum rule. The sum rule is 
given by \cite{Taylor:2010ju,Enss:2010qh}
\be 
\frac{2}{\pi} \int d\omega\, \left[ \eta(\omega)
  -\frac{\cal C}{15\pi\sqrt{m\omega}} \right] 
 = \frac{2}{3}\,{\cal E}\, ,
\ee
where ${\cal C}$ is a short distance coefficient known as the 
contact density, which measures the strength of short range 
correlations \cite{Tan:2005}, and the subtraction term inside
the integral corresponds to the high frequency tail of the spectral 
function \cite{Hofmann:2011qs}. In the high temperature limit 
${\cal C}=4\pi n^2\lambda^2_{\it dB}$, and one can check that the 
high frequency tail smoothly matches kinetic theory for $\omega
\sim T$. We can now identify the relevant scales that limit the 
fluid dynamic and kinetic descriptions, $\omega_{\it fluid}\sim zT$
and $\omega_{\it micro}\sim T$. For $z\ll 1$ we find the expected
hierarchy of scales, but in the strongly correlated regime 
both scales are comparable, and new theoretical methods are 
needed$^{16}$.

\subsection{Flow and viscosity}
\label{sec_flow_nr}

\begin{figure}[t!]
\begin{center}
\includegraphics*[width=15cm]{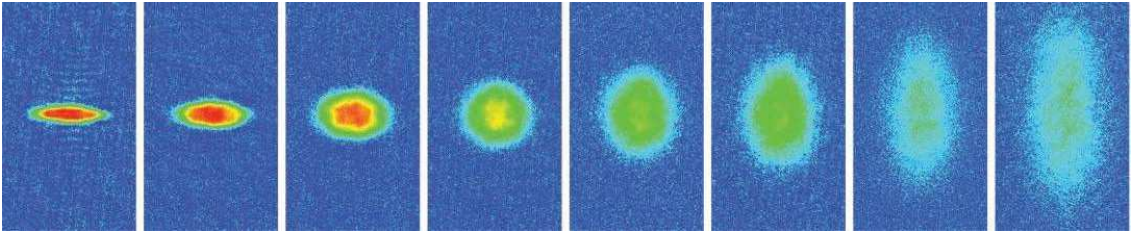}
\end{center}
\caption{\label{fig_cag_flow}
Expansion of a dilute Fermi gas at unitarity \cite{oHara:2002}.
The cloud contains $N\simeq 1.5\cdot 10^5$ $^6$Li atoms at a 
temperature $T\simeq 8\mu$K. 
The figure shows a series of false color absorption images taken 
between $t=(0.1-2.0)$ ms. The scale of the images is the same.
The axial size of the cloud remains nearly constant as the transverse
size is increasing. }
\end{figure}

 Fluid dynamics can be observed in experiments that involve 
releasing the gas from a deformed trap. In typical experiments
the trap corresponds to a harmonic confinement potential $V=
\frac{1}{2}m(\omega_\perp^2 x_\perp^2+\omega_z^2 z^2)$ with 
an aspect ratio $\omega_\perp/\omega_z\sim (20-30)$. In hydrostatic
equilibrium pressure gradients along the transverse direction are 
much larger than pressure gradients along the longitudinal direction. 
Hydrodynamic evolution after the gas is released converts this
difference into different expansion velocities, and during the late 
stages of the evolution the cloud is elongated along the transverse 
direction, see Fig.~\ref{fig_cag_flow}. The observation of this effect 
led to the discovery of nearly perfect fluidity in ultracold gases 
\cite{oHara:2002}. Shear viscosity counteracts the differential 
acceleration and leads to a less deformed final state. The shear 
viscosity can be measured by studying the time evolution of the 
cloud radii \cite{Schaefer:2009px,Cao:2010wa}. 

 An alternative approach is based on recapturing the gas after
release from the trap, which excites a transverse breathing mode. 
Hydrodynamic behavior can be verified by measuring the frequency of 
the collective mode. In an ideal fluid $\omega=\sqrt{10/3}\,
\omega_\perp$, whereas in a weakly collisional gas $\omega=2
\omega_\perp$ \cite{Stringari:2004,Bulgac:2004}. The transition 
from ballistic behavior in the weak coupling limit to hydrodynamic
behavior in the unitary gas has been observed experimentally
\cite{Kinast:2004b,Bartenstein:2004}. In the hydrodynamic regime 
damping of collective modes is governed by dissipative terms.
The rate of energy dissipation is 
\bea
\label{E_dis}
\dot{E} &=& - \int d^3x\, \Bigg\{\frac{1}{2}\,\eta(x)
     \left(\sigma_{ij}\right)^2
   + \zeta(x)\, \langle\sigma\rangle^2
   + \frac{\kappa(x)}{T} (\vec\nabla T)^2\Bigg\} \, . 
\eea
At unitarity the system is scale invariant and the bulk viscosity 
is predicted to vanish \cite{Son:2005tj,Castin:2011}. This prediction 
was experimentally verified in \cite{Elliott:2013}. Thermal conductivity 
does not contribute to damping because the gas is isothermal. As 
a consequence the damping rate is a measure of shear viscosity.

\begin{figure}[t!]
\begin{center}
\includegraphics*[width=9cm]{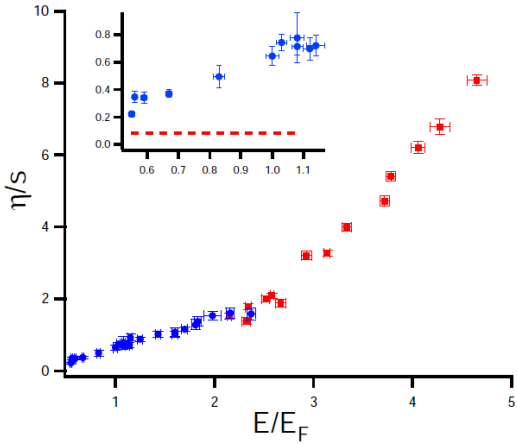}
\end{center}
\caption{\label{fig_eta_nr}
Measurements of $\eta/s$ in the dilute Fermi gas at unitarity
by use of collective modes (blue circles) and elliptic flow (red 
squares), from \cite{Cao:2010wa}. The data are shown as a function 
of the total energy of the clouds in units of $E_F$, the energy of 
a zero temperature Fermi gas with the same number of particles. 
At high temperature $E/E_F$ is proportional to temperature. Note 
that $\eta/s$ in the plot refers to a ratio of trap integrated 
quantities, $\langle\eta\rangle/\langle s\rangle$. }
\end{figure}

 Both the expansion and the collective mode experiments involve
approximate scaling flows$^{17}$. The motion is analogous to the Hubble 
flow in cosmology, and to the Bjorken expansion of a quark gluon plasma
(QGP). Consider the Euler equation for the acceleration of an ideal 
fluid, $\dot{\vec{u}}\simeq -\vec{\nabla}P/\rho=-\vec{\nabla}\mu/m$, 
where we have used the Gibbs-Duhem relation $dP=nd\mu$. Because the 
external potential is harmonic, the chemical potential is harmonic, 
too. As a consequence the velocity field is linear, and the cloud
expands in a self-similar fashion. Because the fluid velocity is 
linear the shear stress $\sigma_{ij}$ is spatially constant and 
the rate of dissipation is sensitive to the spatial integral of 
$\eta(x)$
\be
\label{alpha_n}
\langle \eta \rangle = \int d^3x \, \eta(x)\, .
\ee
Using measurements of the trap integrated entropy we can extract 
the ratio $\langle \eta\rangle/\langle s\rangle$. This analysis 
was originally performed in \cite{Schafer:2007pr,Turlapov:2007}.
A more recent analysis that combines collective mode data at low
$T$ with expansion data at high $T$ is shown in Fig.~\ref{fig_eta_nr} 
\cite{Cao:2010wa}. The high temperature data matches expectations 
from kinetic theory$^{18}$. The viscosity drops with $T$ and the ratio 
of trap averages reaches $\langle \eta\rangle/\langle s\rangle\lsim 
0.4$.

 It is clearly desirable to unfold these measurements and 
determine local values of $\eta/s$. The main difficulty is a reliable
treatment of the low density corona. In this regime $\eta$ is 
independent of density and the integral in equ.~(\ref{alpha_n}) 
is ill defined, signaling the breakdown of fluid dynamics in 
the dilute region. The problem also appears if one applies the
Navier-Stokes equation to an expanding gas cloud. In the dilute 
regime $\eta$ is not a function of density and the viscous stresses 
$\eta\sigma_{ij}$ are independent of position, implying that 
although ideal stresses propagate with the speed of sound, viscous 
stresses propagate with infinite speed. As discussed in 
Sect.~\ref{sec_eft} this problem can be solved by including a 
finite relaxation time \cite{Bruun:2007,Schaefer:2009px}. In the 
low density regime the viscous relaxation time $\tau_R\simeq
\eta/(nT)$ is large. Because the dissipative stresses are zero 
initially, taking a finite relaxation time into account suppresses 
the contribution of the corona$^{19}$. A schematic version of this idea 
was used in Cao et al.~\cite{Cao:2010wa}, but a more systematic
treatment is needed. 

\section{Relativistic Fluids}
\label{sec_rel}
\subsection{The quark gluon plasma}
\label{sec_qgp}

 The QGP is a hot and dense systems of quarks and gluons governed
by the QCD Lagrangian 
\be
\label{l_qcd}
 {\cal L } =  - \frac{1}{4} G_{\mu\nu}^a G_{\mu\nu}^a
  + \sum_f \bar{q}_f ( i\gamma^\mu D_\mu - m_f) q_f\, ,
\ee
where $G_{\mu\nu}^a = \partial_\mu A_\nu^a - \partial_\nu A_\mu^a  + 
gf^{abc} A_\mu^b A_\nu^c$ is the QCD field strength tensor, $g$ is the 
coupling constant and $f^{abc}$ are the $SU(3)$ structure constants. 
The covariant derivative acting on the quark fields is $i D_\mu q = 
( i\partial_\mu + g A_\mu^a \frac{\lambda^a}{2}) q$ and $m_f$ is the 
quark mass. At the temperature scale probed in RHIC or LHC experiments 
the three light flavors, up, down, and strange, are thermally populated, 
whereas the heavy flavors, mainly charm and bottom, are produced 
in hard collisions and can serve as probes of the medium. 

 Asymptotic freedom implies that at very high temperature the 
QGP can be described in terms of quark and gluon quasi-particles.
A typical gluon has a thermal momentum of order $T$. Soft gluons
with momenta much lower than $T$ are modified by the interaction with 
hard particles. As a consequence, electric gluons acquire a Debye
screening mass $m_D\sim gT$. In perturbation theory there is no
static screening of magnetic fields, but magnetic gluons are
dynamically screened for momenta greater than $(m_D^2\omega)^{1/3}$, 
where $\omega$ is the frequency. The static magnetic sector 
of QCD is non-perturbative even if the temperature is very large.
Confinement in three-dimensional pure gauge theory generates
a mass scale of order $g^2T$. This mass scale determines the 
magnetic screening scale in the QGP, $m_M\sim g^2T$. 

 Perturbation theory in the quark gluon plasma is based on the 
separation of scales $m_M\ll m_D \ll T$. Strict perturbation theory 
in $g$ works only for very low values of the coupling constant, 
$g\lsim 1$ \cite{Kajantie:2002wa}. However, quasi-particle 
models that rely on the separation of scales describe the thermodynamics 
of the plasma quite well, even for temperatures close to the phase 
transition to a hadronic gas \cite{Blaizot:2003tw}. 

 The dispersion relation for the bosonic modes in the plasma 
evolves smoothly from quasi-gluons with masses $m\sim m_D$ at 
momenta $q\gsim gT$ to collective oscillations, plasmons, at 
low $q$. The energy of the plasmon in the limit $q\to 0$ is 
$\omega_P=m_D/\sqrt{3}$, and the plasmon width is $\Gamma \sim 
g^2T$ \cite{Braaten:1990it}. The calculation of the collisional
width of quasi-particles with momenta of order $T$ is a complicated,
non-perturbative problem,  but the width remains parametrically 
small, $\Gamma\sim g^2\log(1/g)T$ \cite{Blaizot:1999fq}.

 Momentum diffusion is controlled by binary scattering between
quarks and gluons. The cross section is proportional to $g^4$, and 
the IR divergence due to the exchange of massless gluons is regulated
by dynamic screening. As a consequence the shear viscosity scales
as $\eta\sim T^3/(g^4\log(1/g))$. A detailed calculation$^{20}$ in 
$N_f=3$ QCD gives \cite{Baym:1990uj,Arnold:2000dr,Arnold:2003zc}
\be
\label{eta_qcd}
 \eta = \frac{kT^3}{g^4\log(\mu^*/m_D)},
\ee
where $k=106.67$. The scale inside the logarithm is sensitive 
to bremsstrahlung processes such as $gg\to ggg$. Arnold et al.~found
$\mu^*=2.96T$ \cite{Arnold:2000dr,Arnold:2003zc}. The time scale 
for momentum diffusion is $\eta/(sT)\sim 1/(g^4\log(1/g)T)$. This 
scale is parametrically large, but the precise value is very sensitive 
to the coupling constant. In $N_f=3$ QCD we get $\eta/s\simeq 9.2/
(g^4\log(1/g))$. Using $g\simeq 2$, which corresponds to $\alpha_s
\simeq 0.3$, and $\log(1/g)\gsim 1$ we conclude that $\eta/s\lsim
0.6$. 

 At $T\simeq 150$ MeV the quark gluon plasma undergoes a crossover 
transition to a hadronic resonance gas \cite{Aoki:2006br,Bazavov:2011nk}.
The resonance gas is strongly coupled, but as the temperature is 
reduced further the system evolves to a weakly coupled gas of mostly 
pions, kaons, and nucleons. The viscosity of a pion gas is parametrically
large, $\eta/s \sim (f_\pi/T)^4$, where $f_\pi\simeq 93$ MeV is the 
pion decay constant \cite{Prakash:1993bt}. Similar to the arguments
in the case of cold Fermi gases we therefore expect that $\eta/s$ 
has a minimum in the vicinity of $T_c$. In this regime the only 
reliable theoretical approach is lattice gauge theory. As in the 
case of non-relativistic fermions the calculations are difficult
because one has to extract the viscosity spectral function from 
imaginary time data. In the case of pure gauge theory Meyer finds
$\eta/s=0.102(56)$ at $T= 1.24T_c$ and $\eta/s=0.134(33)$ at $T= 1.65T_c$
\cite{Meyer:2007ic}.

 Useful constraints on the spectral function are provided by sum 
rules. Romatschke and Son showed that \cite{Romatschke:2009ng} 
\be 
\label{qcd_sr}
\frac{2}{\pi} \int d\omega\, \left[ \eta(\omega)
  -\eta_{T=0}(\omega) \right] 
 = \frac{2}{5}\,{\cal E}\, ,
\ee
where $\eta_{T=0}(\omega)$ is the spectral function at zero temperature. 
The high frequency behavior can be studied in perturbation theory. We 
find $\eta(\omega)\sim \omega^3$ at both zero and non-zero temperature. 
Finite temperature effect were studied in \cite{Aarts:2002cc,Zhu:2012be}.
We note that in non-relativistic theories the tail of the spectral 
function is determined by short range correlations, whereas in a 
relativistic theory the high frequency behavior is determined by 
the $gg$ and $q\bar{q}$ continuum. In kinetic theory the shape of the 
spectral function at small frequency is a Lorentzian with a width
proportional to $1/\eta$. The lattice calculation in \cite{Meyer:2007ic} 
does not find a quasi-particle peak, but the resolution is insufficient
to draw a final conclusions. A spectral function that is broadly consistent 
with the existence of quasi-particles was observed in a study of the 
electric conductivity of the quark gluon plasma \cite{Ding:2010ga}.

\subsection{Flow, higher moments of flow, and viscosity}
\label{sec_flow}

\begin{figure}[t!]
\begin{center}
\includegraphics*[width=7.5cm,angle=-90]{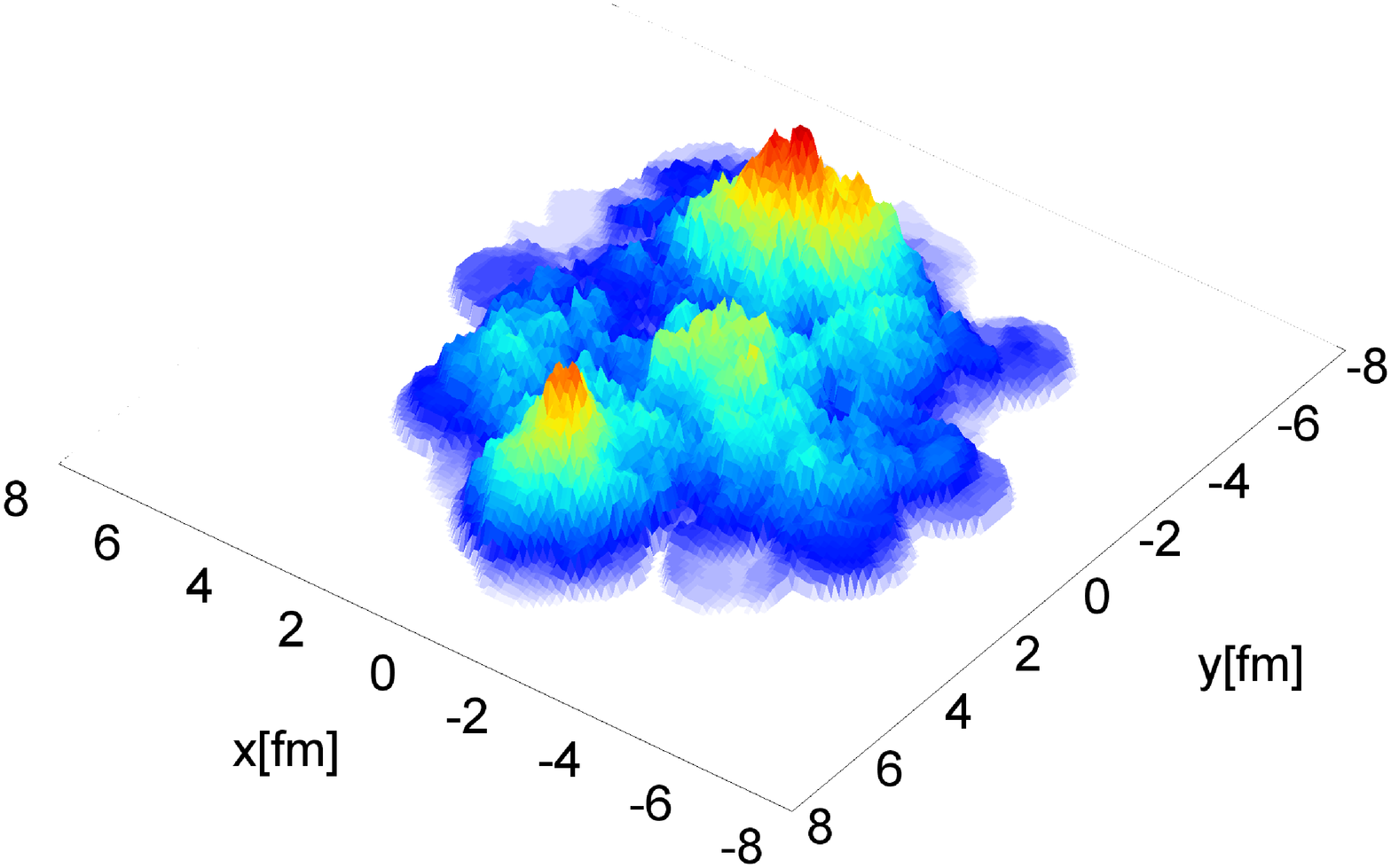}
\end{center}
\caption{\label{fig_init}
Initial energy density in a Au+Au collision at RHIC from the 
Monte-Carlo KLN model, see \cite{Schenke:2012wb,Kharzeev:2000ph}.
This model include the effects from the collision geometry, fluctuations
in the initial position of the nucleons inside the nucleus, and 
non-linear gluon field evolution. More sophisticated versions of the 
model also include quantum fluctuations of the gluon field. }
\end{figure}

 Experimental information about transport properties of the quark
gluon plasma comes from the observation of hydrodynamic flow in heavy 
ion collisions at collider energies \cite{Heinz:2009xj,Teaney:2009qa}.
Several observations support the assumption that heavy ion 
collisions create a locally thermalized system: 

\begin{enumerate}

\item The overall abundances of produced particles is described by 
a simple thermal model that depends on only two parameters, the 
temperature $T$ and the baryon chemical potential $\mu$ at freezeout 
\cite{Cleymans:1992zc,BraunMunzinger:2009zz}.

\item For transverse momenta $p_\perp \lsim 2$ GeV  the spectra
$dN/d^3p$ of produced particles follow a modified Boltzmann distribution
characterized by the freezeout temperature and a collective radial
expansion velocity \cite{Schnedermann:1993ws,Heinz:2009xj}. Radial 
flow manifests itself in the fact that the spectra of heavy hadrons, 
which acquire a larger momentum boost from the collective expansion, 
have a larger apparent temperature than the spectra of light hadrons.

\item In non-central collisions the azimuthal distribution 
of produced particles shows a strong anisotropy termed elliptic 
flow \cite{Ollitrault:1992,Heinz:2009xj}. Elliptic flow represents 
the collective response of the quark gluon plasma to pressure gradients 
in the initial state, which in turn are related to the geometry of 
the overlap region of the colliding nuclei, see Fig.~\ref{fig_init}.

\end{enumerate}

 Analysis of the azimuthal distribution is the main tool for 
constraining the shear viscosity of the plasma. We define
harmonics of the particle distribution 
\be
\label{v_2}
 \left. p_0\frac{dN}{d^3p}\right|_{p_z=0} = v_0(p_T)
 \Big( 1 + 2v_1(p_T)\cos(\phi-\Psi_1)
         + 2v_2(p_T)\cos(2\phi-\Psi_2) +\ldots \Big) ,
\ee
where $p_z$ is the longitudinal (beam) direction, $p_T$ is the 
transverse momentum, and $\phi$ is the angle relative to the impact 
parameter direction. The coefficient $v_2$ is known as elliptic flow, 
and the higher moments are termed triangular, quadrupolar, etc.~flow. 
The angles $\Psi_i$ are known as flow angles. Substantial elliptic 
flow, reaching about $v_2(p_T\!=\!2\,{\rm GeV}) \simeq 20\%$ in 
semi-central collisions, was discovered in the early RHIC data 
\cite{Adler:2003kt,Adams:2004bi} and confirmed at the LHC 
\cite{Aamodt:2010pa}. More recently, it was realized that fluctuations
in the initial energy density generates substantial higher harmonics, 
including odd Fourier moments such as $v_3$ \cite{Alver:2010gr}, 
and fluctuations of the flow angles relative to the impact parameter 
plane \cite{Alver:2008zza}.

 Viscosity tends to equalize the radial flow velocity and suppress
elliptic flow and higher flow harmonics. An estimate of the relevant
scales can be obtained from simple scaling solutions of fluid 
dynamics$^{21}$. The simplest solution of this type was proposed by 
Bjorken, who considered a purely longitudinal expansion \cite{Bjorken:1983}. 
Bjorken assumed that the initial entropy density is independent of 
rapidity, and that the subsequent evolution is invariant under boosts 
along the $z$ axis. The Bjorken solution provides a natural starting 
point for more detailed numerical and analytical studies 
\cite{Heinz:2009xj,Gubser:2010ui}. Bjorken flow is characterized 
by a flow profile of the form $u_\mu = \gamma(-1,0,0,u_z)= (-t/\tau,
0,0,z/\tau)$, where $\gamma=(1-u_z^2)^{1/2}$ is the boost factor and 
$\tau=(t^2-z^2)^{1/2}$ is the proper time. This velocity field solves 
the relativistic Navier-Stokes equation. Energy conservation then 
determines the evolution of the entropy density. We find
\be
\label{bj_ns}
-\frac{\tau}{s}\frac{ds}{d\tau} = 
  1 - \frac{4}{3}\frac{\eta}{sT\tau}\, ,
\ee
where we have neglected bulk viscosity. In ideal hydrodynamics 
$s\sim T^3$ and $T\sim 1/\tau^{1/3}$. The validity of the gradient 
expansion requires that the viscous correction is small 
\cite{Danielewicz:1984ww}
\be
\label{DG}
\frac{\eta}{s} \ll \frac{3}{4}(T\tau)\, .
\ee
It is usually assumed that in the QGP $\eta/s$ is approximately
constant. For the Bjorken solution $T\tau\sim \tau^{2/3}$ increases 
with time, and equ.~(\ref{DG}) is most restrictive during the 
early stages of the evolution. Using an equilibration time $\tau_0
=1$ fm and an initial temperature $T_0=300$ MeV gives $\eta/s\lsim
0.6$.  We conclude that fluid dynamics can be applied to heavy ion 
collisions only if the QGP behaves as a nearly perfect fluid. 

\begin{figure}[t!]
\begin{center}
\includegraphics*[width=10cm]{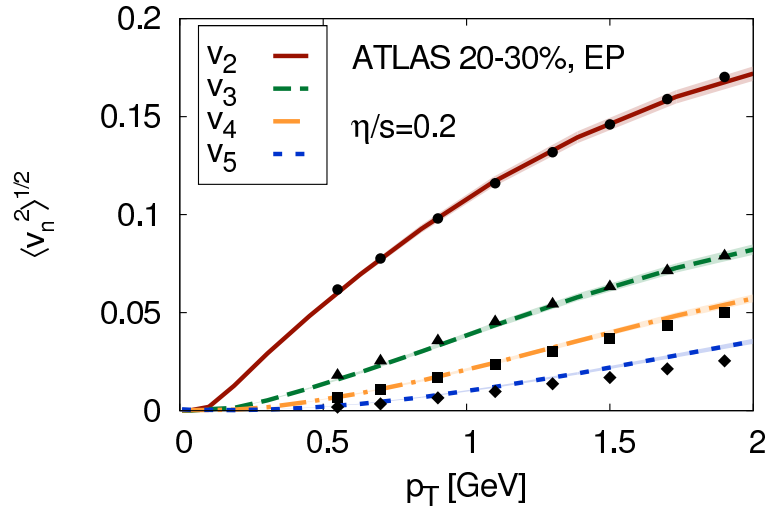}
\end{center}
\caption{\label{fig_vn}
Fourier coefficients $v_2,\ldots,v_5$ of the azimuthal charged
particle distribution as a function of the transverse momentum
$p_T$ measured in $Pb+Pb$ collisions at the LHC \cite{ATLAS:2012at}
The lines show a hydrodynamic analysis performed using $\eta/s=0.2$
\cite{Gale:2012rq}.}
\end{figure}

 At late time the expansion becomes three dimensional and $T\tau$ 
is independent of time. The fluid is composed of hadronic resonances
that have cross sections that reflect hadronic sizes and are 
approximately independent of energy. In that case $\eta\sim T/
\sigma$. Using $s\sim T^3$ and $T\sim 1/\tau$ we find that the 
dissipative correction $\eta/(sT\tau)$ increases with proper time as 
$\tau^2$. This result shows that fluid dynamics also breaks down at late 
times. At RHIC and LHC energies the duration of the fluid dynamic 
phase is 5-10 fm/c, depending on collision energy and geometry. 
We note that in contrast to the situation in heavy ion collisions 
there is no freeze-out in the cold atomic gas experiments. At unitarity 
the mean cross section increases as the temperature drops, and the 
fluid parameter $\eta/(nT\tau)$ is approximately constant during 
the evolution. 

 In heavy ion collisions we can observe only the final distribution
of hadrons. In principle one could imagine reconstructing azimuthal
harmonics of the stress tensor from the measured particle distribution, but 
doing so would require very complete coverage and particle identification, 
and it has not been attempted. In any case, hadrons continue to interact
after the fluid freezes out, and some rearrangement of momentum takes 
place. This means that we need a prescription for converting hydrodynamic
variables to kinetic distribution functions. What is usually done is 
that on the freezeout surface the conserved densities in fluid dynamics
are matched$^{22}$ to kinetic theory \cite{Cooper:1974mv}. 

 In ideal fluid dynamics the distribution functions are Bose-Einstein
or Fermi-Dirac distributions characterized by the local temperature
and fluid velocity. Viscosity modifies the stress tensor, and via 
matching to kinetic theory this modification changes the distribution 
functions $f_p$. The value of $\eta/s$ constrains only the $p_iv_j$ 
moment of the distribution function. The full distribution function
can be reconstructed only if the collision term is specified. Using 
the BGK collision term one obtains a very simple formula for the leading 
correction $\delta f_p$
\be
\label{del_f}
\delta f_p = \frac{1}{2T^3} \frac{\eta}{s}f_0(1\pm f_0)
   p_\alpha p_\beta  \sigma^{\alpha\beta} \, ,
\ee
where the $\pm$ sign refers to Bose/Fermi distributions. This result 
is a reasonable approximation to more microscopic theories 
\cite{Arnold:2000dr}. The shift in the distribution function leads to 
a modification of the single particle spectrum. In the case of the 
Bjorken expansion and at large $p_T$ we find
\be
\label{del_N}
 \frac{\delta (dN)}{dN_0} = \frac{1}{3\tau_fT_f}\frac{\eta}{s}
    \left( \frac{p_T}{T_f} \right)^2 \,,
\ee
where $dN_0$ is the number of particles produced in ideal fluid 
dynamics, $\delta (dN)$ is the dissipative correction, and $\tau_f$ 
is the freezeout time. In a system with strong longitudinal expansion
viscous corrections tend to equalize the momentum flow by pushing 
particles to higher $p_T$. Because the single particle distribution
enters into the denominator of $v_2$ this effect tends to suppress 
$v_2$ at large $p_T$. The effect from the numerator, dissipative 
corrections due to the $\cos(2\phi)$ component of the radial flow,
act in the same direction \cite{Teaney:2003kp}. What is important is 
that corrections to the spectrum are controlled by the same parameter 
$\eta/(s\tau T)$ that governs the derivative expansion in fluid 
dynamics$^{23}$. This reflects the fact that in the regime in which 
kinetic theory can be matched to fluid dynamics we have ${\it Kn}\sim 
{\it Re}^{-1}$.

 We obtain several simple predictions that have been confirmed
by experiment \cite{Heinz:2013th}: Dissipative corrections increase 
with $p_T$, they are larger in small systems that freeze out 
earlier, and they are larger for higher harmonics that are more 
sensitive to gradients of the radial flow profile. Quantitative 
predictions that provide not only bounds on $\eta/s$ but also reliable 
measurements of transport properties of the plasma require a number 
of ingredients \cite{Gale:2013da}:

\begin{enumerate}

\item An initial state model that incorporates the nuclear geometry 
and fluctuations in the initial energy deposition. The simplest 
possibility is a Monte-Carlo implementation of the Glauber model 
\cite{Miller:2007ri}, but some calculations also include saturation 
effects, quantum fluctuations of the initial color field, and 
pre-equilibrium evolution of the initial field \cite{Gale:2012rq}. 
Alternatively, one may try to describe the pre-equilibrium stage 
using kinetic theory \cite{Martinez:2010sc,Petersen:2008dd} or 
the AdS/CFT correspondence \cite{vanderSchee:2013pia}. At the end 
of the initial stage the stress tensor is matched to fluid dynamics.

\item Second order dissipative fluid dynamics in 2+1 (boost invariant)
or 3+1 dimensions. Calculations must include checks to ensure
insensitivity to poorly constrained second order transport 
coefficients$^{24}$ and a realistic equation of state (EOS). A realistic
EOS has to match lattice QCD results at high temperature, and a 
hadronic resonance gas below $T_c$ \cite{Huovinen:2009yb}. The
resonance gas EOS must allow for chemical non-equilibrium effects
below the chemical freezeout temperature $T_{\it chem}\simeq T_c$.

\item Kinetic freezeout and a kinetic afterburner. At the kinetic
freezeout temperature the fluid is converted to hadronic distribution
functions. Ideally, these distribution functions are evolved further
using a hadronic cascade \cite{Bass:2000ib,Hirano:2005xf}, but at a 
minimum one has to include feed-down from hadronic resonance decays.

\end{enumerate}
 
 Initial estimates of $\eta/s$ from the RHIC data have been obtained in 
\cite{Romatschke:2007mq,Dusling:2007gi,Song:2007ux}. A more recent 
analysis of LHC data is shown in Fig.~\ref{fig_vn} \cite{Gale:2012rq}. 
The authors found $\eta/s\simeq 0.2$ at the LHC, and $\eta/s\simeq 0.12$ 
from a similar analysis of RHIC data. Similar results were obtained by 
other authors. Song et al.~reported an average value of $\eta/s\simeq (0.2
-0.24)$ at the LHC and $\eta/s\simeq 0.16$ at RHIC \cite{Song:2011qa}. 
Luzum and Ollitrault tried to constrain the allowed range of $\eta/s$, 
obtaining $0.07\leq\eta/s\leq 0.43$ at RHIC \cite{Luzum:2012wu}. Given the 
complexity of the analysis, uncertainties are difficult to quantify. A 
survey of the main sources of error in the determination of $\eta/s$ can 
be found in \cite{Song:2008hj}. Interestingly, the extracted 
values of $\eta/s$ are lower at RHIC than they are at the LHC, as one 
would expect based on asymptotic freedom. We emphasize, however, 
that given the uncertainties it is too early to make this statement with 
high confidence.

\section{Frontiers}
\label{sec_out}

 In absolute units the shear viscosity of the ultracold Fermi gas 
and the quark gluon plasma differ by more than 25 orders of magnitude 
\cite{Schafer:2009dj}. The approximate universality of $\eta/s$ in 
strongly coupled fluids, and the near agreement with the value predicted 
by the AdS/CFT correspondence in the strong coupling limit of a large
class of field theories is quite remarkable$^{25}$. Much work remains
to be done in order to determine to what extent this observation
can be made precise, and what it implies about the structure 
of strongly correlated quantum systems. In this outlook we can 
only give a very brief summary of some of these issues. 

\subsection{Transport coefficients} There is an ongoing effort to 
map out the full density and temperature dependence of $\eta/s$
in both the ultracold gases and the quark gluon plasma, and to
determine other transport coefficients, like the bulk viscosity
and diffusion coefficients. There are a number of experimental
puzzles that remain to be addressed$^{26}$. In the case of heavy ion 
collisions, nearly ideal flow is even more pervasive than one 
would expect. Strong flow is also observed in photons, electrons
from heavy quark decays, and hadrons emitted in high-multiplicity
$p+Pb$ collisions at LHC energies, see \cite{Huovinen:2013wma} for 
a recent summary and original references. In the case of cold atomic 
gases we now have very accurate data for the dependence of $\langle
\eta\rangle$ on the total energy of the cloud \cite{Elliott:2013b}. 
These data have not been unfolded. It was observed that the scaling 
of $\langle\eta\rangle$ with the total energy is remarkably simple, 
$\langle\eta\rangle/\langle n \rangle \sim aE+bE^3$ for all energies 
above the critical point, but the origin of this scaling behavior 
is not understood. 

\subsection{Quasi-particles} We would like to understand whether nearly 
perfect fluidity, $\eta/s\sim 1/(4\pi)$, necessarily implies the 
absence of quasi-particles, as is the case in the AdS/CFT correspondence.
The most direct way to study this issue is to determine the spectral 
function. Since the only local probe of the stress tensor is the 
graviton, this will likely require numerical studies. We are also 
interested in pushing weak coupling descriptions into the regime 
where the quasi-particle picture breaks down, for example by 
using the renormalization group. 

\subsection{Viscosity bound} Whether there is a fundamental lower 
limit for $\eta/s$ is unknown. Part of the issue may well be that we need
to define more carefully what we mean by a fluid, and that we need to
understand how these defining characteristics are reflected in microscopic
theories. We would also like to know what kinds of theories have 
holographic duals, and what aspects of the field theory lead to 
the emergence of certain universal features, such as a shear viscosity 
to entropy density ratio that saturates the holographic bound $\eta/s
=1/(4\pi)$.

\subsection{Other strongly correlated fluids} In addition to the two 
fluids discussed in this review several other 
systems may be of interest. One interesting class
is two dimensional fluids, for example the electron gas in 
graphene \cite{Mueller:2009}, and the so-called strange metal 
phase of the high $T_c$ superconductors \cite{Guo:2010}. 

\subsection{Equilibration at strong and weak coupling} Empirical evidence
suggests that equilibration in heavy ion collisions takes place
on a very short time scale, $\tau_{\it eq}\sim 1$ fm. Rapid equilibration 
is natural in holographic theories \cite{Chesler:2010bi}, but it is 
difficult to make contact with asymptotic freedom and the well-established 
theory and phenomenology of parton distribution functions. Understanding
equilibration in weak coupling is a complicated problem that involves 
many competing scales, and even establishing the parametric dependence 
of the equilibration time on $\alpha_s$ is difficult, see
\cite{Kurkela:2011ti} for a recent overview. 

\subsection{Anomalous hydrodynamics} Several novel hydrodynamic
effects have been discovered in recent years. An example is the 
chiral magnetic effect. Topological charge fluctuations in the initial 
state of a heavy ion collision, combined with the magnetic field
generated by the highly charged ions, can manifest themselves in electric 
charge fluctuations in the final state  \cite{Kharzeev:2007jp}. This effect
is now understood as part of a broader class of anomalous hydrodynamic 
effects \cite{Kharzeev:2010gr}. Anomalous transport coefficients were 
originally discovered in the context of holographic dualities in 
\cite{Erdmenger:2008rm}, and interpreted using general arguments 
based on fluid dynamics in \cite{Son:2009tf}.

\vspace*{0.3cm}
Acknowledgments: This work was supported in parts by the US 
Department of Energy grant DE-FG02-03ER41260. We would like to 
acknowledge useful discussions with Peter Arnold, Harvey Meyer,
Guy Moore, Dam Son, Derek Teaney, and John Thomas.


\newpage
\noindent
{\bf\large Endnotes}

\vspace*{0.6cm}
\addcontentsline{toc}{section}{Endnotes: Fluid Dynamics }
\noindent
{\bf\large 1 Fluid dynamics}

\vspace*{0.4cm}
\noindent
{\bf\large 1.1 Fluid dynamics as an effective theory}

\vspace*{0.3cm}
\addcontentsline{toc}{subsection}{Hydrodynamic variables}
\noindent
{\it 1. Hydrodynamic variables:} In addition to the conserved charges 
there are two mores classes of hydrodynamics variables, Goldstone modes 
associated with spontaneously broken global symmetries, and order 
parameters near second order phase transitions. The simplest example 
of a Goldstone mode is the phase of the order parameter in a superfluid. 
In the dilute Fermi gas discussed in Sect.~\ref{sec_nr} the order 
parameter is $\langle\psi\psi\rangle = \rho e^{i\varphi}$. The low energy 
effective theory can be expressed in terms of gradients of $\varphi$.
The corresponding hydrodynamic variable is the superfluid velocity 
$\vec{u}_s=(\vec{\nabla}\varphi)/m$. The hydrodynamic description of 
he superfluid then involves two velocity fields, the normal velocity
$\vec{u}_n$ and the superfluid velocity $\vec{u}_s$. The momentum
density can be written as $\vec{\pi}=\rho_n \vec{u}_n+\rho_s\vec{u}_s$, 
where $\rho=\rho_n+\rho_s$ is the total mass density of the fluid. 
The theory of superfluid (two fluid) hydrodynamics was developed by 
Landau and Khalatnikov \cite{Khalatnikov:1965}. A new ingredient in 
hydrodynamic theories involving broken symmetries is the role of 
non-trivial commutation relations between the order parameter and the 
conserved charges. These commutators are implemented in fluid dynamics 
as non-trivial Poisson brackets \cite{Dzyaloshinski:1980}, which constrain 
the equation for the Goldstone modes.

 In QCD chiral symmetry is broken and in the limit that quarks are massless
the pion is a Goldstone mode. The hydrodynamic theory of pions is described 
in \cite{Son:1999pa,Son:2002ci}, but the theory is of somewhat limited value 
because the mass of the pion, $m_\pi\simeq 135$ MeV, is comparable to the 
breakdown scale of hydrodynamics. 

  Near a continuous phase transition fluctuations of the order parameter are 
large and the magnitude of the order parameter also becomes a hydrodynamic 
variable. Hydrodynamic theories near a second order phase transition can 
be classified according to the symmetries of the order parameter, and possible 
non-trivial Poisson brackets. The resulting theories are known as model A-J 
in the classification of Hohenberg and Halperin \cite{Hohenberg:1977ym}. The 
superfluid transition in the cold Fermi gas is described by model F, which 
also governs the lambda point in liquid Helium. A possible tri-critical point 
in QCD can be analyzed in terms of model H  \cite{Son:2004iv}, which also 
describes the endpoint of the liquid-gas transition in water.

\vspace*{0.4cm} 
\addcontentsline{toc}{subsection}{Mass current, momentum density, and 
relativistic fluids}
\noindent
{\it 2. Mass current, momentum density, and relativistic fluids:} In 
equ.~(\ref{hydro1},\ref{hydro2}) we have used that the mass current 
$\vec{\jmath}_\rho=\rho\vec{u}$, which appears in the conservation law 
$\partial_t\rho=-\vec{\nabla}\cdot\vec{\jmath}_\rho$, is equal to the 
momentum density, $\vec{\pi}$. This identification follows from very 
general arguments \cite{Landau:fluid}. It implies that there are no 
diffusive terms in the mass current, and provides an important constraint 
for quasi-particle theories, see equ.~(\ref{pi_constr}). 

 In relativistic hydrodynamics there need not be a conserved particle
number current. In this case the fluid four velocity $u_\mu$ is defined 
in terms of the energy current. In particular, we define $u_\mu$ to be 
the velocity of the frame in which the ideal stress tensor is diagonal. 
The ideal stress tensor is 
\be 
\Pi_{\mu\nu} = ({\cal E}+P)u_\mu u_\nu + P g_{\mu\nu}\, ,
\ee
where we use the convention $u^2=-1$. More formally, we can define $u^\mu$ 
through the condition $u^\mu\Pi_{\mu\nu}={\cal E}u_\nu$. This relation implies
that the energy current in the rest frame does not receive any dissipative 
corrections, $\Pi_{0i}=0$. The energy and momentum conservation laws are 
expressed through the relation $\nabla^\mu T_{\mu\nu}=0$. We can split 
this equation into longitudinal and transverse parts using the projectors
\be
\Delta_{\mu\nu}^{||}=-u_\mu u_\nu\, ,\hspace{0.5cm}
\Delta_{\mu\nu} = g_{\mu\nu}+u_\mu u_\nu\, . 
\ee
The longitudinal and transverse projections of $\nabla^\mu T_{\mu\nu}=0$
can be viewed as the equation of energy (or entropy) conservation and
the relativistic Euler equation, respectively. We get 
\be 
\nabla^\mu\left(su_\mu\right)=0\, ,\hspace{0.5cm}
Du_\mu = -\frac{1}{{\cal E}+P}\nabla_\mu^\perp P\, , 
\ee
where $D=u^\mu\nabla_\mu$ and $\nabla_\mu^\perp=\Delta_{\mu\nu}\nabla^\nu$.

 There are two basic possibilities for defining the fluid velocity in 
a theory with a conserved particle current $n_\mu$, such as the baryon current 
in QCD. The first option, called the Landau frame, is to define the fluid 
velocity in terms of the energy current. In this case there are dissipative 
corrections to the baryon current
\be 
 n_\mu = nu_\mu + \delta n_\mu\, ,
\ee
where, at leading order in the gradient expansion, $\delta n_\mu$ is 
related to the thermal conductivity \cite{Landau:fluid}. This choice is
convenient in the relativistic domain, but the non-relativistic limit is 
somewhat subtle. The other option, known as the Eckardt frame, corresponds
to defining the fluid velocity in terms of the particle current. In this 
case $n_\mu$ is non-diffusive, and the energy current contains dissipative 
corrections, in particular the thermal conductivity. 

\vspace*{0.4cm} 
\addcontentsline{toc}{subsection}{Second order fluid dynamics}
\noindent
{\it 3. Second order fluid dynamics:} The most general form of the 
stress tensor of a non-relativistic scale invariant fluid at second 
order in the gradient expansion was determined in \cite{Chao:2011cy}. 
The result is 
\bea 
\delta\Pi_{ij} &=& -\eta\sigma_{ij}
   + \eta\tau_R\left[
    g_{ik}\dot\sigma^{k}_{\; j} + u^k\nabla_k \sigma_{ij}
    + \frac{2}{3} \langle \sigma\rangle \sigma_{ij} \right] 
    + \lambda_1 \sigma_{\langle i}^{\;\;\; k}\sigma^{}_{j\rangle k} 
    + \lambda_2 \sigma_{\langle i}^{\;\;\; k}\Omega^{}_{j\rangle k}\nonumber\\
   && \mbox{} 
    + \lambda_3 \Omega_{\langle i}^{\;\;\; k}\Omega^{}_{j\rangle k}  
    + \gamma_1 \nabla_{\langle i}T\nabla_{j\rangle}T
    + \gamma_2 \nabla_{\langle i}P\nabla_{j\rangle}P
    + \gamma_3 \nabla_{\langle i}T\nabla_{j\rangle}P  \nonumber \\[0.1cm]
   \label{del_pi_fin}
   && \mbox{}
    + \gamma_4 \nabla_{\langle i}\nabla_{j\rangle}T 
    + \gamma_5 \nabla_{\langle i}\nabla_{j\rangle}P
    + \kappa_R  R_{\langle ij\rangle}\, . 
\eea
Here, ${\cal O}_{\langle ij\rangle}=\frac{1}{2}({\cal O}_{ij}+{\cal O}_{ji}
-\frac{2}{3}g_{ij}{\cal O}^k_{\;\;k})$ denotes the symmetric traceless 
part of a tensor ${\cal O}_{ij}$, $\Omega_{ij} = \nabla_iu_j-\nabla_ju_i$ 
is the vorticity tensor, and $R_{ij}$ is the Ricci tensor. This term
vanishes in flat space, but it is needed to establish the general form 
of the response function even in flat space, see equ.~(\ref{Kubo_2}).  
We note that equ.~(\ref{del_pi_fin}) contains 10 second order transport 
coefficients. This number is larger than the number of second order 
coefficients in the Burnett equation \cite{Garcia:2008}, despite the 
fact that we have imposed conformal symmetry. This is related to the 
fact that the Burnett equations were derived from kinetic theory, and 
that some transport coefficients allowed by the symmetries vanish in 
this framework. 

 For phenomenological applications it is useful to rewrite the second 
order equations as a relaxation equation for the viscous stress $\pi_{ij}
\equiv\delta\Pi_{ij}$. For this purpose we use the first order relation 
$\pi_{ij}=-\eta\sigma_{ij}$ and rewrite equ.~(\ref{del_pi_fin}) as 
\bea
\label{pi_rel}
\pi_{ij} &=& -\eta\sigma_{ij}
   - \tau_R\left[
    g_{ik}\dot\pi^{k}_{\; j} + V^k\nabla_k \pi_{ij}
    + \frac{5}{3} \langle \sigma\rangle \pi_{ij} \right] 
 \\
 & & \hspace{0.2cm}\mbox{}
    + \frac{\lambda_1}{\eta^2} \pi_{\langle i}^{\;\;\; k}\pi^{}_{j\rangle k} 
    - \frac{\lambda_2}{\eta} \pi_{\langle i}^{\;\;\; k}\Omega^{}_{j\rangle k}
    + \lambda_3 \Omega_{\langle i}^{\;\;\; k}\Omega^{}_{j\rangle k}  
    + \ldots \, , \nonumber 
\eea
where $\ldots$ refers to the terms proportional to $\gamma_i$ and $\kappa_R$. 
Note that this reformulation is not unique, and that it does not represent
a formal improvement over the second order equations. Theories of this
type can be derived in kinetic theory, and they can be used to restore 
causality and stability for perturbations of all wavelengths, not just 
wavelengths longer than the inverse breakdown scale of the hydrodynamic 
description.

 It is interesting to study the physical meaning of the different
terms in equ.~(\ref{pi_rel}). We begin with the term proportional to
$\tau_R$. Consider a non-zero strain $\sigma_{ij}$ which arises at 
time $t=0$. For simplicity assume the local rest frame and vanishing
bulk stress $\langle\sigma\rangle$. Then the stress tensor is $\pi_{ij}
= -\eta\sigma_{ij}(1-\exp(-t/\tau_R))$, which shows that $\tau_R$ is 
the time scale for dissipative stresses to relax to the Navier-Stokes 
value. The relaxation time also ensures that the front of a shear wave
propagates with a finite speed, see equ.~(\ref{v_max}).
In a periodically driven system $\tau_R$ determines the phase lag 
between the strain $\sigma_{ij}$ and the stress $\pi_{ij}$. Note that
the dissipated energy is proportional to $\sigma_{ij}\pi_{ij}$, and 
a non-zero phase lag reduces the amount of energy dissipated by 
viscous stresses. 

 The terms proportional to $\lambda_i$ describe non-linearities in
the stress-strain relation. Consider a fluid moving in the $x$ 
direction sheared between two parallel plates in the $xz$ plane. The 
first order term $\pi_{ij}=-\eta\sigma_{ij}$ describes Newton's law of 
friction, $F_x/A=\eta\nabla_y u_x$. At second order we also find a 
normal force, $F_y/A=-\lambda_1(\nabla_y u_x)^2$. The $\lambda_2$ term
describes the coupling between shear and vorticity, and the 
$\lambda_3$ term implies that in a rotating fluid confined in a 
cylindrical container there is a normal force on the walls of the
container. We note that this force is not dissipative. 

\vspace*{0.4cm} 
\addcontentsline{toc}{subsection}{Second order relativistic fluid dynamics}
\noindent
{\it 4. Second order relativistic fluid dynamics:} In relativistic
fluid dynamics we define the shear tensor $\sigma^{\mu\nu}$ using the 
projection operator $\Delta_{\mu\nu}$. We have
\be 
\label{sig_vis}
\sigma^{\mu\nu} = 
  \Delta^{\mu\alpha}\Delta^{\nu\beta} 
  \left(\nabla_\alpha u_\beta+\nabla_\beta u_\alpha 
  - \frac{2}{3}\eta_{\alpha\beta}\nabla\cdot u \right) \, . 
\ee
Note that in rest frame of the fluid this expression reduces to 
the non-relativistic result. At second order in the gradient expansion
the stress tensor of a scale invariant fluid is \cite{Baier:2007ix}
\bea
\label{del_Pi_2}
 \delta \Pi^{\mu\nu} &=& -\eta\sigma^{\mu\nu}
 +\eta\tau_{R} \left[ ^\langle D\sigma^{\mu\nu\rangle}
 +\frac{1}{3}\,\sigma^{\mu\nu} (\nabla\cdot u) \right] \\
& & \hspace{0.2cm}\mbox{}
 +\lambda_1\sigma^{\langle\mu}_{\;\;\;\lambda} \sigma^{\nu\rangle\lambda}
 +\lambda_2\sigma^{\langle\mu}_{\;\;\;\lambda} \Omega^{\nu\rangle\lambda}
 +\lambda_3\Omega^{\langle\mu}_{\;\;\;\lambda} \Omega^{\nu\rangle\lambda}
 + \kappa_R\left[ R^{\langle\mu\nu\rangle}
           -2u_\alpha u_\beta R^{\alpha\langle\mu\nu\rangle\beta} \right] \, ,
\nonumber 
\eea
where $D=u\cdot\nabla$ and ${\cal O}^{\langle\mu\nu\rangle}= \frac{1}{2}
\Delta^{\mu\alpha}\Delta^{\nu\beta} ( {\cal O}_{\alpha\beta}+{\cal O}_{\beta\alpha} 
- \frac{2}{3}\Delta^{\mu\nu}\Delta^{\alpha\beta}{\cal O}_{\alpha\beta})$ 
denotes the transverse traceless part of ${\cal O}^{\alpha\beta}$. The 
relativistic vorticity tensor is 
$\Omega^{\mu\nu} = \frac{1}{2}\Delta^{\mu\alpha}\Delta^{\nu\beta} 
\left(\partial_\alpha u_\beta - \partial_\beta u_\alpha \right)$,
and $R_{\mu\nu\alpha\beta}$ is the Riemann tensor.  
We note that the number of terms is smaller than in the non-relativistic 
case. This is related to the fact that without a conserved baryon current
the number of independent hydrodynamical variables is smaller. We also note 
that the numerical coefficient in front of $\sigma_{\mu\nu}(\nabla\cdot u)$ 
is different. This is due to the difference between relativistic and 
non-relativistic scale transformations. To second order accuracy the 
stress tensor is equivalent to the relaxation equation
\bea
\label{del_Pi_3}
 \pi^{\mu\nu} &=&
 -\eta\sigma^{\mu\nu} 
 -\tau_{R} \left[ ^\langle D\pi^{\mu\nu\rangle}
 +\frac{4}{3}\,\pi^{\mu\nu} (\nabla\cdot u) \right] \\
& & \hspace{0.2cm}\mbox{}
 +\frac{\lambda_1}{\eta^2}
        \pi^{\langle\mu}_{\;\;\;\lambda} \pi^{\nu\rangle\lambda}
 -\frac{\lambda_2}{\eta}
        \pi^{\langle\mu}_{\;\;\;\lambda} \Omega^{\nu\rangle\lambda}
 +\lambda_3
        \Omega^{\langle\mu}_{\;\;\;\lambda} \Omega^{\nu\rangle\lambda}
 + \kappa_R\left[ R^{\langle\mu\nu\rangle}
           -2u_\alpha u_\beta R^{\alpha\langle\mu\nu\rangle\beta} \right] \, ,
\nonumber 
\eea
where $\pi^{\mu\nu}=\delta \Pi^{\mu\nu}$ is the dissipative contribution
to the stress tensor. 

\vspace*{0.4cm} 
\addcontentsline{toc}{subsection}{Hydrodynamics as an effective field theory}
\noindent
{\it 5. Hydrodynamics as an effective field theory:} Can hydrodynamics
be formulated not only as an effective theory, but as an effective 
field theory? The standard response is that this is not possible
\cite{Leutwyler:1993gf}, because dissipative effects cannot be described
by a local lagrangian. There are, however, at least a few situations in
which hydrodynamics can be reformulated as an effective field theory. 
The simplest case is the non-dissipative flow of a superfluid at zero 
temperature, see \cite{Greiter:1989qb,Son:2002zn,Son:2005rv}. The basic 
observation is that if one implements the full Galilean (or Lorentz) and 
gauge invariance of the microscopic theory then the effective action of the 
Goldstone mode will necessarily contain the non-linear terms needed to 
recover the equations of fluid mechanics. Consider the dilute Fermi gas
in the superfluid phase, see Sect.~\ref{sec_uni}. The effective 
lagrangian for the Goldstone mode $\varphi$ is 
\be
\label{L_hydro}
{\cal L} = P(X)\, , \hspace{0.25cm}
X = \mu - \partial_0\varphi - \frac{(\vec{\nabla}\varphi)^2}{2m}\, , 
\ee
where $P(\mu)$ is the pressure and $\mu$ is the chemical potential. The 
form of the variable $X$ is determined by $U(1)$ and Galilean invariance, 
and the relation ${\cal L} = P(X)$ ensures that we obtain the correct
thermodynamic potential for constant fields $X=\mu$. Note that 
equ.~(\ref{L_hydro}) is the leading term in a low energy expansion
where we treat $\nabla\varphi\sim O(1)$ but $\nabla^2\varphi\ll\nabla\varphi$
\cite{Son:2005rv}. This expansion is useful because it respects 
Galilean and $U(1)$ symmetry exactly order by order in the low energy
expansion, and as a consequence it is equivalent to superfluid 
hydrodynamics. 

Expanding equ.~(\ref{L_hydro}) in powers of $(\partial\varphi)/\mu$ 
reproduces the conventional low energy expansion of the effective field 
theory for the Goldstone mode. We find, in particular, that the velocity 
of the Goldstone mode is the speed of sound, 
\be 
 {\cal L} = \frac{f^2}{2} \left[ \big(\partial_0\varphi\big)^2
 - c_s^2 \big(\vec{\nabla}\varphi\big)^2 + \ldots \right]\, , 
\ee
with $f^2=(\partial n)/(\partial\mu)$ and $c_s^2=(\partial P)/
(\partial\rho)$. Note that for the dilute Fermi gas at unitarity 
dimensional analysis implies that $P(\mu)\sim m^{3/2}\mu^{5/2}$. 

 We can write equ.~(\ref{L_hydro}) in terms of hydrodynamic variables
by introducing the superfluid velocity $\vec{v}_s =(\vec{\nabla}\varphi)
/m$. The equation of motion for the field $\varphi$ leads to
\be
\label{hyd_cont}
  \partial_0 \bar{n} + \frac{1}{m} 
    \vec\nabla \left(\bar{n}\vec\nabla \varphi\right) =0,
\ee
where we have defined $\bar{n}=P'(X)$. Equ.~(\ref{hyd_cont}) is the 
continuity equation for the current $\vec\jmath=\bar{n}\vec{u}_s$. We 
can derive a second equation by using the identity $dP=nd\mu$. We get 
\be
 \label{euler_sfl}
  \partial_0 \vec{u}_s +\frac{1}{2} \vec\nabla u_s^2 = 
  -\frac{1}{m}\vec{\nabla}\mu.
\ee
which is the Euler equation in a superfluid. We note that higher 
derivative corrections to equ.~(\ref{L_hydro}) correspond to 
non-dissipative higher order terms in the equations of fluid dynamics.
Next-to-leading order (NLO) terms have been determined \cite{Rupak:2008xq}. 
They lead to non-linearities in the dispersion law for sound waves, and 
to corrections to the equation of hydrostatic equilibrium in small systems.

\begin{figure}[t!]
\begin{center}
\includegraphics*[width=11cm]{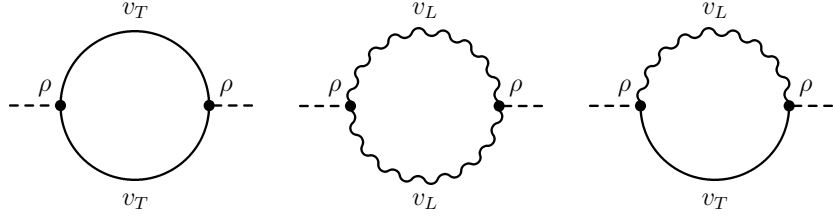}
\end{center}
\caption{\label{fig_loop}
Diagrammatic representation of the leading contribution of 
thermal fluctuations to the stress tensor correlation function.
Solid lines labeled $v_T$ denote the transverse velocity correlator, 
dominated by the shear pole, and wavy lines labeled $v_L$ denote the 
longitudinal velocity correlator, governed by the sound pole and
the diffusive heat mode.}   
\end{figure}

A different situation in which the fluid dynamic expansion can 
be written as an effective field theory is the systematic calculation
of retarded correlation functions including noise and dissipation
\cite{Martin:1973zz,DeDominicis:1977fw,Khalatnikov:1983ak,Kovtun:2012rj,Kovtun:2014hpa}. 
We can write the effective action as a functional integral over the 
noise, the hydrodynamic variables, and suitable Lagrange multipliers 
that enforce the linearized equations of motion. This representation 
can be used to derive a set of Feynman rules for the retarded correlation 
functions. The diagrammatic approach is particularly powerful as 
a method for computing non-analytic terms in the correlation function
induced by thermal fluctuations, see Fig.~\ref{fig_loop} and 
equ.~(\ref{G_R_1l}).

 More recent ideas about hydrodynamics and effective field theory 
can be found in \cite{Jensen:2012jh,Dubovsky:2011sj,Torrieri:2011ne}.

\vspace*{0.6cm}
\noindent
{\bf\large 1.2 Models of fluids: Kinetic theory}

\vspace*{0.4cm} 
\addcontentsline{toc}{subsection}{Conserved charges in kinetic theory}
\noindent
{\it 6. Conserved charges in kinetic theory:} For completeness we 
give the complete definition of the conserved charges in non-relativistic
kinetic theory. We have 
\bea
\label{rho_kin}
 \rho\left(\vec{x},t\right) &=& \int d\Gamma_p\,  mf_p\left(\vec{x},t\right)\\
\label{j_kin}
 \vec\jmath_\rho (\vec{x},t)  &=& \int d\Gamma_p\,  m\vec{v}
                              f_p\left(\vec{x},t\right)\, , \\
\label{pi_ij_kin_2}
\Pi_{ij}\left(\vec{x},t\right) &=& 
   \int d\Gamma_p\, p_iv_j  f_p\left(\vec{x},t\right)
    + \delta_{ij} \left( \int d\Gamma_p \, E_p f_p\left(\vec{x},t\right)
       - {\cal E}\left(\vec{x},t\right)\right)\, ,
\eea
where $\vec{v}=\vec{\nabla}_p E_p$ and ${\cal E}$ is the energy density. 
Note that in equilibrium we view $E_p$ and ${\cal E}$ as functions of the 
thermodynamic variables, but in kinetic theory we must consider these 
quantities as functionals of the distribution function $f_p$. In a weakly 
interacting gas we have $E_p=p^2/(2m)$ and ${\cal E}=\int d\Gamma_p\, E_pf_p$, 
but in general these relations are modified by interactions. The dependence
of ${\cal E}$ and $E_p$ on $f_p$ is constrained by conservation laws. 
Momentum conservation requires \cite{Baym:1962,Schaefer:2013oba}
\be 
\label{E_p_var}
 E_p =\frac{\delta{\cal E}}{\delta f_p}\, ,
\ee
and the equality of the mass current $\vec{\jmath}_\rho$ and the momentum 
density $\vec{\pi}$ implies that 
\be 
\label{pi_constr}
\int d\Gamma_p m \vec{v}\, f_p =\int d\Gamma_p\, \vec{p}\, f_p \, . 
\ee
These conditions are quite non-trivial to satisfy. Microscopic theories 
that are consistent with the constraints are discussed in 
\cite{Baym:1962,Schaefer:2013oba,Levin:2013}. The condition given in 
equ.~(\ref{E_p_var}) also holds in relativistic theories, see 
\cite{Jeon:1995zm}. Another difficulty in constructing quasi-particle 
models of the thermodynamic properties of the many-body system is to 
find an explicit expression for ${\cal E}[f_p]$. This problem can be 
avoided by focusing on the enthalpy
\be 
{\cal E}+P = \int d\Gamma_p\, \left(\frac{1}{3}\vec{v}\cdot\vec{p}
 + E_p \right)f_p(\vec{x},t)\, . 
\ee
The same observation applies to relativistic theories. In a relativistic
fluid we can use ${\cal E}+P=sT$ (for $\mu=0$) to construct a quasi-particle
model for the entropy density. 

\vspace*{0.4cm} 
\addcontentsline{toc}{subsection}{Linearized collision operator}
\noindent
{\it 7. Linearized collision operator:} The relaxation of hydrodynamic
variables near equilibrium is determined by the linearized collision 
term. We write the distribution function as $f_p=f_p^0(1+\chi_p/T)$
where $f_p^0$ is the equilibrium distribution, see equ.~(\ref{f_p_0}). 
The linearized collision operator corresponding to binary $2\to 2$ is 
scattering is $C[f_p]\equiv (f^0_p/T) C_L[\chi_p]$ with 
\be 
 C_L[\chi_{p_1}] = -\int \Big(\prod_{i=2}^4d\Gamma_{i}\Big)
   w(1,2;3,4) f^0_{p_2}
   \left[\chi_{p_1}+\chi_{p_2}-\chi_{p_3}+\chi_{p_4}\right]\, . 
\ee
The transition rate $w(1,2;3,4)$ is given by 
\be 
w(1,2;3,4) = (2\pi)^4\delta^3\Big(\sum_i \vec{p}_i\Big)
  \delta \Big( \sum_i E_{i}\Big) \left| {\cal M}\right|^2\, ,
\ee
and ${\cal M}$ is the scattering amplitude. The scattering amplitude
for low energy $s$-wave scattering is given in equ.~(\ref{M_nr}).
We can define an inner product for distribution functions
\be 
\langle \chi|\psi\rangle = \int d\Gamma_p \,f_p^0 \chi_p\psi_p\, .
\ee
Detailed balance and the symmetries of the transition rate imply
that $C_L$ is a hermitean, negative semi-definite operator. Zero 
eigenvalues of $C_L$ correspond to the conservation laws for 
particle number, momentum, and energy, $\chi^{(0)}_i\sim 1,\vec{p},E_p$.
In the space orthogonal to the zero modes $C_L$ can be written 
as 
\be
 C_L = -\sum_i \frac{|\chi_i\rangle\langle \chi_i|}{\tau_i}\, . 
\ee
The  BGK (or relaxation time) model is based on the assumption that 
that the collision term, or, more accurately, its inverse, is dominated 
by the longest collision time, 
\be
C_L \simeq -\frac{|\chi_0\rangle\langle \chi_0|}{\tau_0}
    \simeq -\frac{1}{\tau_0}\, .
\ee
Here, we also assume that $C_L$ acts on a distribution function 
that has a large component along $\chi_0$. This is a reasonable 
approximation at late times. 

 Near the hydrodynamic limit the left hand side of the Boltzmann
equation can be expanded in derivatives of the hydrodynamic variables 
$T,\mu,\vec{u}$. This is known as the Chapman-Enskog expansion 
\cite{Chapman:1970}. The linearized Boltzmann equation is of the 
form 
\be 
 |X\rangle = C_L|\psi\rangle \, , 
\ee
where the driving term $X$ arises from the gradient expansion and
$\psi_p$ is the off-equilibrium distribution induced by the external
stress. Consider a distribution $f^0(T,\mu,\vec{u})$ describing a 
pure shear flow $u_y(x)$. We find 
\be 
\label{BE_CE}
 \left( \frac{\partial}{\partial t} 
  + \vec{v} \cdot \vec{\nabla}_x
  + \vec{F}  \cdot \vec{\nabla}_p \right) f^0(T,\mu,u_y(x)) 
= -\frac{f^0}{T}v_xp_y \nabla_x u_y\, , 
\ee
and $X=v_xp_y \nabla_x u_y \equiv X_0 \nabla_x u_y$. Using the 
definition of the stress tensor in kinetic theory, equ.~(\ref{pi_ij_kin_2}), 
we get 
\be 
\label{eta_me}
\eta = -\frac{1}{T}\, \langle X_0 | \psi_0 \rangle 
     =  \frac{1}{T}\, \langle X_0 | (-C_L^{-1})|X_0\rangle\, , 
\ee
where we have defined $\psi=\psi_0\nabla_x u_y$ and used the linearized 
Boltzmann equation. This result shows that the shear viscosity is positive. 
We can also establish a variational bound on $\eta$. The Cauchy-Schwarz 
inequality implies
\be 
\label{eta_var}
 \eta \geq \frac{\langle X_0|\psi_{\it var}\rangle^2}
                {\langle \psi_{\it var} | (-C_L)|\psi_{\it var}\rangle}\, ,
\ee
which is valid for any variational distribution function $|\psi_{\it var}
\rangle$. We note that this is not a fundamental bound for $\eta$. Instead, 
equ.~(\ref{eta_var}) provides a bound on $\eta$ in the context of a given 
collision term and quasi-particle dispersion relation. Similar bounds can 
be derived for other transport coefficients. Finally, we note that in the 
BGK approximation the solution of the Boltzmann equation is given by 
$|\psi_0\rangle = -\tau_0 |X_0\rangle$. Equation (\ref{eta_me}) then leads 
to the simple result $\eta=\tau_0 P$. Using $P=nT$ and $\langle mv^2\rangle 
= 3T$ we can write this as $\eta = \frac{1}{3}nl_{\it mfp}\bar{p}$. More 
systematic calculations of $\eta$ are based on expanding $\psi_{\it var}$ 
in a complete set of polynomials $L^{(k)}(x)$, 
\be 
 \psi_{\it var}(\vec{p}) = p_xp_y \sum^{N-1}_{k=0} 
           c_kL^{(k)}\left(\frac{p^2}{mT}\right)\, . 
\ee
and truncate the expansion at order $N$. A convenient choice in 
non-relativistic kinetic theory is the set of generalized Laguerre (Sonine) 
polynomials \cite{Landau:kin}. This expansion typically converges rapidly. 
The result for $\eta$ given in equ.~(\ref{eta_aa}) is based on using $N=1$, 
but higher order corrections are known to be quite small, on the order 
of 2\% \cite{Bruun:2006}.

\vspace*{0.4cm} 
\addcontentsline{toc}{subsection}{Knudsen expansion}
\noindent
{\it 8. Knudsen expansion:} The Chapman-Enskog method provides an
expansion of $\delta f_p$ in the Knudsen number ${\it Kn}=l_{\it mfp}/L$. 
This expansion corresponds to the gradient expansion in hydrodynamics. 
Schematically, $\delta f_p = \delta f_p^1 \tau_0 (\nabla u) + \delta 
f_p^2 \tau^2_0 (\nabla^2 u) + \ldots$, where $\tau_0$ is the relaxation
time and $\nabla u$ is a shorthand for derivatives of the hydrodynamic
variables. The first term, $\delta f_p^1$ determines the viscosity
and thermal conductivity, the second term determines second order 
transport coefficients, and so on. Each of these term has an expansion
in powers of the density of the gas. In the case of the shear viscosity
\be 
\eta = \eta_0 \left[ 1 + \eta_1 \left(n\lambda_{\it dB}^3\right) 
+ \eta_2 \left(n\lambda_{\it dB}^3\right)^2 
+ \ldots \right]\, ,
\ee 
where $n$ is the density and $\lambda_{\it dB}$ is the de Broglie wave
length. Note that $\eta_1$ may contain terms of order $l_{\it mfp}/
\lambda_{\it dB}$, but not terms of order $l_{\it mfp}/L$. Higher 
order terms in the density arise from a number of sources. The first is 
that the Boltzmann equation for the single particle distribution arises 
from truncating a set of classical or quantum equations for $N$-body 
distribution functions \cite{Landau:kin,Baym:1962}. At leading order 
in the density the Boltzmann equation only contains two-body collisions, 
but at higher order it also includes collisions between three and more 
particles. The second source of corrections is the density expansion 
of the equation of state and the quasi-particle properties. In the 
case of the equation of state, the resulting expression is the well 
known virial expansion. 

 It was found that the expansion in density breaks down at the 
level of four-body collisions, and that resummation beyond effects
already summed by the Boltzmann equation is required \cite{Ernst:1997}.
This leads to the appearance of terms that are logarithmic in the density, 
and to a breakdown of the Knudsen expansion. The latter can be
traced to hydrodynamic modes, and is equivalent to the appearance 
of non-analytic terms in the gradient expansion. 

 In relativistic theories $n\lambda_{\it dB}^3\sim 1$ and the only 
expansion parameter is the coupling constant. The structure of the 
perturbative expansion is not well understood. Only the heavy 
quark diffusion constant has been determined beyond leading order 
in the coupling constant \cite{CaronHuot:2008uh}. The shear viscosity 
has been determined beyond leading logarithmic accuracy, i.e. 
the numerical coefficient inside the logarithm of $g$ was
computed \cite{Arnold:2003zc}.  

 Finally, we note that it was recently argued that one can organize 
kinetic theory in terms of separate power series expansion in 
${\it Re}^{-1}$ and ${\it  Kn}$ \cite{Denicol:2012cn}. This corresponds 
to a situation where we view the Israel-Stewart model (or similar 
relaxation schemes) not only as practical implementation of second 
order hydrodynamics, but as resummed hydrodynamic theories that can 
be used in cases where the inverse Reynolds number is not small.

\vspace*{0.6cm}
\noindent
{\bf\large 1.3 Matching and Kubo relations}

\vspace*{0.4cm}
\addcontentsline{toc}{subsection}{Linear response and general covariance}
\noindent
{\it 9. Linear response and general covariance:} In order to study 
linear response we have to couple the stress tensor $\Pi_{ij}$ (or 
$\Pi_{\mu\nu}$ in the relativistic theory) to an external tensor 
field. From the symmetries of the stress tensor it is clear that 
this tensor transforms like the metric. We can therefore perform
the analysis by considering fluid dynamics in a curved background. 
There is a large amount of literature on relativistic fluid dynamics
in curved space \cite{Weinberg:1972}. The method can be extended to 
non-relativistic fluid dynamics using the formalism developed in 
\cite{Son:2005rv}. Consider a three-dimensional metric $g_{ij}(t,
\vec{x})$. A non-relativistic diffeomorphism is a time-dependent 
change of coordinates $x^i\to x^i+\xi^i(\vec{x},t)$ that transforms
the  metric as $\delta g_{ij}=-g_{ik}\nabla_j\xi^k-g_{kj}\nabla_i
\xi^k$.

 The generally covariant Navier-Stokes equation is 
\be 
\frac{1}{\sqrt{g}} \frac{\partial}{\partial t}
  \left(\sqrt{g}\rho u_i\right) +\nabla_k\Pi^{k}_{i} = 0\, , 
\ee
where $g=\det(g_{ij})$ and $\nabla_k$ is the covariant derivative 
associated with the metric $g_{ij}$. The stress tensor is $\Pi_{ij} = 
\Pi_{ij}^0+\delta\Pi_{ij}$, where $ \Pi_{ij}^0 = \rho u_i u_j +P g_{ij}$
is the ideal fluid part, and $\delta\Pi_{ij}$ is the viscous correction. 
At one-derivative order we have $\delta\Pi_{ij}=-\eta\sigma_{ij}-\zeta 
g_{ij}\langle\sigma\rangle$ with \cite{Son:2005tj}
\bea
\label{def_sig}
 \sigma_{ij} &=& \nabla_{i}u_{j}+\nabla_{j}u_{i}+\dot{g}_{ij}
         -\frac{2}{3}g_{ij}\langle\sigma\rangle \, , \\
\label{def_s}
 \langle\sigma\rangle &=& \nabla\cdot u+\frac{\dot{g}}{2g}\, .
\eea
The structure of the extra terms involving time derivatives of the 
metric is dictated by diffeomorphism invariance \cite{Chao:2011cy}. 

 Consider a ``pure shear'' perturbation $g_{ij}(\vec{x},t)=\delta_{ij}
+h_{ij}(\vec{x},t)$ where the only non-vanishing component of $h_{ij}$ 
is $h_{xy}(z,t)$. From the linearized Euler equation we can see that 
this perturbation does not induce a shift in the density, temperature, 
or velocity. This means that we can directly read off $\delta\Pi_{ij}$ 
from equ.~(\ref{def_sig}) and (\ref{del_pi_fin}). The induced stress
determined the retarded correlation function via the linear response
relation
\be
\label{G_R_lin}
\delta \Pi^{ij} =  - \frac{1}{2} G_R^{ijkl}h_{kl} \, . 
\ee
We find
\be 
\label{Kubo_2}
 G_R^{xyxy}(\omega,k) = P -i\eta \omega +\tau_R\eta \omega^2
  - \frac{\kappa_R}{2} k^2 + O(\omega^3,\omega k^2)\, , 
\ee
which is the Kubo relation quoted in the text. The analogous result in 
the relativistic theory is \cite{Son:2007vk,Baier:2007ix}
\be 
\label{Kubo_rel}
 G_R^{xyxy}(\omega,k) = P -i\eta \omega +\tau_R\eta \omega^2
  - \frac{\kappa_R}{2} \left(k^2-\omega^2\right) + O(\omega^3,\omega k^2)\, , 
\ee
The expansion in powers of $\omega$ and $k$ maps onto the derivative 
expansion in hydrodynamics. We note, however, that the two point function
$G_R^{xyxy}$ only describes higher derivative terms that are linear 
in the hydrodynamic variables. There are a number of second order terms 
that encode non-linearities in the relation between stress and strain, 
in particular the coefficients $\lambda_{1,2,3}$ in 
equ.~(\ref{del_pi_fin},\ref{del_Pi_2}). Kubo relations for these 
transport coefficients can be derived by considering higher order 
terms in the response, which are related to retarded three point
functions \cite{Moore:2010bu}.

The idea of embedding the theory in curved space is also useful 
for computing the spectral function in kinetic theory. The Boltzmann 
equation in a four-dimensional curved space is \cite{Cercignani:2002}
\be
\label{BE_4d_cur}
\frac{1}{p^{0}}\left(p^{\mu}\frac{\partial}{\partial x^{\mu}}
  - \Gamma^{i}_{\alpha\beta}p^{\alpha}p^{\beta}
     \frac{\partial}{\partial p^{i}}\right)
      f(t,x,p) = C[f]\, ,
\ee
where $\Gamma^\alpha_{\mu\nu}$ is the Christoffel symbol associated 
with the four-dimensional covariant derivative $\nabla_\mu$, $i,j,k$ 
are three-dimensional indices and $\mu,\alpha,\beta$ are four-dimensional 
indices. In the non-relativistic limit this equation reduces to  
\be
\label{BE_3d_cur}
  \left( \frac{\partial}{\partial t} 
  + \frac{p^{k}}{m}\frac{\partial}{\partial x_{k}}
  - \frac{\Gamma^{i}_{jk}p^{j}p^{k}}{m}\frac{\partial}{\partial p^{i}}
  - g^{il}\dot{g}_{lk}p^{k}\frac{\partial}{\partial p^{i}}
   \right)f(t,x,p) = C[f]\, .
\ee
The term involving $\dot{g}_{ij}$ carries the information about the 
leading response to a time dependent shear strain. Using the BGK
approximation to the collision operator gives the simple result for 
$\delta f_p$ quoted in equ.~(\ref{del_f_w}). Together with the linear 
response relation (\ref{G_R_lin}) we obtain the spectral function
\be 
\label{eta_Lorentz}
 \eta(\omega) = -\frac{1}{\omega}\, {\it Im}G_R(\omega,0) 
  = \frac{\eta(0)}{1+\omega^2\tau_0^2}\, ,
\ee
with $\eta(0)=\tau_0 nT$. A more detailed calculation that takes into 
account the momentum dependence can be found in \cite{Braby:2010tk},
and a calculation using a T-matrix approximation can be found in 
\cite{Enss:2010qh}. A study of the QCD shear spectral function in 
kinetic theory is presented in \cite{Hong:2010at}.

\vspace*{0.4cm} 
\addcontentsline{toc}{subsection}{Fluctuations and the ``breakdown'' of 
second order fluid dynamics}
\noindent
{\it 10. Fluctuations and the ``breakdown'' of second order fluid dynamics:}
Non-analytic terms in the low energy expansion can be found by 
computing the low frequency behavior of the retarded correlator 
in fluid dynamics. In practice it is convenient to begin with the 
symmetrized correlation function 
\be 
 G_S^{xyxy}(\omega,\vec{k}) = 
    \int d^3x\int dt\, e^{i(\omega t -\vec{k}\cdot\vec{x})} 
  \left\langle \frac{1}{2}
     \left\{ \Pi_{xy}(t,\vec{x}), \Pi_{xy}(0,0)\right\}
  \right\rangle \, ,
\ee
and use the fluctuation-dissipation theorem. In the limit $\omega\to 0$ 
we have 
\be 
\label{f-dis}
 G_S(\omega,\vec{k}) \simeq -\frac{2T}{\omega} 
          {\rm Im}\, G_R(\omega,\vec{k})\, . 
\ee
At leading order in the gradient expansion $\Pi_{xy}=\rho u_x u_y$.
We expand the hydrodynamic variables around their mean values, $\rho=
\rho_0+\delta\rho$ etc., and use the Gaussian approximation to write 
expectation values of products of fluctuating fields as products of 
two point functions. The leading contribution is 
\be
\label{G_S_1l}
 G_S^{xyxy}(\omega,0) = \rho_0^2 \int \frac{d\omega'}{2\pi}
   \int \frac{d^3k}{(2\pi)^3} 
   \Big[ \Delta_S^{xy}(\omega',\vec{k}) \Delta_S^{yx}(\omega-\omega',\vec{k})
       + \Delta_S^{xx}(\omega',\vec{k}) \Delta_S^{yy}(\omega-\omega',\vec{k})
   \Big] \, . 
\ee
where $\Delta_S^{ij}$ is the symmetrized velocity correlation function
\be 
\label{del_s_def}
 \Delta_S^{ij}(\omega,\vec{k}) = 
   \int d^3x\int dt\, e^{i(\omega t -\vec{k}\cdot\vec{x})} 
      \left\langle \frac{1}{2}
           \left\{ u^i(t,\vec{x}), u^j(0,0)\right\}
      \right\rangle \, . 
\ee
We can view equ.~(\ref{G_S_1l}) as a one-loop diagram composed of two 
propagators of hydrodynamic modes, see Fig.~\ref{fig_loop}. Using the 
low frequency limit of the fluctuation-dissipation relation we can write 
the one-loop contribution to the retarded correlation function as
\bea
 G_R^{xyxy}(\omega,0) &=& \rho_0^2 \int \frac{d\omega'}{2\pi}
   \int \frac{d^3k}{(2\pi)^3} 
   \Big[ \Delta_R^{xy}(\omega',\vec{k}) \Delta_S^{yx}(\omega-\omega',\vec{k})
       + \Delta_S^{xy}(\omega',\vec{k}) \Delta_R^{yx}(\omega-\omega',\vec{k})
\nonumber \\[0.1cm]
 & & \hspace*{0.5cm}\mbox{}
       + \Delta_R^{xx}(\omega',\vec{k}) \Delta_S^{yy}(\omega-\omega',\vec{k})
       + \Delta_S^{xx}(\omega',\vec{k}) \Delta_R^{yy}(\omega-\omega',\vec{k})
   \Big] \, . 
\label{G_R_1l}
\eea
This result can be generalized. Retarded correlation functions of 
hydrodynamic variables have diagrammatic expansions in terms of 
retarded and symmetrized correlation functions, see 
\cite{Martin:1973zz,Hohenberg:1977ym,DeDominicis:1977fw,Khalatnikov:1983ak}.
The velocity correlation function can be decomposed into longitudinal 
and transverse parts
\be 
\label{vv_cor_lt}
 \Delta_{S,R}^{ij}(\omega,\vec{k}) = 
 \left( \delta^{ij}-\hat{k}^i\hat{k}^j\right) 
             \Delta_{S,R}^{T}(\omega,\vec{k})
 + \hat{k}^i\hat{k}^j \Delta_{S,R}^L(\omega,\vec{k})\, . 
\ee
The transverse part is purely diffusive. The symmetrized correlation 
function is given by \cite{Landau:smII}
\bea
\label{Del_S_T}
\Delta_S^T(\omega,\vec{k}) &=& 
 \frac{2T}{\rho} \frac{D_\eta k^2}{\omega^2+\left(D_\eta k^2\right)^2}\, ,
\\
\label{Del_R_T}
\Delta_R^T(\omega,\vec{k}) &=&
 \frac{1}{\rho} \frac{-D_\eta k^2}{-i\omega+D_\eta k^2}\, ,
\eea
where $k=|\vec{k}|$ and $D_\eta=\eta/\rho$ is the momentum diffusion
constant. The longitudinal part describes propagating sound modes
and diffusive heat modes. The complete calculation of the one-loop 
diagram is described in \cite{Chafin:2012eq}. Here, we briefly outline 
the computation of the contribution due to shear modes. We use the 
propagators given in equ.~(\ref{Del_S_T},\ref{Del_R_T}) and 
perform the integral over $\omega'$ by contour integration. We get
\be 
\label{G_R_sh}
\left. G_R^{xyxy}(\omega,0)\right|_{\it shear}= -\frac{7T}{30\pi^2} 
  \int dk\, \frac{k^4}{k^2-i\omega/(2D_\eta)}\, . 
\ee
This integral is divergent in the UV. We regulate the divergence 
by introducing a momentum cutoff $\Lambda_K$. We then expand the 
retarded correlation function in the limit $\omega\to 0$. We 
find 
\be 
\label{G_R_sh_exp}
\left. G_R^{xyxy}(\omega,0)\right|_{\it shear}=
  - \frac{7}{90\pi^2}T\Lambda_K^3
  - i \omega\,\frac{7T\Lambda_K}{60\pi^2D_\eta} 
  + (1+i)\omega^{3/2} \frac{7T}{240\pi D_\eta^{3/2}}
  + O(\omega^{5/2})\; . 
\ee
Including the contribution of sound modes changes the coefficient 
of the $i\omega$ term to $17/120$, and the coefficient of the 
$\omega^{2/3}$ term to $(7+(\frac{3}{2})^{3/2})/240$. Comparing 
equ.~(\ref{G_R_sh_exp}) to the Kubo formula we observe that the 
first term is a contribution to the pressure, the second renormalizes
the viscosity, and the the third is a non-analytic term not captured
by the classical linear response formula. 

 Equation (\ref{G_R_sh_exp}) has a number of interesting aspects. 
First, we note that the fluctuation contribution to the shear viscosity 
scales inversely with the bare shear viscosity. This leads to the bound 
on the shear viscosity discussed in the text, see equ.~(\ref{eta_eff}). 
The fluctuation contribution depends on the cutoff. This is consistent 
with the idea that fluid dynamics is renormalizable in the effective 
theory sense, because the dependence on $\Lambda_K$ can be absorbed 
into the cutoff dependence of the bare viscosity. We also note that 
the non-analytic term is independent of the cutoff. This is important 
because there are no bare parameters in the hydrodynamic description 
that could be used to absorb the cutoff dependence of the $\omega^{3/2}$ 
term. Finally, we emphasize that the presence of a non-analytic term 
does not imply a breakdown of hydrodynamics. It only implies that 
beyond the Navier-Stokes order in three dimensions, and beyond 
ideal hydrodynamics in two dimensions, fluctuations have to be 
included. 

 The structure of the retarded correlation function also implies 
a low energy theorem for the spectral function. Taking into account 
both shear and sound modes we get 
\be 
\label{eta_non_analyt}
\eta(\omega) = \eta(0) - \sqrt{\omega}\,T\, 
  \frac{7+\left(\frac{3}{2}\right)^{3/2}}
       {240\pi D_\eta^{3/2}}\, . 
\ee
This prediction is reliable in the regime of validity of fluid dynamics, 
which implies $\omega\ll nT/\eta$. A similar non-analytic structure 
also appears in the relativistic theory, see \cite{Kovtun:2011np}.

\vspace*{0.6cm}
\noindent
{\bf\large 1.4 Models of fluids: Holography}

\vspace*{0.4cm} 
\addcontentsline{toc}{subsection}{Strong coupling results}
\noindent
{\it 11. Strong coupling results:} The leading correction to 
the infinite coupling limit of $\eta/s$ in ${\cal N}=4$ SUSY Yang Mills 
theory is \cite{Buchel:2004di,Buchel:2008ac,Buchel:2008sh,Myers:2008yi}
\be 
\label{eta_s_l32}
\frac{\eta}{s} = \frac{1}{4\pi} \left\{ 
 1 +  \frac{15\zeta(3)}{\lambda^{3/2}} + \ldots \right\} \, . 
\ee
where $\lambda=g^2N_c$ is the 't Hooft coupling. We observe that the 
$\lambda^{-2/3}$ correction is positive, consistent with the idea that
$\eta/s$ evolves smoothly from strong to weak coupling. At weak 
coupling $\eta/s\sim 1/[\lambda^2\log(\lambda)]$, see \cite{Huot:2006ys}.
It is not clear how $\lambda$ should be chosen in order to make 
predictions for the quark gluon plasma in the vicinity of $T_c$. 
For $\alpha_{SYM}\sim 0.3$ and $N_c=3$ we get $\lambda\sim 10$ and 
the next order term increases $\eta/s$ by about 50\%.

 The AdS/CFT correspondence has been used to compute the second
order hydrodynamic coefficients defined in equ.~(\ref{del_Pi_2}). 
The result is \cite{Baier:2007ix,Bhattacharyya:2008jc,Arnold:2011ja}
\be
\label{2nd_order_ads}
\tau_R = \frac{2 - \log 2}{2\pi T}\, , 
       \qquad   \lambda_1 = \frac{\eta}{2\pi T}\, ,   
       \qquad   \lambda_2 =-\frac{\eta\log 2}{\pi T}\, , 
       \qquad   \lambda_3 = 0\, ,
       \qquad   \kappa_R = \frac{\eta}{\pi T}\, ,  
\ee
The coefficients $\tau_R$ and $\kappa_R$ can be determined from the 
retarded two-point function, see equ.~(\ref{G_R_w_low}). The remaining
coefficients have been computed using the fluid-gravity correspondence
\cite{Bhattacharyya:2008jc}, as well as using the Kubo formula combined
with the three-point function for the stress tensor in AdS/CFT 
 \cite{Arnold:2011ja}.

The relaxation time is very short, $\tau_R\sim 1/T$, but $(\tau_RT)/
(\eta/s)$ is not very different from the result in perturbative QCD, 
see equ.~(\ref{2nd_order_QCD}). The coefficient $\lambda_3$ corresponds
to the vorticity squared term in the stress tensor. In kinetic theory 
this term does not appear because the second order terms are induced 
by the first order stresses $\delta\Pi_{\mu\nu}\sim \sigma_{\mu\nu}$.
As a result, we can get $D\sigma_{\mu\nu}$, $\sigma_{\mu\lambda}
\sigma^{\lambda}_{\;\nu}$ and $\sigma_{\mu\lambda}\Omega^{\lambda}_{\;\nu}$,
but not $\Omega_{\mu\lambda}\Omega^{\lambda}_{\;\nu}$. The sign of 
$\lambda_1$ and $\lambda_2$ can also be understood in kinetic theory
\cite{York:2008rr}. In AdS/CFT there is no obvious reason why 
$\lambda_1>0$, $\lambda_2<0$ and $\lambda_3=0$.

\begin{figure}[t!]
\begin{center}
\includegraphics*[width=7.5cm]{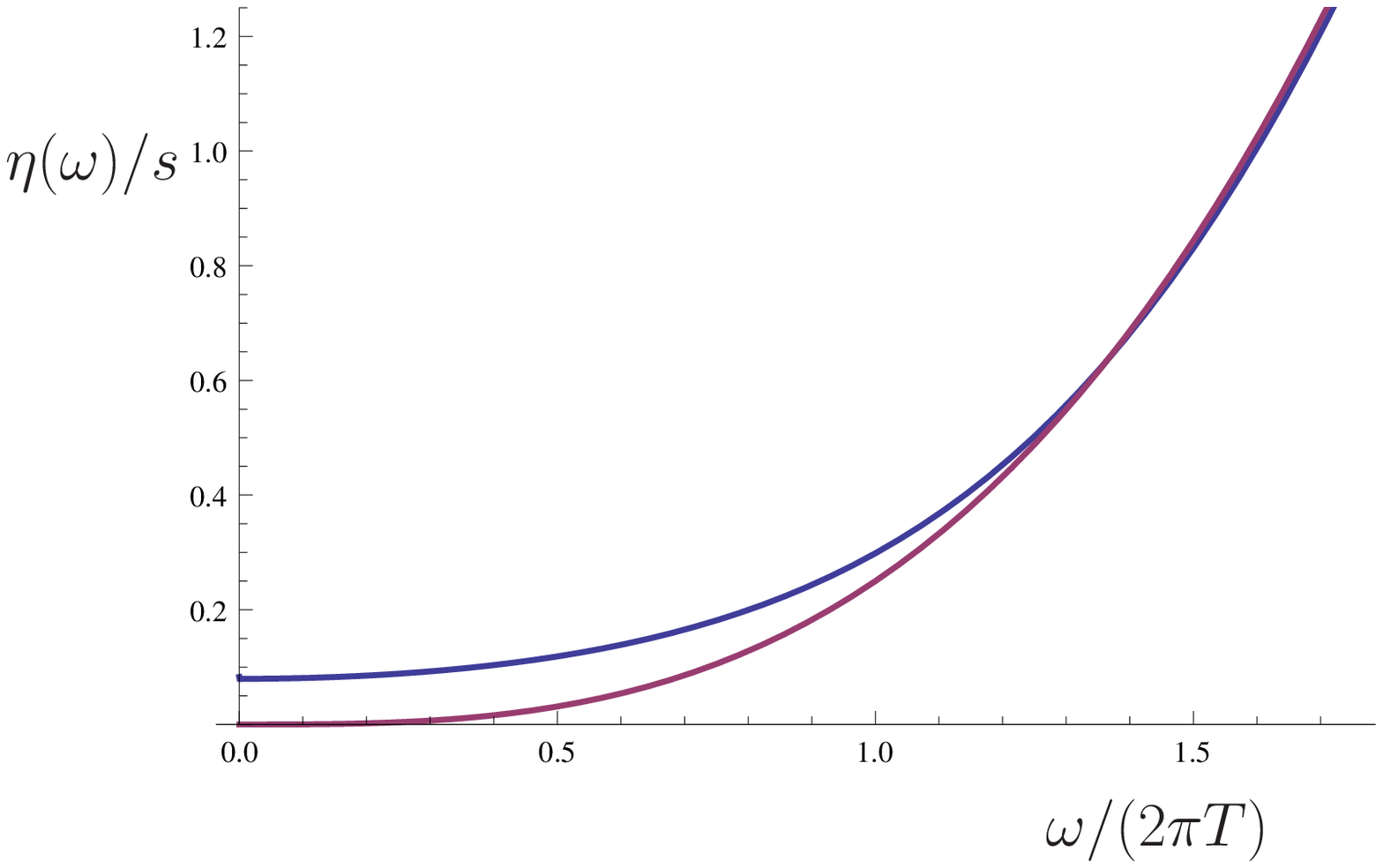}
\includegraphics*[width=7.5cm]{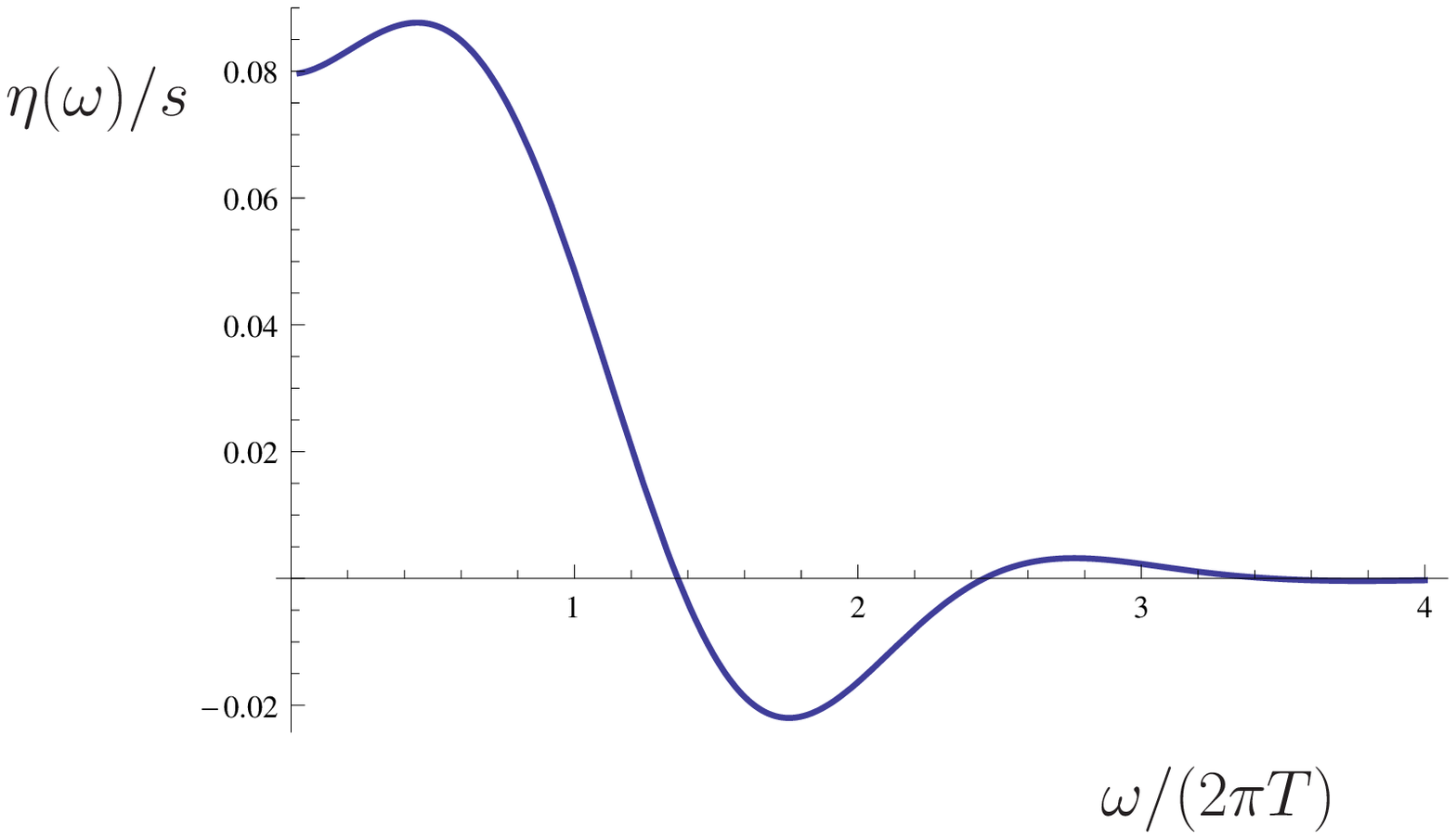}
\end{center}
\caption{\label{fig_ads_spec_fct}
Viscosity spectral function in the large $N_c$ limit of strongly 
coupled ${\cal N}=4$ SUSY Yang Mills theory, computed using the 
AdS/CFT correspondence. The left panel shows $\eta(\omega)/s$ (blue) 
and $\eta_{T=0}(\omega)/s$ (red) as a function of $\omega$. The right 
panel shows the finite temperature contribution $[\eta(\omega)-
\eta_{T=0}(\omega)]/s$.}
\end{figure}

${\cal N}=4$ SUSY Yang Mills theory has a conserved gauge invariant 
density called R-charge. Son and Starinets computed the shear viscosity 
and entropy density as a function of the R-charge chemical potential 
$\mu$. They find that both $\eta$ and $s$ depend on $\mu$, but the ratio 
$\eta/s$ does not \cite{Son:2006em}. They also determine the thermal 
conductivity 
\be 
\label{kappa_AdSCFT}
\kappa = \frac{8\pi^2T}{\mu^2}\, \eta \, .
\ee
The scaling $\kappa\sim 1/\mu^2$ is related to the definition of $\kappa$ 
in the Landau frame and also appears in kinetic theory. The diffusion 
constant of a heavy  test quark in the SUSY Yang Mills plasma was calculated 
in \cite{Herzog:2006gh,CasalderreySolana:2006rq,Gubser:2006bz}. The result is 
\be 
\label{D_AdSCFT}
D=\frac{2}{\pi T} \frac{1}{\sqrt{\lambda}},
\ee
which depends on the value of the coupling $\lambda$, and goes to 
zero in the strong coupling limit. The functional dependence on 
$\lambda$ is unusual from the point of view of perturbation theory,
which would suggest that $D$ scales as $1/\lambda^2$, and that 
$D$ is proportional to the momentum diffusion constant $\eta/(sT)$.

 ${\cal N}=4$ SUSY Yang Mills theory is scale invariant and the 
bulk viscosity of the SUSY plasma vanishes. Non-conformal generalizations
of the AdS/CFT correspondence have been studied. Buchel proposed that 
in holographic models there is a lower bound on the bulk viscosity, 
$\zeta\geq 2(\frac{1}{3}-c_s^2)\eta$, where $c_s$ is the speed of sound 
\cite{Buchel:2007mf}. This is in contrast to the weak coupling result
$\zeta\sim (\frac{1}{3}-c_s^2)^2\eta$ \cite{Weinberg:1972}. Gubser and 
collaborators considered a number of holographic models tuned to 
reproduce the QCD equation of state, and find that $\zeta/s$ has 
a maximum near the critical temperature where $\zeta/s\simeq 0.05$ 
\cite{Gubser:2008sz,DeWolfe:2011ts}.

\vspace*{0.4cm} 
\addcontentsline{toc}{subsection}{Spectral function and quasi-normal modes}
\noindent
{\it 12. Spectral function and quasi-normal modes:} The viscosity spectral 
function in the strong coupling limit of ${\cal N}=4$ SUSY Yang Mills 
theory can be computed from the solution of wave equation $AdS_5$, see 
equ.~(\ref{lingrav}). We have defined $\delta g_x^y=\phi_k(u)e^{ikx-
i\omega t}$. The infalling solution can be written as 
\begin{figure}[t!]
\begin{center}
\includegraphics*[width=7.5cm]{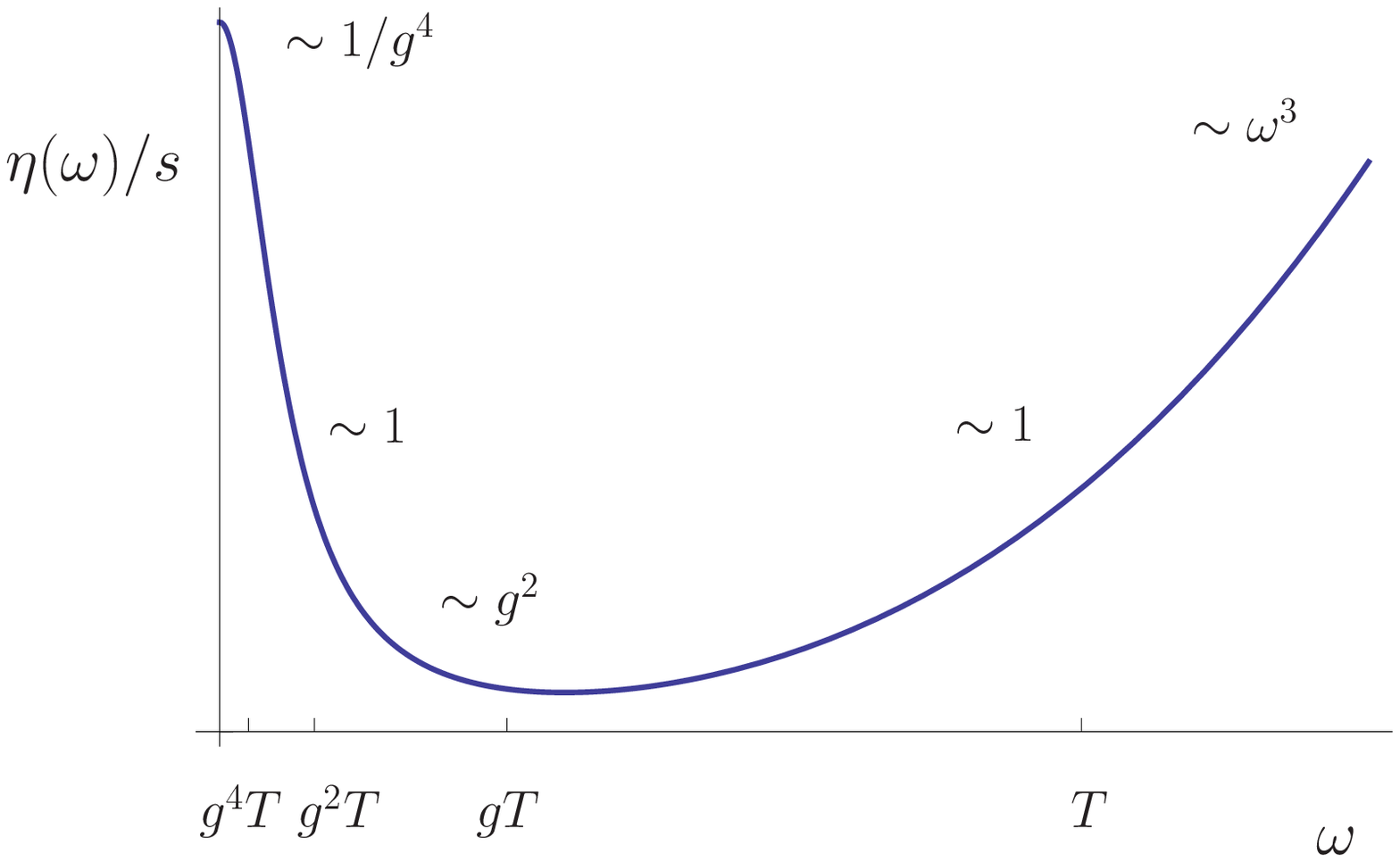}
\includegraphics*[width=7.5cm]{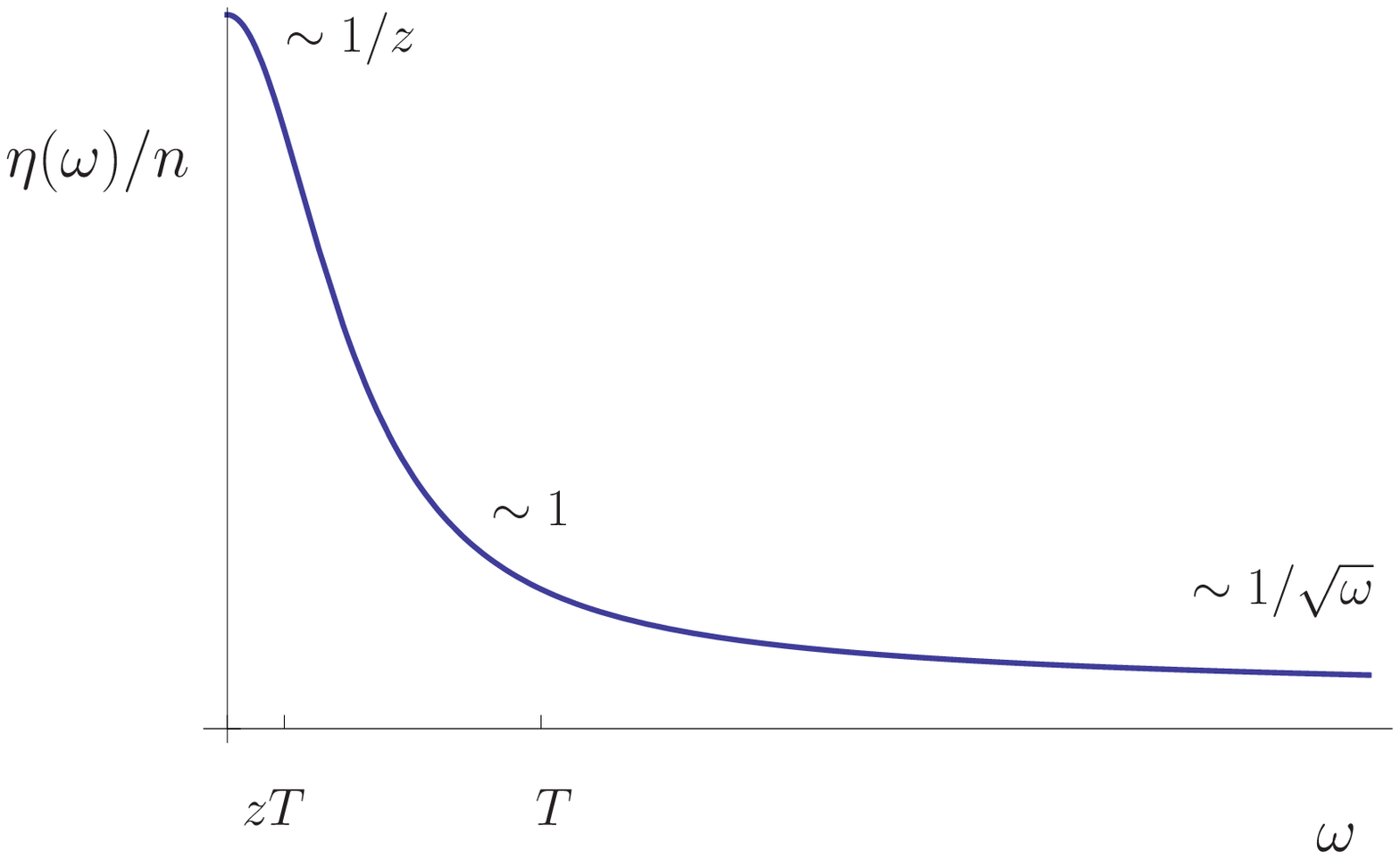}
\end{center}
\caption{\label{fig_spec_fct}
Schematic behavior of the viscosity spectral function in QCD
(left panel) and a dilute Fermi gas (right panel). In QCD we plot 
$\eta(\omega)/s$ as a function of $\omega$. The relevant scales
are $g^4T$(momentum relaxation) $\ll g^2T$(magnetic screening) $\ll 
gT$(electric screening) $\ll T$, where $g$ is the coupling constant.
In the dilute Fermi gas $\eta$ is normalized to the density $n$, 
and the momentum relaxation scale $zT\ll T$ involve powers of the 
fugacity $z$. } 
\end{figure}
\be 
\label{phi_k_ans}
\phi_k(u) = (1-u) ^{-i\mathfrak{w}/2}F_k(u)
\ee
where $\mathfrak{w}=\omega/(2\pi T)$ and we have factored out the
near horizon behavior. The function $F_k(u)$ can be determined 
as an expansion in $\mathfrak{w}$ and $\mathfrak{k}=k/(2\pi T)$. 
At order $O(\mathfrak{w}^2,\mathfrak{k}^2)$ the solution is 
\cite{Policastro:2002se}
\be 
F_k(u) = 1-\frac{i\mathfrak{w}}{2}\log\left(\frac{1+u}{2}\right)
  +\frac{\mathfrak{w}^2}{8} \left\{
     \left[ 8 - \frac{8\mathfrak{k}^2}{\mathfrak{w}^2}
              + \log\left(\frac{1+u}{2}\right)\right]
                              \log\left(\frac{1+u}{2}\right)
   - 4{\it Li}_2\left(\frac{1-u}{2}\right)\right\}\, . 
\ee
The wave equation can also be solved analytically in the limit 
of large $\mathfrak{w},\mathfrak{k}$ \cite{Teaney:2006nc}. For 
$\mathfrak{k}=0$ we get 
\be 
\phi_k(u) = \pi\mathfrak{w}^2\frac{u}{\sqrt{1-u^2}}
 \left[ iJ_2\left(2\mathfrak{w}\sqrt{u}\right)
         -Y_2\left(2\mathfrak{w}\sqrt{u}\right)\right]\, . 
\ee
For intermediate values of $\mathfrak{w}$ and $\mathfrak{k}$ the wave 
equation can be solved numerically, for example by starting from the 
near horizon behavior given in equ.~(\ref{phi_k_ans}) and integrating 
outwards towards the boundary. The retarded correlation function is 
determined by the variation of the boundary action with respect to 
the field. The relevant term in the action is
\be 
 S = -\frac{\pi^2N^2T^4}{8}\int du\int d^4x\, 
    \frac{f(u)}{u}  \left(\partial_u\phi\right)^2 + \ldots \, .  
\ee
This is the quadratic part of the Einstein-Hilbert action, where 
we have used the AdS/CFT correspondence to express Newton's constant
in terms of gauge theory parameters. The boundary action can be derived 
by integrating by parts. The retarded Green function is given by the 
second variational derivative with respect to the boundary value of 
the field \cite{Policastro:2002se,Son:2006em}, 
\be 
\label{G_R_Action}
G_R(\mathfrak{w},\mathfrak{k})= -\frac{\pi^2N^2T^4}{4}
  \left[ \frac{f(u)\partial_u \phi_k(u)}{u\phi_k(u)}
     \right]_{u\to 0}\, .
\ee
\begin{figure}[t!]
\begin{minipage}{0.48\hsize}
\begin{center}
\includegraphics*[width=5cm,angle=-90]{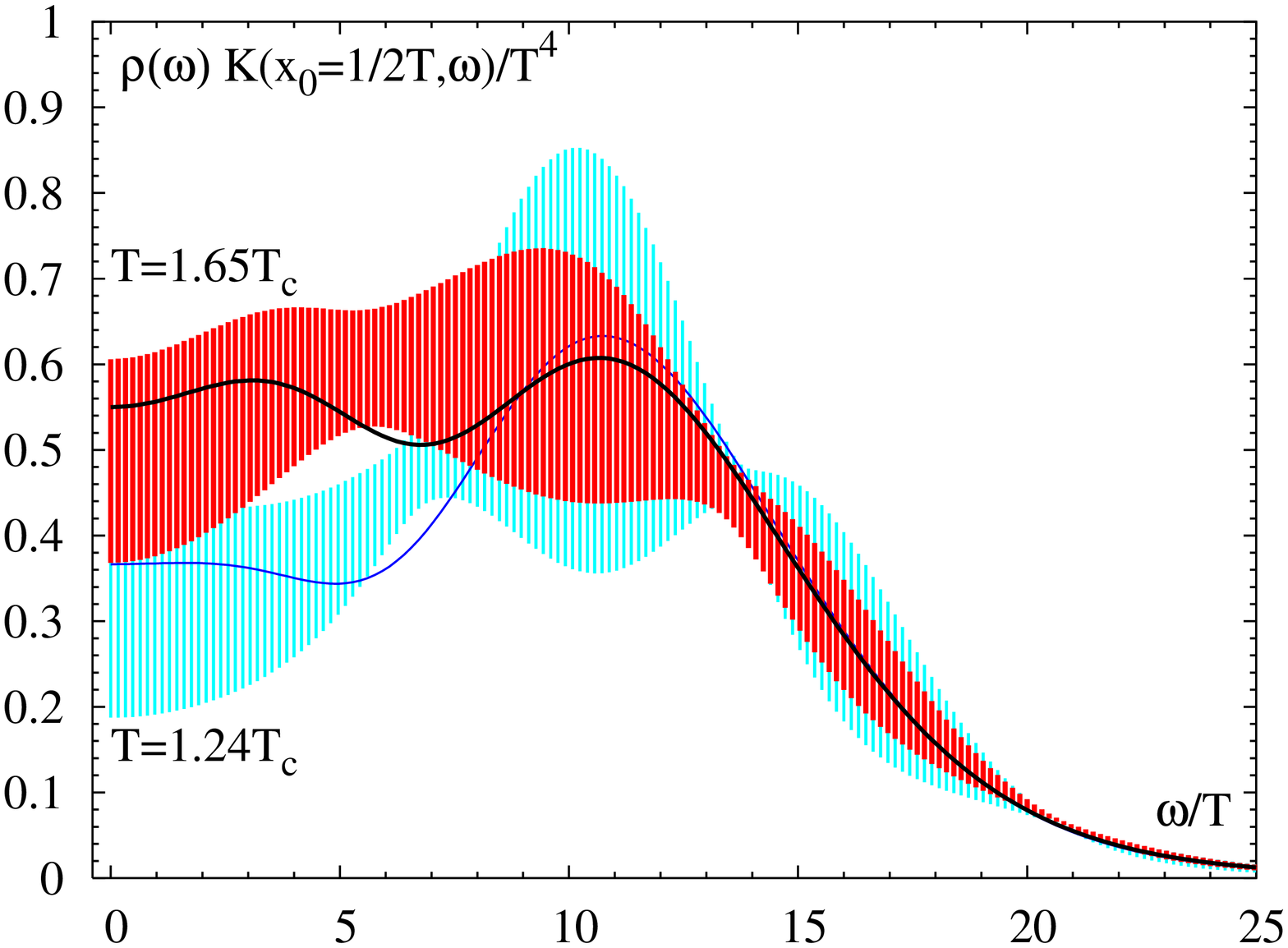}
\end{center}
\end{minipage}\begin{minipage}{0.48\hsize}
\begin{center}
\includegraphics*[width=7.5cm]{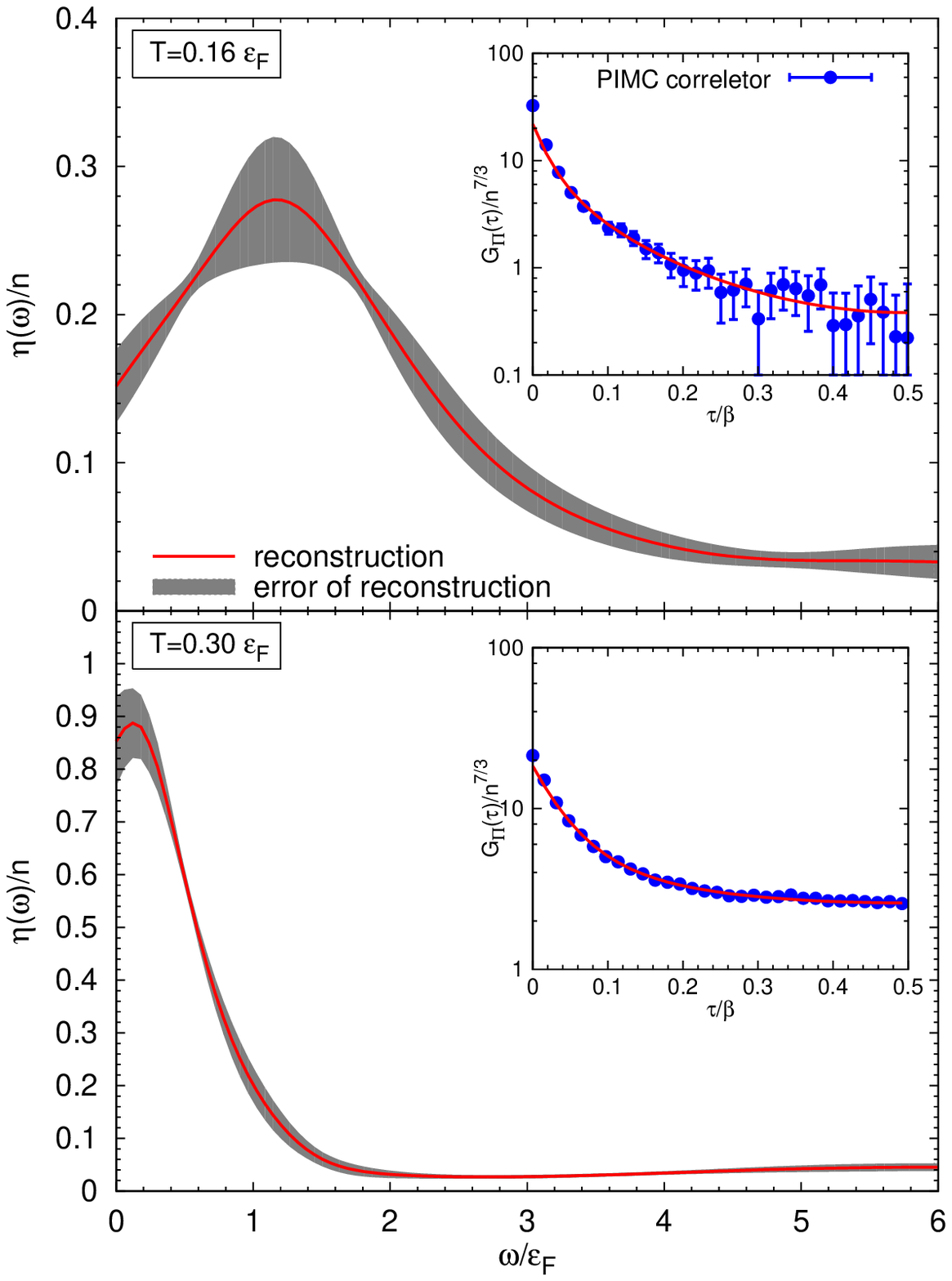}
\end{center}
\end{minipage}
\caption{\label{fig_spec_fct_latt}
Numerical determination of the viscosity spectral function in QCD
\cite{Meyer:2007ic} (left panel) and a dilute Fermi gas 
\cite{Wlazlowski:2013owa} (right panel). In the QCD case the plot
shows $\rho(\omega)/[\sinh(\beta\omega/2)T^4]$ with $\rho(\omega)=
\omega\eta(\omega)$. The method for determining the error band is 
explained in \cite{Meyer:2007ic}. The intercept at $\omega=0$ 
corresponds to the values of $\eta/s$ quoted in Sect.~\ref{sec_rel}. 
The spectral function $\eta(\omega)$ of the unitary gas is normalized 
to the density $n$ and computed at two different temperatures $T=0.16
\epsilon_F$ and $T=0.3\epsilon_F$, where $\epsilon_F=k_F^2/(2m)$ with 
$k_F=(3\pi^2n)^{1/3}$ is the Fermi energy. See \cite{Wlazlowski:2013owa} 
for a discussion of the error bands shown in gray. The insets shows the 
underlying imaginary time correlation function.}
\end{figure}
In the low frequency, low momentum limit \cite{Baier:2007ix}
\be 
\label{G_R_w_low}
G_R(\mathfrak{w},\mathfrak{k}) = 
   - \frac{\pi^2N^2T^4}{4} \left[ -\frac{1}{2} 
  + i\mathfrak{w} - \mathfrak{w}^2 \left(1-\log(2)\right)
  + \mathfrak{k}^2 \right] + \ldots \, .
\ee
Comparing to the Kubo relation (\ref{Kubo_rel}) we obtain the 
relaxation time in equ.~(\ref{2nd_order_ads}).
The spectral function $\eta(\omega)=-\omega^{-1}{\it Im}\,G_R(\omega,
k\!=\!0)$ is shown in the left panel of Fig.~\ref{fig_ads_spec_fct}.
In the right panel we show the finite temperature contribution 
$\eta(\omega)-\eta_{T=0}(\omega)$, and in the left panel of 
Fig.~\ref{fig_spec_fct} we show the qualitative behavior of the 
spectral function in the weak coupling limit. The AdS/CFT result
has a number of interesting features:

\begin{enumerate}
\item The spectral function does not have a quasi-particle peak. 
The low energy limit $\eta(0)=s/(4\pi)$ is smoothly connected to
the high energy limit $\eta(\omega)\sim T^3$. 

\item The tail of the finite temperature part of the spectral function
oscillates in sign. The result is consistent with the sum rule
given in equ.~(\ref{qcd_sr}), and there is no fundamental requirement
that $\eta(\omega)-\eta_{T=0}(\omega)$ has to be positive. 

\item The curvature of the spectral function near the origin is 
positive. This is different from the result in kinetic theory. 
Note that in kinetic theory the downward curvature is determined
by the viscous relaxation time $\tau_R$. In particular, the decrease
in $\eta(\omega)$ can be understood as resulting from the lag 
between the strain $\sigma_{xy}$ and the viscous stress $\delta
\Pi_{xy}$. However, in general there is no direct relation 
between $\tau_R$ and the curvature of $\eta(\omega)$. The Kubo
formula relates $\tau_R$ to the $\omega^2$ term in ${\it Re}\,G_R$,
whereas the curvature is determined by the $\omega^3$ term in 
${\it Im}\,G_R$.

\end{enumerate}

\begin{figure}[t!]
\begin{center}
\includegraphics*[width=7.5cm]{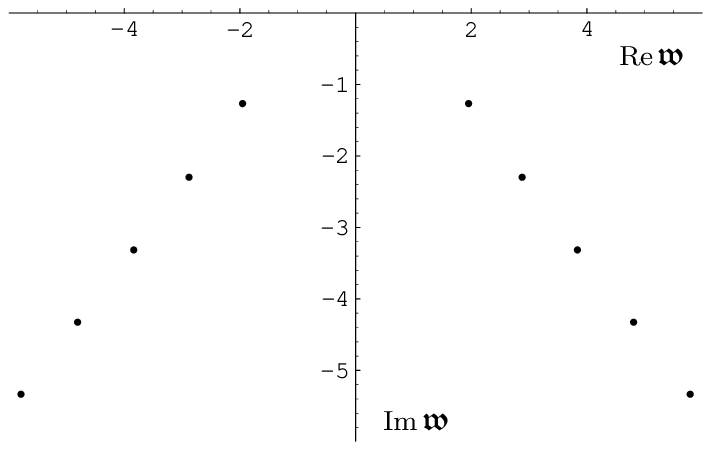}
\includegraphics*[width=7.5cm]{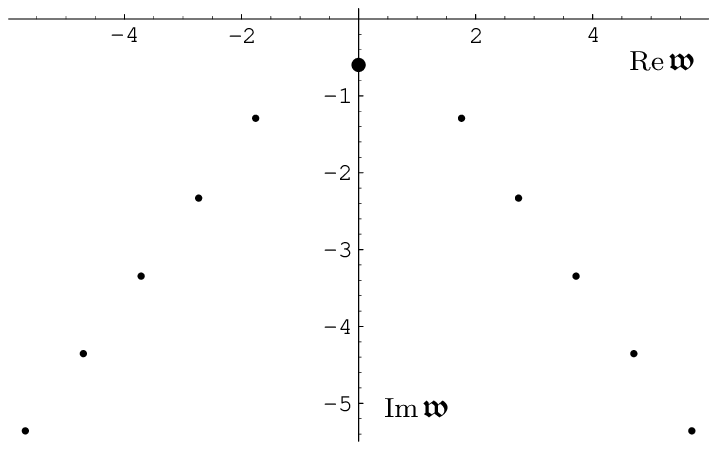}
\end{center}
\caption{\label{fig_ads_QNM}
Quasi-Normal modes of gravitational fluctuations around the $AdS_5$ 
black hole solution, from \cite{Kovtun:2005ev}. The left panel shows 
the correlator in the scalar channel $G_R^{xyxy}$ (for $\vec{k}=k\hat{z}$), 
and the right panel shows $G_R^{xzxz}$. The plots show the location of 
poles in the complex $\mathfrak{w}$ plane for $\mathfrak{k}=1$. Note 
that $G_R^{xzxz}$ has a hydrodynamic pole at $\mathfrak{w}\simeq 
-i\bar{\gamma}\mathfrak{k}^2$ with $\bar\gamma=2\pi\eta/s$.}
\end{figure}

 Whether these features are present in the QCD spectral function
near $T_c$ is unclear. The stress tensor spectral function is not
directly accessible in experiment, and the determination of $\eta(\omega)$
on the lattice is difficult because of the finite resolution of the 
lattice and the need for analytic continuation. The spectral function
extracted in \cite{Meyer:2007ic} is shown in the left panel of 
Fig.~\ref{fig_spec_fct_latt}. There is no quasi-particle peak, 
but the resolution is not very good, and the continuum is strongly
modified by cutoff effects. 

 We can also study the location of the poles of $G_R(\omega)$ in the 
complex plane. Poles of the retarded correlator that approach the 
origin as $k\to 0$ are related to hydrodynamic modes, and poles that 
remain at a finite distance from the origin determine corrections 
to hydrodynamic behavior. Near the boundary $u=0$ the function 
$\phi_k(u)$ can be written as
\be 
 \phi_k(u) = {\cal A}(\omega,k)\big[ 1 + \ldots \big]
   + {\cal B}(\omega,k) \big[ u^2 + \ldots \big]\, . 
\ee
Equ.~(\ref{G_R_Action}) implies that $G_R(\omega,k)\sim {\cal B}
(\omega,k)/ {\cal A}(\omega,k)$, and poles of $G_R(\omega,k)$
correspond to zeros of ${\cal A}(\omega,k)$. In this case $\phi_k(u)$
satisfies a Dirichlet problem on the boundary, and infalling conditions
on the horizon. The corresponding frequencies $\omega$ are known as 
quasi-normal modes. 

\begin{figure}[t!]
\begin{center}
\includegraphics*[width=7.5cm]{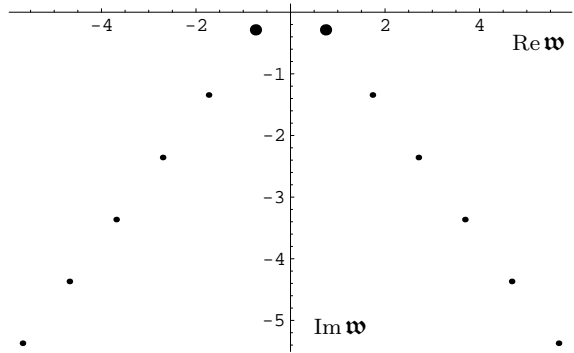}
\end{center}
\caption{\label{fig_ads_QNM_2}
Quasi-Normal modes of gravitational fluctuations around the $AdS_5$ 
black hole solution. This figure shows the correlation function 
$G_R^{zzzz}$ (for $\vec{k}=k\hat{z}$). The plots show the location 
of poles in the complex $\mathfrak{w}$ plane for $\mathfrak{k}=1$. 
The correlator has a hydrodynamic pole at $\mathfrak{w}\simeq \pm 
c_s\mathfrak{k}-i\bar{\gamma}_s\mathfrak{k}^2$ with $\bar{\gamma}_s
=\frac{2}{3}\bar{\gamma}$. }
\end{figure}

\begin{figure}[t!]
\begin{center}
\includegraphics*[width=7.5cm]{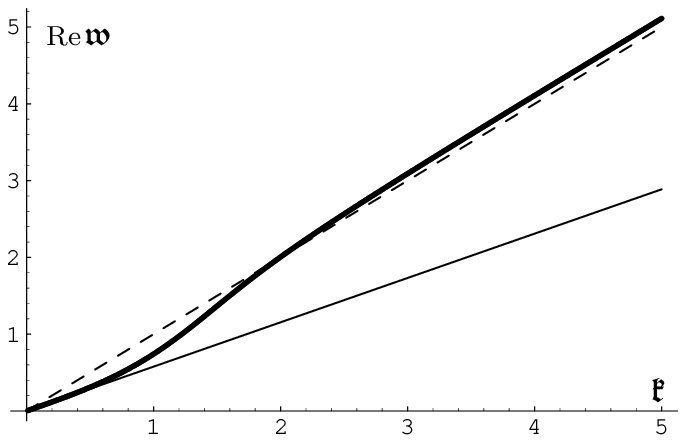}
\includegraphics*[width=7.5cm]{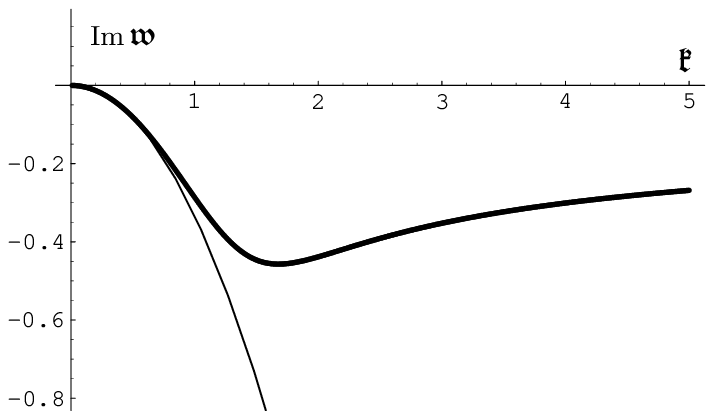}
\end{center}
\caption{\label{fig_sound_dis}
Real and imaginary parts of the sound wave frequency as a function 
of the sound wave momentum in the strong coupling limit of ${\cal N}=4$ 
SUSY Yang Mills theory, computed using the AdS/CFT correspondence. 
Light curves correspond to the 
hydrodynamic approximation for small $\mathfrak{k}$. The dashed 
line is $\mathfrak{w} = \mathfrak{k}$. Note that in the regime 
where the imaginary part is large there is a range of momenta for 
which $\partial{\rm Re}(\mathfrak{w})/\partial\mathfrak{k}>1$. }
\end{figure}

 The quasi-normal mode spectrum of the $AdS_5$ black hole was determined
in \cite{Starinets:2002br,Nunez:2003eq,Kovtun:2005ev}. We show some 
of the results in Fig.~\ref{fig_ads_QNM} and \ref{fig_ads_QNM_2}.
Poles of $G_R^{xyxy}(\omega,k)$ are shown in the left panel of 
Fig.~\ref{fig_ads_QNM}. We observe that quasi-normal modes occur in 
pairs. Asymptotically, the position of the poles in the limit 
$\mathfrak{k}=0$ is given by 
\be
\mathfrak{w}^\pm_n= (\pm 0.607 - 0.389i) \pm n(1\mp i)\, , 
\ee 
for integer $n$ \cite{Starinets:2002br}. The implies that the 
regime of validity of the hydrodynamic expansion is indeed given by 
$\omega\lsim \pi T$. The right panel of Fig.~\ref{fig_ads_QNM} and 
Fig.~\ref{fig_ads_QNM_2} also show the hydrodynamic modes
\be 
\mathfrak{w}\simeq -i\bar{\gamma}\mathfrak{k}^2\, , 
\hspace{0.5cm}
\mathfrak{w}\simeq \pm c_s\mathfrak{k}-i\bar{\gamma}_s
\mathfrak{k}^2\, , 
\ee
with $\bar{\gamma}=2\pi\eta/s$ and $\bar{\gamma}_s=\frac{2}{3}\bar{\gamma}$. 
The first mode is a diffusive shear mode, and the second is a sound wave.
The AdS/CFT correspondence can be used to follow the sound mode beyond 
the hydrodynamic regime. For large $\mathfrak{k}\gg 1$ the speed of 
sound goes to one, and sound attenuation is small, see 
Fig.~\ref{fig_sound_dis} \cite{Kovtun:2005ev}. 

\vspace*{0.6cm}
\noindent
{\bf\large 1.5 Viscosity bounds}

\vspace*{0.4cm}
\addcontentsline{toc}{subsection}{Dimensionless ratios: $\eta/s$ or $\eta/n$?}
\noindent
{\it 13. Dimensionless ratios: $\eta/s$ or $\eta/n$?} It is not 
immediately obvious which dimensionless ratio we should consider in 
connection with possible bounds for the shear viscosity
\cite{Schafer:2009dj,Liao:2009gb}. The kinetic theory argument establishes 
a possible bound for $\eta/n$, but the AdS/CFT correspondence and the 
theory of hydrodynamic fluctuations establish limits on $\eta/s$. We 
cannot resolve this question here, as none of the proposed bounds have 
been rigorously proven. We note, however, that the ratio $\eta/s$ is 
well defined for all fluids, whereas $\eta/n$ can only be defined for 
fluids with a conserved particle number. Even though $\eta/s$ was initially 
introduced for relativistic fluids, it has a smooth non-relativistic limit. 
Indeed, holographic dualities provide examples of non-relativistic fluids 
with $\eta/s=1/(4\pi)$ \cite{Adams:2008wt,Herzog:2008wg}.

 If we accept the idea that the basic measure of fluidity is $\eta/s$, 
then we have to address the possibility of driving $\eta/s$ to zero
by increasing the entropy per particle. This can be done, for example, 
by considering a dilute gas composed of a large number of different
species \cite{Kovtun:2004de,Cohen:2007qr,Cherman:2007fj}. We first 
note that in practice it is quite difficult to reduce $\eta/s$ in this 
way, because increasing $s/n$ by a factor $\xi$ requires the number of 
species to grow by a factor $e^\xi$ \cite{Dobado:2007tm}. We also note that 
a dilute gas composed of an exponentially large number of species is a 
very unusual fluid \cite{Son:2007xw}, as the time to reach mechanical 
equilibrium via diffusion of momentum is much shorter than the time 
required to reach thermal equilibrium. In particular, it takes an 
exponentially long time for the system to reach the equilibrium entropy 
starting from a generic non-equilibrium state. 

 Despite these caveats there is no obstacle that prevents us from 
constructing a fluid with $e^\xi$ degrees of freedom in non-relativistic
quantum mechanics. Whether these models can be embedded in a relativistic
field theory is not clear. An ingenious construction was suggested in 
\cite{Cohen:2007qr}, but the proposed system is not stable on time
scales required to observe the large mixing entropy \cite{Son:2007xw}.

\vspace*{0.6cm}
\addcontentsline{toc}{section}{Endnotes: Nonrelativistic fluids}
\noindent
{\bf\large 2 Nonrelativistic fluids}

\vspace*{0.4cm}
\noindent
{\bf\large 2.1 The unitary Fermi gas}

\vspace*{0.4cm} 
\addcontentsline{toc}{subsection}{Transport properties of the dilute Fermi
gas}
\noindent
{\it 14. Transport properties of the dilute Fermi gas:} In kinetic theory 
we can not only compute the shear viscosity of the dilute Fermi gas at 
unitarity, but also other transport properties like the thermal conductivity, 
the spin diffusion constant, and the bulk viscosity. Suitable ratios
of these quantities provide additional information on quasi-particle
properties. The thermal conductivity is \cite{Braby:2010ec}
\be 
 \kappa = \frac{225}{128\sqrt{\pi}}\, m^{1/2}T^{3/2}\, .
\ee
The relative magnitude of thermal and momentum diffusion is characterized
by the Prandtl number ${\it Pr} = c_P\eta/(\rho\kappa)$, where $c_P$ is 
the specific heat at constant pressure. The Prandtl number determines, 
for example, the relative importance of shear viscosity and thermal 
conductivity in sound attenuation. In the high temperature limit
we find ${\it Pr}=2/3$, which is equal to the Prandtl ratio of a weakly
interacting gas. If the shear viscosity of the gas is known the thermal
conductivity can be extracted from the sound attenuation length. The 
speed of sound has been measured by a number of groups \cite{Joseph:2006}, 
but the sound attenuation length has not been measured. 

The spin diffusion constant is defined by Fick's law,
\be
  \vec{\jmath}_{s} = -D_s\vec{\nabla} M \, ,
\ee
where $\vec{\jmath}_s$ is the spin current, and $M=n_\uparrow-n_\downarrow$ 
is the polarization. A calculation of the diffusion constant in kinetic
theory gives \cite{Bruun:2011}
\be
 D_s = \frac{3}{16\sqrt{\pi}}\frac{(mT)^{3/2}}{mn} \, .
\ee
The spin diffusion constant decreases as the temperature is lowered. 
Near the critical temperature $D_s$ is expected to approach the 
universal value $D_s\sim \hbar/m$, where we have reinstated Planck's 
constant. Quantum limited spin diffusion was observed experimentally
in \cite{Sommer:2011}, see also \cite{Bruun:2011b}. The experiment
is based on observing the late time relaxation of two colliding 
clouds of spin up and down fermions. It is interesting to compare the
result $D_s\sim \hbar/m$ to the observed shear viscosity near $T_c$. 
The momentum diffusion constant is $D_\eta = \eta/(mn)$. In the vicinity
of $T_c$ we have $\eta/s\simeq 0.5 \hbar/k_B$ and $s/n\simeq k_B$. 
These numbers imply $D_\eta\simeq 0.5 \hbar/m$, and we conclude that 
the spin and momentum diffusion constants are comparable. A similar
correlation between the heavy quark and momentum diffusion constants
can be studied in the quark gluon plasma, see below. 

 The dilute Fermi gas at unitarity is scale invariant and the bulk 
viscosity vanishes \cite{Son:2005tj}. The leading contribution to the 
bulk viscosity near $a=\infty$ can be computed systematically in the
high temperature limit. The result is \cite{Schaefer:2013oba}
\be 
\zeta = \frac{1}{96\pi^{5/2}} (mT)^{3/2} 
   \Big(\frac{z\lambda_{\it dB}}{a}\Big)^2\, , 
\ee
where $z$ is the fugacity and $\lambda_{\it dB}$ is the de Broglie 
wave length. This result is consistent with the assumption that 
the bulk viscosity scales as the shear viscosity multiplied by the 
square of the departure from scale invariance in the equation of state,
$\zeta \sim \eta (P-\frac{2}{3}{\cal E})^2$. 

 The physical mechanism for generating bulk viscosity is somewhat subtle. 
Bulk viscosity can arise in elastic two-body collisions provided the 
quasi-particle self energy has a momentum dependent contribution that 
violates scale invariance.  In this case the equilibrium distribution
function is not only a function of $p^2/(mT)$. As the gas expands 
two-body collisions are needed to reestablish the correct equilibrium
distribution. Since the collisions rate is finite the resulting lag 
will lead to a non-equilibrium contribution to the pressure and a non-zero
bulk viscosity.

 Second order transport coefficients are given by \cite{Schaefer:2014xma}
\be 
\label{2nd_order_cag}
\eta\tau_R=\frac{\eta^2}{P}\, , \hspace{0.3cm}
\lambda_1 =  \frac{15\eta^2}{14P}\, , \hspace{0.3cm}
\lambda_2 = -\frac{\eta^2}{P}\, , \hspace{0.3cm}
\lambda_3 = 0 \, ,
\ee
where $\tau_R$ is the viscous relaxation time, and $\lambda_{123}$ are 
the coefficients of non-linear terms defined in equ.~(\ref{del_pi_fin}).
The result for $\tau_R$ shows that the expansion parameter of the gradient 
expansion is indeed $\omega/\omega_{\it fl}$ with $\omega_{\it fl}=P/\eta$. 
The expressions for the second order coefficients can be compared to the 
analogous results for a quark gluon plasma, see equ.~(\ref{2nd_order_QCD}). 
We observe that, in units of $\eta^2/P$, the results are very similar.

\vspace*{0.4cm} 
\addcontentsline{toc}{subsection}{Spectral function at unitarity}
\noindent
{\it 15. Spectral function:} The schematic behavior of the shear viscosity
spectral function in the high temperature limit is shown in the right 
panel of Fig.~\ref{fig_spec_fct}. The low frequency behavior is obtained
in kinetic theory, see equ.~(\ref{eta_Lorentz}). The high frequency behavior
$\eta(\omega)\sim 1/\sqrt{\omega}$ was first determined, up to an overall 
factor, using the high frequency behavior of the f-sum rule 
\cite{Taylor:2010ju}. The correct prefactor was computed in \cite{Enss:2010qh}
based on a T-matrix approach. 

 A more general method for studying the high frequency behavior of spectral
functions is based on the operator product expansion (OPE) 
\cite{Braaten:2008uh,Hofmann:2011qs} (see \cite{Romatschke:2009ng} for 
an OPE study of the viscosity spectral function in QCD). The basic idea 
can be explained using the current correlation function as an example. 
Indeed, since the transverse current correlator has a diffusive pole, it 
is possible to extract $\eta(\omega)$ from the current correlation function. 
Consider the operator product
\be 
 A_{ij}^{\sigma\sigma'}(\omega,\vec{q}) = \int dt \int d^3r\int d^3R\, 
  e^{i(\omega t-\vec{q}\cdot\vec{r})}\, 
  T\left[ \jmath_{i}^{\sigma}\left(\vec{R}+\frac{\vec{r}}{2},t\right)
          \jmath_{i'}^{\sigma'}\left(\vec{R}-\frac{\vec{r}}{2},0\right)
   \right]\, , 
\ee
where $\jmath_{i}^{\sigma}=-i/(2m)\psi^\dagger_\sigma \stackrel{\leftrightarrow}
{\nabla}\psi_\sigma$ is the current operator and $T$ is the time ordering
symbol. The OPE proceeds by expanding the operator product in a series 
of local operators \cite{Hofmann:2011qs}, 
\be 
\label{ope_nr}
 A_{ij}^{\sigma\sigma'}(\omega,\vec{q}) = \sum_{k,\alpha}
  \frac{1}{\omega^{\Delta_k/2-3/2}}
 c_{ij\alpha}^{(k)} \left(
   \frac{q^2}{2m\omega},\frac{a^{-1}}{\sqrt{m\omega}} \right) 
 \int d^3R\, {\cal O}^{(k)}_\alpha(\vec{R})
\ee
where ${\cal O}^{(k)}_\alpha$ is an operator labeled by $k$, and $\alpha$ 
is a set of indices that the operator may carry. $\Delta_k$ denotes the 
scaling dimension of the operator defined by ${\cal O}^{(k)}_\alpha(\lambda 
\vec{x},\lambda^2t)=\lambda^{-\Delta_k} {\cal O}^{(k)}_\alpha(\vec{x},t)$. 
Current correlation functions are determined by taking thermal averages 
of equ.~(\ref{ope_nr}). This implies that the frequency and momentum 
dependence is determined by the coefficient functions $c_{ij\alpha}^{(k)}$,
and the density and temperature dependence is carried by the expectation
values of the local operators ${\cal O}^{(k)}_\alpha$. The simplest local 
operator is the density $n(\vec{x},t)$ with $\Delta_n=3$. Other one-body
operators are the current $\jmath_i$ and the stress tensor $\pi_{ij}=1/(2m)
\psi^\dagger_\sigma \stackrel{\leftrightarrow}{\nabla}_i 
\stackrel{\leftrightarrow}{\nabla}_j\psi_\sigma$.

 Short range correlations are described by two-body operators. The 
simplest operator is the contact density \cite{Tan:2005}
\be 
\label{C_def}
 \hat{\cal C} = m^2C_0^2 \psi^\dagger_\uparrow\psi^\dagger_\downarrow
                    \psi_\downarrow\psi_\uparrow\, . 
\ee
The contact density has scaling dimension $\Delta_{\cal C}=4$, which 
agrees with naive dimensional analysis. The crucial observation is 
that ${\cal C}$ as defined in equ.~(\ref{C_def}) has UV finite matrix 
elements even though $C_0$ and $\psi^\dagger_\uparrow\psi^\dagger_\downarrow
\psi_\downarrow\psi_\uparrow$ are divergent. This can be seen, for example, 
using the effective lagrangian give in equ.~(\ref{l_4f}). This lagrangian 
can be written in a partially bosonized form as 
\be 
\label{l_4f_bos}
{\cal L} = \psi^\dagger \left( i\partial_0 +
 \frac{{\vec\nabla}^2}{2m} \right) \psi 
 + \Big[ \psi_\uparrow\psi_\downarrow \Phi^\dagger + {\it h.c.}\Big]
 + \frac{1}{C_0} \Big(\Phi\Phi^\dagger\Big) \, . 
\ee
We note that the equation of motion for the bosonic field is $\Phi=
-C_0 \psi_\uparrow\psi_\downarrow$, so the contact density is $\hat{\cal C}
=m^2\Phi^\dagger\Phi$. At zero chemical potential we can compute the 
propagator for $\Phi$ exactly, see for example \cite{Schaefer:2013oba}. 
We get
\be 
\label{phi_prop}
D(\omega,\vec{q})=\frac{4\pi}{m^{3/2}} \, 
   \frac{i}{\sqrt{\omega-\frac{q^2}{4m}+i\epsilon}}\, . 
\ee
The scaling dimension of ${\cal C}$ can be extracted from the Fourier
transform of this result. We find $\Delta_{\cal C}=4$.

 Having identified the relevant operators we can now study the 
OPE for the current correlation function
\be
 G_{ij}(\omega,\vec{q}) = \frac{i}{2} \sum_{\sigma\sigma'}
   \left\langle A^{\sigma\sigma'}_{ij}(\omega,\vec{q})\right\rangle \, . 
\ee
Using the equation of motion for the momentum density, $\partial_t\pi_{i}
=-\nabla_j\Pi_{ij}$, we can relate the shear viscosity to the retarded 
transverse correlation function, 
\be 
 \eta(\omega) = m^2 \lim_{q\to 0} \frac{\omega}{q^2}
   {\it Im}\,G_T(\omega,\vec{q})\, ,
\ee
where $G_T$ is defined as in equ.~(\ref{vv_cor_lt}). The behavior of 
$G_T$ at large $\omega$ is determined by the lowest dimension operator 
in the OPE. We note, however, that one-body operators like the density 
lead to diagrams in which all the momentum flows through a single fermion 
line. This means that the imaginary part is a delta-function. The tail of 
the spectral function is therefore dominated by the lading two-body operator, 
which is the contact density. Using $\Delta_{\cal C}=4$ we get
\be 
\label{eta_tail}
 \eta(\omega) \sim  \frac{\cal C}{\sqrt{m\omega}}\, ,
\ee
where ${\cal C}=\langle\hat{\cal C}\rangle$. The appearance of a non-analytic 
dependence on $\omega$ is interesting. The numerical coefficient in 
equ.~(\ref{eta_tail}) is $1/(15\pi)$, see \cite{Enss:2010qh,Hofmann:2011qs}. 
The expectation value ${\cal C}$ is a non-perturbative quantity that can 
be measured experimentally \cite{Sagi:2012}, or extracted from quantum Monte 
Carlo calculations \cite{Drut:2010yn}. In the high temperature limit 
${\cal C}$ can be computed using the virial expansion, ${\cal C}=4\pi 
n^2\lambda_{\it dB}^2$ \cite{Yu:2009}.

 Knowledge of the large frequency behavior of $\eta(\omega)$ is important 
for quantum Monte Carlo studies of the shear viscosity, and for identifying 
consistent many-body approaches to transport theory. It is not clear how the 
large frequency behavior of $\eta(\omega)$ can be measured. However, the 
analogous tail in the dynamic structure factor can be studied experimentally, 
see \cite{Braaten:2010if}.

\vspace*{0.4cm} 
\addcontentsline{toc}{subsection}{Nonrelativistic AdS/CFT correspondence}
\noindent
{\it 16. Nonrelativistic AdS/CFT correspondence:} There have been a 
number of attempts to extend the AdS/CFT correspondence to non-relativistic
fluids. The idea proposed in \cite{Son:2008ye,Balasubramanian:2008dm}
is to embed a $d+1$ non-relativistic theory into a $d+2$ dimensional 
relativistic theory. Consider the Minkowski metric in light cone 
coordinates $(x^+,x^-,x^i)$ with $x^\pm=(x^0\pm x^{d+1})/\sqrt{2}$ for
$i=1,\ldots,d$. We have 
\be 
 ds^2 = \eta_{\mu\nu} dx^\mu dx^\nu 
      = -2dx^+dx^- + dx^idx^i\, .
\ee 
The equation of motion of a massless scalar field is given by
\be 
\label{KG_lc}
 \left( -2 \frac{\partial}{\partial x^-}
           \frac{\partial}{\partial x^+} 
         + \frac{\partial^2}{\partial x_i^2}
               \right) \phi(x) = 0\, . 
\ee
We now compactify the theory on a light-like circle, $\phi(x^-)=
\phi(x^-+2\pi/m)$. The winding number one mode is given by $\phi(x)\sim 
e^{-imx^-}\psi(x^+,x_i)$ where $\psi$ satisfies the Schr\"odinger equation
\be 
  \left( i \frac{\partial}{\partial x^+} 
        + \frac{\vec\nabla^2}{2m} \right) \psi(x^+,x_i) = 0\, .
\ee
In terms of symmetries this construction shows that the non-relativistic
conformal group in $d+1$ dimensions, the Schr\"odinger group $Sch(d)$
\cite{Hagen:1972pd,Nishida:2007pj}, can be embedded in $SO(d+2,2)$, 
the conformal group in $d+2$ dimensions. 

 This idea can be applied to spaces that are asymptotically AdS. The 
specific proposal described in \cite{Son:2008ye,Balasubramanian:2008dm} 
is that the $Schr(d)$ symmetry of a non-relativistic $d+1$ dimensional 
conformal field theory can be mapped onto the isometries of the $d+3$ 
dimensional metric 
\be 
\label{ds2_schr}
ds^2 = r^2\left( -2dx^+dx^- - \beta^2 r^2(dx^+)^2 +(dx^i)^2\right)
   + \frac{dr^2}{r^2},
\ee
which reduces to the metric of $AdS_{d+3}$ in the limit $\beta\to 0$. 
This metric can be obtained in type IIB string theory starting from 
geometries of the form $AdS_{d+3}\times {\cal X}$, where ${\cal X}$
is a compact manifold \cite{Herzog:2008wg,Adams:2008wt,Maldacena:2008wh}. 
The construction can be extended to AdS Schwarzschild black holes. If 
we start from $AdS_5$ we obtain a strongly coupled $2+1$ dimensional 
conformal field theory. This theory has an unusual equation of state,
$P\sim T^4/\mu^2$ \cite{Adams:2008wt}, but it can be shown that the 
fluid saturates the KSS bound, $\eta/s=1/(4\pi)$ 
\cite{Herzog:2008wg,Adams:2008wt}. The method of light-like 
compactifications can also be used to establish a non-relativistic
version of the fluid gravity correspondence \cite{Rangamani:2008gi}.
The equations of conformal fluid dynamics obtained in this way
obey constraints that go beyond those that follow from Galilean
invariance and conformal symmetry alone \cite{Chao:2011cy}, which 
suggest that the light cone method is too restrictive. 

 A new idea for constructing non-relativistic holographic theories is based 
on Horava-Lifshitz gravity, see \cite{Janiszewski:2012nf,Janiszewski:2014ewa}.
Another proposal is based on Vasiliev theory, a gravitational theory with 
higher spin gauge fields \cite{Bekaert:2011cu}. These theories are very 
interesting, but it remains to be seen whether they provide more realistic 
models of non-relativistic fluids. We should note that 
it may be difficult to realize a holographic dual of the dilute Fermi gas at 
unitarity. One reason is that the unitary gas does not have a smooth limit 
as the number of internal degrees of freedom is taken to infinity. Unitary 
Fermi gases with three or more spin states are thermodynamically unstable
because of the existence of deeply bound three-body states, although it is
possible to construct $1/N$ expansions for thermodynamics observables 
based on a $Sp(2N)$ invariant interaction \cite{Nikolic:2007zz,Veilette:2007}.
Another reason is that the unitary gas is just one member of a family of 
non-relativistic conformal field theories. For example, one can construct 
conformal fluids with different thermodynamic and transport properties by 
varying the mass ratio $m_\uparrow/m_\downarrow$ of the two spin states 
\cite{Nishida:2007mr}.  This means that the value of $\eta/s$ at unitarity
is not completely fixed by the symmetries of the unitary gas.

\vspace*{0.6cm}
\noindent
{\bf\large 2.2 Viscosity and flow}

\vspace*{0.4cm} 
\addcontentsline{toc}{subsection}{Nonrelativistic scaling flows}
\noindent
{\it 17. Nonrelativistic scaling flows:} There are two types of experiments 
that have been used to estimate the shear viscosity of an ultracold Fermi 
gas. The first is based on collective oscillations, and the second studies
the expansion after release from a harmonic trap. Both of these involve 
approximate scaling flows. Consider a time dependent density profile of 
the form 
\be 
\label{n_scale}
 n(x,t) =  \frac{1}{b_x(t)b_y(t)b_z(t)}\,
        F\left(\frac{x^2}{b_x^2(t)} +\frac{y^2}{b_y^2(t)}
        +\frac{\lambda^2 z^2}{b_z^2(t)}\right)\, , 
\ee
where $\lambda=\omega_z/\omega_\perp$ is the trap deformation, $F(x)$ is 
an arbitrary function and the scale parameters $b_i(t)$ satisfy the 
initial condition $b_i(0)=1$. At $t=0$ equ.~(\ref{n_scale}) is consistent
with hydrostatic equilibrium, which requires that the density is only a 
function of the local chemical potential $\mu(\vec{x})=\mu-V(\vec{x})$. 

 For time dependent solutions this ansatz satisfies the continuity equation 
provided the velocity field is given by $u_i(x,t)=\alpha_i(t)x_i$ with 
$\alpha_i=\dot{b}_i/b_i$. It is fairly straightforward to find solutions
to the equations of ideal hydrodynamics. In that case entropy is conserved
and we only have to solve the Euler equation, which can be written as a 
coupled set of differential equations for $b_i(t)$. In the case of free
expansion We find
\cite{Schaefer:2009px}
\be 
\ddot{b}_i = \frac{\omega_i^2}{\left(\prod_i b_i\right)^{2/3}}
 \frac{1}{b_i}\, . 
\ee
In a strongly deformed trap the transverse and axial motion approximately 
decouple. For $t\gsim\omega_\perp^{-1}$ the transverse scale parameter
is $b_\perp \sim \sqrt{3/2}\, \omega_\perp t$. The cloud becomes spherical
after a time of order $\sqrt{2/3}\,\omega_z^{-1}$, and then continues
to expand in the transverse direction.

It is more difficult to find solutions to the equation of dissipative 
fluid dynamics. In the case of collective oscillations we can, as a 
first approximation, ignore the increase in entropy. The increase in 
entropy is due to viscous heating which converts the kinetic energy of 
the collective mode to heat, and leads to a slow increase in the temperature
and mean radius of the cloud. The change in the mean radius does not 
directly back-react on the damping rate, which can be computed from 
the viscous force in the Navier-Stokes equation. The result is equivalent
to the calculation of the damping rate from the rate of energy dissipation,
see equ.~(\ref{E_dis}).

 In the case of an expanding system we cannot ignore the increase in
entropy, because viscous heating increases the thermal energy and
therefore also the pressure of the cloud. Pressure drives the expansion
of the cloud, and viscous heating partially compensates for the effects
of viscous friction. Semi-analytical solutions to the hydrodynamic
equations can be found if the viscosity scales like the density of the
system, $\eta(x)=\alpha_n n(x)$, and the equation of state is that of a 
free gas, $P=nT$. For a scaling solution to exist the 
force $f_i=(\nabla_i P)/n$ must be linear in the coordinates. We use
the ansatz $f_i=a_i x_i$ together with the velocity field $u_i=\alpha_i 
x_i$ introduced above. The continuity equation requires $\alpha_i=\dot{b}_i
/b_i$. The Navier-Stokes equation and conservation of energy give a 
set of coupled equations for the scale parameters $a_i$ and $b_i$, 
\bea
\label{ns_for_1} 
\frac{\ddot b_\perp}{b_\perp}  &=&  a_\perp
   -  \frac{2\beta\omega_\perp}{b^2_\perp}
      \left( \frac{\dot b_\perp}{b_\perp} 
                - \frac{\dot b_x}{b_x} \right)\, ,  \\
\label{ns_for_2}
\frac{\ddot b_z}{b_z}  &=& a_z
   +  \frac{4\beta\lambda^2\omega_\perp}{b^2_z}
      \left( \frac{\dot b_\perp}{b_\perp} 
                - \frac{\dot b_z}{b_z} \right)\, ,\\
\label{ns_for_3}
\dot{a}_\perp  &=& 
 \mbox{}-\frac{2}{3}\,a_\perp
   \left(5\,\frac{\dot{b}_\perp}{b_\perp} + \frac{\dot{b}_z}{b_z}\right)
 + \frac{8\beta\omega_\perp^2}{3b_\perp}
  \left(\frac{\dot{b}_\perp}{b_\perp} - \frac{\dot{b}_z}{b_z}\right)^2
  \, , \\
\label{ns_for_4}
 \dot{a}_z  &=& 
 \mbox{}-\frac{2}{3}\,a_z
   \left(4\,\frac{\dot{b}_z}{b_z} + 2\, \frac{\dot{b}_\perp}{b_\perp}\right)
 + \frac{8\beta\lambda^2\omega_\perp^2}{3b_z^2}
  \left(\frac{\dot{b}_\perp}{b_\perp} - \frac{\dot{b}_z}{b_z}\right)^2
  \, , 
\eea
where $\beta=\alpha_nN\omega_\perp/E_0$ and $E_0$ is the total (internal
and potential) energy of the gas cloud. The initial conditions are 
$b_\perp(0)=b_z(0)=1$, $\dot{b}_\perp(0)=\dot{b}_z(0)=0$ as before, and 
$a_\perp(0)=\omega_\perp^2$, $a_z(0)=\omega_z^2$. Terms proportional to 
$\beta$ in eqns.~({\ref{ns_for_1},\ref{ns_for_2}) are linear in $\dot{b}_i$ 
and correspond to viscous friction, whereas dissipative terms in 
eqns.~({\ref{ns_for_3},\ref{ns_for_4}) are quadratic in $\dot{b}_i$ 
and are related to viscous heating. The effect of the viscous terms 
is to slow down the transverse expansion of the cloud. We find, in 
particular, that the delay in the time at which the cloud becomes
spherical is $(\Delta t)/t\sim \beta$. 

 The scaling solution described by eqns.~({\ref{ns_for_1}-\ref{ns_for_4})
was compared to numerical solutions in \cite{Schafer:2010dv}, and it 
forms the basis of the experimental measurements presented in 
\cite{Cao:2010wa,Elliott:2013,Elliott:2013b}. These experiments address,
in the order listed, the high temperature behavior of the shear viscosity,
exact scale invariance at unitarity, and the dependence of the shear 
viscosity on $1/a$ near the unitary limit. 

\begin{figure}[t!]
\bc\includegraphics*[width=8.5cm]{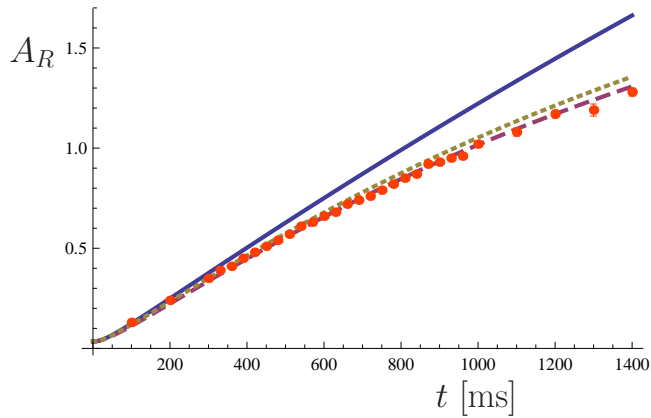}\ec
\caption{\label{fig_relax_vs_hydro}
This figure shows the matching between kinetic theory and Navier-Stokes 
hydrodynamics. We show the evolution of the aspect ratio $A_R=R_z/R_\perp$, 
where $R_z$ and $R_\perp$ are the longitudinal and transverse radii, as a 
function of time. The solid points are data taken at an initial energy 
$E/E_F=3.61$ \cite{Cao:2010wa}. The solid line shows a solution of the 
Euler equation, the long dashed line is a solution of the Navier-Stokes 
equation where the viscosity coefficient $\alpha_n=22.1$ ($\eta=\alpha_n n$)
was adjusted to reproduce the data, and the short-dashed line is a solution 
to the Boltzmann equation in the relaxation time approximation with 
$\tau=\alpha_n/T$.}   
\end{figure} 

\vspace*{0.4cm} 
\addcontentsline{toc}{subsection}{Corona and ballistic limit}
\noindent
{\it 18. Corona and ballistic limit:} The regime of validity of the 
hydrodynamic expansion can be established by computing the Knudsen number 
${\it Kn}=l_{\it mfp}/L$ of the trapped atomic gas. Consider a deformed 
trap containing $N$ atoms. We use $l_{\it mfp}=1/(n\sigma)$, where 
$\sigma$ is a thermal average of the cross section. We also take $L$ to 
be the short axis of the cloud, and use the density at the center of the 
cloud. We find
\be
{\it Kn} = \frac{3\pi^{1/2}}{4(3\lambda N)^{1/3}}
 \left(\frac{T}{T_F}\right)^2\, ,
\ee
where $T_F$ is the Fermi temperature of the cloud. The Fermi temperature
is defined by $k_BT_F=\epsilon_F$, where $\epsilon_F=(3N)^{1/3}\bar{\omega}$ 
with $\bar{\omega}=(\omega_z\omega_\perp^2)^{1/3}$ is the Fermi energy of 
$N$ non-interacting fermions in a harmonic trap. For the conditions probed 
in experiments ${\it Kn}\ll 1$ corresponds to $T\lsim 5T_F$ 
\cite{Adams:2012th}. In this regime the center of the cloud is hydrodynamic.
Note that because of scale invariance any dimensionless scale describing 
the gas is only a function of $T/\mu$. For an ideal scaling expansion 
$T/\mu$ is constant in a comoving fluid element. This means that the 
center of the cloud remains hydrodynamic even as the gas is expanding. 

 In the dilute corona the cloud the mean free path is large and 
hydrodynamics is not applicable. In this regime we can study the expansion 
of the cloud using kinetic theory and the Boltzmann equation. In order to 
understand the connection to hydrodynamics we consider the case that the 
whole cloud is in the kinetic regime. For simplicity we consider solutions 
of the Boltzmann equation (\ref{B_eq}) with the BGK collision term given 
in equ.~(\ref{BGK}). We follow 
\cite{Guery:1999,Pedri:2002,Menotti:2002,Dusling:2011dq} 
and use a scaling ansatz for the distribution function
\be 
\label{f_scal_ans}
 f\left(\vec{x},\vec{v},t\right) = \Gamma(t) 
     f_0\left(\vec{R}(t),\vec{U}(t)\right)\, ,\hspace{0.5cm}
   R_i = \frac{x_i}{b_i}\, ,\hspace{0.5cm}
   U_i = \frac{v_i-\frac{\dot{b}_i}{b_i}x_i}{\theta_i^{1/2}}\, ,\hspace{0.5cm}
    \Gamma = \prod_j\frac{1}{b_j\theta_j^{1/2}}\, ,
\ee
where $b_i,\theta_i$ are functions of $t$ and $f_0(r,v)$ is a solution 
of the Boltzmann equation in equilibrium. In the present case $f_0$ is 
a Maxwell-Boltzmann distribution at temperature $T$ and chemical potential
$\mu=\mu_0-V(x)$. 

 The scaling ansatz (\ref{f_scal_ans}) breaks local thermal equilibrium 
only through the anisotropy of the temperature parameters $\theta_i$. 
The corresponding local equilibrium distribution $f_{\it le}$ can be found 
by replacing $\theta_i\to \bar{\theta}=(\sum_i\theta_i)/3$. This distribution
function is characterized by having the same mean kinetic energy as the 
non-equilibrium distribution $f$. 

 We can obtain a differential equation for the parameters $b_i(t)$ and 
$\theta_i(t)$ by taking moments of the Boltzmann equation. Integrating the 
Boltzmann equation over $\int d^3U\,d^3R\, U_jR_j$ (no sum over $j$) gives 
\cite{Pedri:2002}
\be 
\label{BE_b_j}
\ddot{b}_j + \omega_j^2 b_j 
   - \omega_j^2 \frac{\theta_j}{b_j} = 0 \, .
\ee
Note that the second term is due to the external potential and is not 
present if we consider free expansion. Integrating over $\int d^3U\,d^3R\, 
U_jU_j$ gives
\be 
\label{BE_th_j}
\dot{\theta}_j + 2\frac{\dot{b}_j}{b_j}\theta_j = 
 - \frac{1}{\tau_0} \left(\theta_j-\bar\theta\right) \, . 
\ee
Moments of the Boltzmann equation weighted with $R_jR_j$ do not provide 
additional constraints. Together with the initial conditions $b_j(0)=
\theta_j(0)=1$ and $\dot{b}_j(0)=0$ the two equations (\ref{BE_b_j}) and
(\ref{BE_th_j}) describe the evolution of an expanding cloud.

 In the free streaming limit $\tau_0\to\infty$ equ.~(\ref{BE_b_j}) 
provides an exact solution of the Boltzmann equation. We get $\theta_i
=1/b_i^2$ and $b_i=(1+\omega_i^2t^2)^{1/2}$. In the opposite limit, $\tau_0
\to 0$, we get $\theta_i=\bar\theta$ with $\bar\theta=(\prod_i b_i)^{-2/3}$
and equ.~(\ref{BE_b_j}) is equivalent to the Euler equation. Keeping leading 
order corrections in $1/\tau_0$ leads to a solution of the Navier-Stokes 
equation with $\zeta=0$ and $\eta = n\tau_0 T_{\it le}$ \cite{Dusling:2011dq},
see Fig.~\ref{fig_relax_vs_hydro}. We have therefore obtained a kinetic model 
that interpolates between the ballistic and Navier-Stokes limits. The 
shortcoming of the model is that we have assumed that $\tau_0$ is a constant 
which is independent of the density and temperature. From the matching 
condition between $\tau_0$ and $\eta$ we observe that this implies that 
the shear viscosity is proportional to density, which is at variance with 
the expected behavior in the low density limit.

 It is possible to consider a more general behavior for $\tau_0$, 
for example by allowing $\tau_0$ to be a functional of the distribution
function. In order to obtain $\eta\sim {\it const}$ we have to assume 
that $1/\tau\sim \int d^3v\, f(\vec{x},\vec{v},t)$ \cite{Dusling:2011dq}. 
Matching quadratic moments of the Boltzmann equation to the Navier-Stokes
equation gives an effective density dependent shear viscosity $\eta(x)
=\lambda^3 n(x)/\bar{n}$, where $\bar{n}$ is the average density. 
The problem is that this solution does not automatically reduce to
the free streaming limit in the dilute part of the cloud. 

\begin{figure}[t!]
\bc\includegraphics[width=8cm]{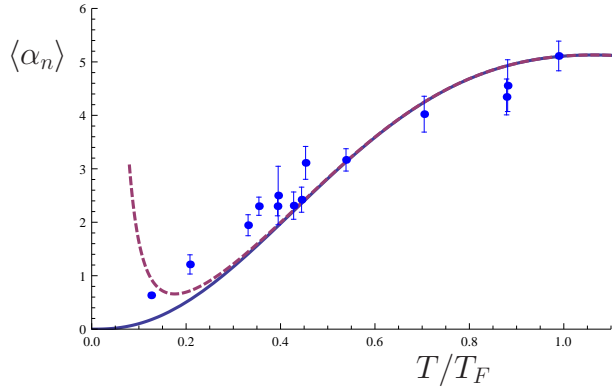}
\ec
\caption{\label{fig_eta_trap}
Trap averaged shear viscosity to density ratio $\langle\alpha_n\rangle$. 
We show $\langle\alpha_n\rangle$ as a function of $T/T_F^{\it trap}$, where
$T_F^{\it trap}=(3\lambda N)^{1/3}\omega_\perp$ is the Fermi temperature of 
the trap. We have chosen $N=2\cdot 10^5$ and $\lambda=0.045$ as in 
\cite{Kinast:2005}. The solid line shows the kinetic theory result, 
the dashed line includes fluctuation corrections to the shear viscosity.
The data are from \cite{Cao:2011fh}, which is a reanalysis of the
results reported in \cite{Kinast:2005}. }   
\end{figure}

\vspace*{0.4cm} 
\addcontentsline{toc}{subsection}{Transient fluid dynamics and the kinetic
limit}
\noindent
{\it 19. Transient fluid dynamics and the kinetic limit:} In the 
case of collective oscillations an improved matching to kinetic 
theory can be obtained by considering transient hydrodynamics as in 
equ.~(\ref{pi_relax}). For a system with harmonic time dependence 
the relaxation time equation is solved by $\delta\Pi_{ij}=-\eta(
\omega)\sigma_{ij}$ with a frequency dependent shear viscosity 
$\eta(\omega)=\eta/(1+i\omega\tau_R)$. The trap integrated shear 
viscosity is 
\be 
\langle \eta \rangle = 
  \int d^3x\, \frac{\eta(\vec{x})}{1+\omega^2\tau_R(\vec{x})^2}\, . 
\ee
Using $\tau_R(\vec{x})=\eta(\vec{x})/(n(\vec{x})T)$ we observe that the 
trap average is well defined even in the dilute corona. In this
regime the relaxation time is large, and the viscous stresses 
never reach the Navier-Stokes value. We can now use the method
described in Sect.~\ref{sec_flow_nr} to compute the damping rate
of collective modes. For the transverse breathing mode the velocity 
profile is $\vec{u}\sim (x,y,0)$ and the damping constant is given
by 
\be 
 \Gamma =  \frac{\langle \alpha_n\rangle}{(3N\lambda)^{1/3}}
    \frac{\omega_\perp}{(E_0/[N\epsilon_F])}\, .
\ee
Here, $E_0$ is the total (potential and internal) energy of the trapped
gas, $\epsilon_F = (3N\lambda)^{1/3}\omega_\perp$ is the Fermi energy
of the trapped system, and $\langle\alpha_n\rangle=\langle\eta\rangle 
/N$. A typical result is shown in Fig.~\ref{fig_eta_trap}.

 The advantage of this formalism is that it reproduces the hydrodynamic
and kinetic theory results in certain limits. For a density dependent 
shear viscosity of the form $\eta \sim n$ it reduces, up to corrections
of second order in the gradient expansion, to the Navier-Stokes
result. For $\eta\sim \lambda^{-3}_{\it dB}$ the result reproduces, in 
the high temperature limit, the damping rate obtained from solutions
of the Boltzmann equation in the Knudsen limit \cite{Bruun:2007}.

 An analysis of the collective mode data reported in \cite{Kinast:2005} 
using this method can be found in \cite{Schaefer:2009px}. For $\eta
\sim \lambda^{-3}_{\it dB} \sim (mT)^{3/2}$ we find that $\Gamma\sim 
T^3$ at low temperature, and $\Gamma\sim 1/T$ at high temperature. 
The fact that the damping rate decreases in the high temperature 
limit even though the viscosity is growing is related to the increase
in the relaxation time $\tau_R\simeq \eta/(nT)$. As the relaxation
time grows the strain $\sigma_{ij}$ and the induced stress $\pi_{ij}$
are increasingly out of phase, and the dissipated energy is reduced.
This implies that careful measurements of the damping rate in the 
regime where the $T^3$ behavior starts to break down can be used to 
measure $\tau_R$. A similar transition from hydrodynamic to kinetic 
behavior is also seen in the dependence of $\Gamma$ on the particle 
number. At low temperature the damping rate scales as $\Gamma\sim 
N^{-1/3}$, and at high temperature the scaling law changes to $\Gamma
\sim N^{1/3}$.

\vspace*{0.6cm}
\addcontentsline{toc}{section}{Endnotes: Relativistic fluids}
\noindent
{\bf\large 3 Relativistic fluids}

\vspace*{0.4cm}
\noindent
{\bf\large 3.1 The quark gluon plasma}

\vspace*{0.4cm}
\addcontentsline{toc}{subsection}{Transport properties of the quark gluon 
plasma}
\noindent
{\it 20. Transport properties of the QGP:} Comparing the shear viscosity 
to other transport properties of the quark gluon plasma provides 
additional information on the existence and properties of quasi-particles,
and on the mechanism for charge and momentum transport in the plasma. 
In the high temperature limit the full set of transport coefficients 
has been computed in kinetic theory, and there are exploratory 
measurements of shear and bulk viscosity, heavy quark diffusion as
well as electric conductivity on the lattice \cite{Meyer:2011gj}.

 Weak coupling results for transport coefficients at second order  
in the hydrodynamic expansion are given by \cite{York:2008rr}
\be
\label{2nd_order_QCD}
\eta\tau_R = (5\cdots 5.9)\frac{\eta^2}{sT}\, , \hspace{0.25cm}
    \lambda_1 = (4.1\cdots 5.2)\frac{\eta^2}{sT}\, , \hspace{0.25cm}
    \lambda_2 =-2\eta\tau_R\, ,                      \hspace{0.25cm}
    \lambda_3 = 0\, ,                                \hspace{0.25cm}
    \kappa_R = \frac{5s}{8\pi^2 T}\, ,  
\ee
where the numerical ranges correspond to the variation of the numerical
coefficients with the coupling constant. The coefficients $\tau_R$ and 
$\lambda_i$ scale inversely with $g$ and were determined using kinetic 
theory \cite{York:2008rr}. The quantity $\kappa_R$, which governs the 
curvature term in the stress tensor is independent of $g$, and was 
determined using the Kubo relation \cite{Romatschke:2009ng}. Kubo 
relations also show that, in general, $\lambda_3$ is not zero 
\cite{Moore:2010bu}. The results given in equ.~(\ref{2nd_order_QCD})
can be compared to the AdS/CFT predictions in equ.~(\ref{2nd_order_ads}),
and to the non-relativistic results in equ.~(\ref{2nd_order_cag}).

The bulk viscosity of the quark gluon plasma was calculated 
by Arnold, Dogan, and Moore \cite{Arnold:2006fz}. The result is  
\be 
\label{zeta_qcd}
\zeta = \frac{A\alpha_s^2T^3}{\log(\mu^*/m_D)}\, , 
\ee
where $A=0.443$ and $\mu^*=7.14\, T$ in pure gauge QCD, and $A=0.657$, 
$\mu^*=7.77\, T$ in QCD with $N_f=3$ light quark flavors. The dependence
of $\zeta$ on $\alpha_s$ can be understood from the simple estimate 
$\zeta \sim ({\cal E}-3P)^2\eta$ with ${\cal E}-3P\sim \alpha_s^2$ 
and $\eta\sim 1/\alpha_s^2$. The thermal conductivity of a quark 
gluon plasma is a somewhat subtle quantity. At zero chemical potential
one cannot distinguish between energy and particle transport, and 
the thermal conductivity is not defined. In the limit of small chemical
potential, $T\gg\mu$, the relaxation time approximation gives $\kappa
\sim T^4/(\alpha_s^2\mu^2)$ \cite{Danielewicz:1984ww}. This result
appears to be singular in the limit $\mu\to 0$, but the dissipative 
contribution to the energy and baryon currents are finite. The 
behavior of $\eta$, $\zeta$ and $\kappa$ in the limit $\mu\gg T$ 
is reviewed in \cite{Alford:2007xm}.

 The heavy quark diffusion constant is \cite{Svetitsky:1987gq,Moore:2004tg}
\be 
\label{D_QCD}
 D=\frac{36\pi}{C_Fg^4T}\left[
 N_c\left(\log\left(\frac{2T}{m_D}\right)+c\right)
 +\frac{N_f}{2}\left(\log\left(\frac{4T}{m_D}\right)+c\right)
 \right]^{-1},
\ee
where $C_F=(N_c^2-1)/(2N_c)$ and $c=0.5-\gamma_E+\zeta'(2)/\zeta(2)$.
Comparing this result with the shear viscosity given in equ.~(\ref{eta_qcd}) 
we observe that heavy quark and momentum diffusion are related. In the 
relevant range of coupling constants one finds $DT\simeq 6(\eta/s)$
\cite{Moore:2004tg}. This relation provides a test whether transport
is dominated by quasi-particles, because in the strong coupling limit
of the AdS/CFT correspondence we find $DT\ll (\eta/s)$. 

\vspace*{0.6cm}
\noindent
{\bf\large 3.2 Flow, higher moments of flow, and viscosity}

\vspace*{0.4cm}
\addcontentsline{toc}{subsection}{Scaling flows, from Bjorken to Gubser}
\noindent
{\it 21. Scaling flows, from Bjorken to Gubser:} The Bjorken flow 
discussed in Sect.~\ref{sec_flow} is an exact solution of the 
Navier-Stokes solution with longitudinal boost invariance and 
no dependence on the transverse coordinates. The Bjorken solution 
is most easily described using a set of coordinates $(\tau,\eta,
r,\phi)$, where $\tau=(t^2-z^2)^{1/2}$ is proper time, $\eta=
(1/2)\log[(t+z)/(t-z)]$ is rapidity, and $(r,\phi)$ are polar
coordinates. The metric is 
\be 
ds^2 = -d\tau^2 + \tau^2d\eta^2 + dr^2 + r^2d\phi^2\, . 
\ee
Bjorken flow corresponds to a velocity field $u^\mu=(1,0,0,0)$.
Energy density and pressure are functions of $\tau$ only and scale 
invariance requires $\epsilon(\tau)=P(\tau)/3$. In ideal fluid 
dynamics ${\cal E}(\tau)\sim 1/\tau^{4/3}$. If dissipation is included
${\cal E}(\tau)$ is determined by equ.~(\ref{bj_ns}).

 Gubser discovered a generalization of Bjorken flow that includes
transverse expansion, and therefore serves as a much more realistic
model of a heavy ion collision \cite{Gubser:2010ze,Gubser:2010ui}.
The solution was inspired by the fluid-gravity correspondence, but 
it can be described purely as a solution to the relativistic Euler
and Navier-Stokes equations for a scale invariant fluid. Scale 
invariance implies that ${\cal E}=3P$ and $\eta=H_0T^3$. The 
velocity profile is 
\be 
u_\mu = \left(\cosh(\kappa),0,\sinh(\kappa),0\right)\, , 
  \hspace{0.25cm}
\kappa = \mathrm{arctanh}\left(\frac{2q^2 \tau r}
                             {1 + q^2 \tau^2 + q^2 r^2}\right) \, ,
 \ee
where $q$ is a parameter. This solution has a hidden $SO(3)$ 
symmetry that can be made manifest by switching to another set 
of coordinates, see \cite{Gubser:2010ze}. Consider first the 
ideal case. The energy density can be written as 
\be
{\cal E} = \frac{\hat{\cal E}(g)}{\tau^\alpha} \, ,
 \hspace{0.4cm}
\alpha=4\, , 
 \hspace{0.4cm}
g = \frac{1 - q^2 \tau^2 + q^2 r^2}{2q\tau}\, . 
\ee
The solution of the Euler equation is 
$ \hat{\cal E} = \hat{\cal E}_0/(1+g^2)^{4/3}$, which corresponds to 
\be
  {\cal E} = \frac{\hat{\cal E}_0}{\tau^{4/3}} 
     \frac{(2q)^{8/3}} 
          {\left[ 1 + 2q^2 (\tau^2 + r^2) 
                    +  q^4 (\tau^2 - r^2)^2 \right]^{4/3}} \, .
\ee
Taking $q\to 0$ with $\hat{\cal E}_0q^{8/3}$ constant we recover the 
Bjorken solution. As in the Bjorken case the flow profile is not
modified by dissipative effects. The evolution of the energy density 
is most easily described in terms of $\hat{T} = \hat{\cal E}^{1/4}$.
We get 
\be
\hat{T} = 
     \frac{\hat{T}_0}{(1+g^2)^{1/3}} 
     +  \frac{ H_0 g}{(1+g^2)^{1/2}} 
    \left[ 1 - \left(1+g^2\right)^{1/6}
       \mbox{}_2F_1\left(\frac{1}{2},\frac{1}{6};\frac{3}{2};-g^2 \right)
       \right] \, ,
\ee
where $\mbox{}_2F_1(\alpha,\beta;\gamma;\delta)$ is a hypergeometric
function. Gubser also studied the evolution of small fluctuations
around this solution \cite{Gubser:2010ui}, see also 
\cite{Shuryak:2009,Staig:2010,Staig:2011,Kapusta:2011}. He finds
that modes with wave number $k$ are suppressed by 
\be 
 P_k =\exp\left(-\frac{2}{3}\frac{\eta}{s}\frac{k^2t}{T}\right)\, . 
\ee
The Glauber model gives an approximately flat spectrum of initial 
perturbations \cite{Mocsy:2011}, so this formula predicts that higher flow
harmonics are exponentially damped. This is in rough agreement with 
the data \cite{Lacey:2013qua}, although the details are more complicated. 
In particular, lower moments of the initial energy deposition depend
on the geometry and the initial state model, and there is some amount 
of mode mixing in the hydrodynamic response 
\cite{Gardim:2011,Teaney:2012,Floerchinger:2013tya}. 

 Bjorken flow arises naturally in weak coupling approaches to
thermalization \cite{Blaizot:1987nc,Baier:2000sb}. In strong coupling
calculations, based on the collision of shock waves in $AdS_5$, more 
complicated flow profiles are obtained \cite{Casalderrey-Solana:2013aba}.
An interesting parameterization, termed ``complex deformation of Bjorken 
flow'', of these flow profiles was recently suggested in \cite{Gubser:2012gy}.
Consider cartesian coordinates $(t,\vec{x})$ and 
\bea 
 u_\mu^{\mathbb{C}} &=& \frac{1}{\sqrt{(t+\mathfrak{t}_3)^2-x_3^2}}
  \left(-( t+\mathfrak{t}_3),0,0,x_3\right)\, ,  \nonumber \\
{\cal E}^{\mathbb{C}} &=& \frac{ {\cal E}_0^{\mathbb{C}} }
      {[(t+\mathfrak{t}_3)^2-x_3^2]^{2/3}} \, ,  \\
 \Pi_{\mu\nu}^{\mathbb{C}} &=& {\cal E}^{\mathbb{C}} 
         u_\mu^{\mathbb{C}} u_\nu^{\mathbb{C}}
   + \frac{ {\cal E}^{\mathbb{C}}}{3} 
    \left( g_{\mu\nu} +  u_\mu^{\mathbb{C}} u_\nu^{\mathbb{C}} \right)\, .  
\nonumber 
\eea
If $\mathfrak{t}_3$ is a real parameter then this flow profile is just a 
time translation of the Bjorken solution. However, if $\mathfrak{t}_3$ is 
a complex parameter then we obtain something new. Note that $\Pi_{\mu\nu}
\equiv {\rm Re}\, \Pi_{\mu\nu}^{\mathbb{C}}$ satisfies the conservation
laws, but $u_\mu\equiv  {\rm Re}\, u_\mu^{\mathbb{C}}$ and ${\cal E}
\equiv  {\rm Re}\,{\cal E}^{\mathbb{C}}$ are not solutions of the 
Euler equation, because they do not satisfy the constitutive equation. 
Nevertheless, for suitable choices of the phases, in particular for 
${\rm arg}\,\mathfrak{t}_3=\pi/2$ and ${\rm arg}\,{\cal E}^{\mathbb{C}}=
\pi/3$, interesting flow profiles are obtained. These flows are 
Landau-like at early time, glasma-like ($P_L\simeq -{\cal E}$) near 
the light cone, and Bjorken-like at late time and in the central 
rapidity slice.

\vspace*{0.4cm} 
\addcontentsline{toc}{subsection}{From kinetics to hydrodynamics in 
relativistic heavy ion collisions}
\noindent
{\it 22. From kinetics to hydrodynamics in relativistic heavy ion 
collisions:} 
The transition from kinetic theory to hydrodynamic behavior 
in relativistic heavy ion collisions has been studied by a number 
of authors. Kolb et al.~compared scaling relations in the ballistic
and hydrodynamic limits to the data at the SPS and RHIC \cite{Kolb:2000fh}.
They found clear indications for hydrodynamic behavior at RHIC. The 
conditions for achieving the hydrodynamic limit in in a kinetic model 
of the quark gluon plasma were studied by Molnar and Gyulassy 
\cite{Molnar:2007}. They find that obtaining hydrodynamic behavior
in a model that includes elastic $2\leftrightarrow 2$ scattering only 
requires rather extreme assumptions concerning the initial parton density 
or the parton cross section. More recently, it was shown that the inclusion 
of $2\leftrightarrow 3$ processes leads to a more rapid approach to 
hydrodynamics \cite{Xu:2004mz,Xu:2007jv}. We should note, however, that 
the correct implementation of $2\leftrightarrow 3$ scattering is still 
being discussed \cite{Chen:2013,Fochler:2013epa}. 

 In the case of expanding Fermi gases we emphasized the need to find 
a transport model that smoothly interpolates between hydrodynamics 
in the center of the cloud and free streaming in the dilute corona. 
In the case of a heavy ion collision we would like to make contact 
with longitudinal free streaming at early times, and with both 
longitudinal and transverse free streaming at late times. This can 
be accomplished by using a kinetic framework in which the longitudinal 
and transverse temperatures are allowed to differ 
\cite{Florkowski:2010,Martinez:2010}. Using moments of the Boltzmann 
equation one can derive a set of fluid dynamic equations that involve 
additional, non-hydrodynamic, modes. If the mean free path is short 
these modes relax quickly and one recovers the usual Navier-Stokes 
equation. In the opposite limit the equations reproduce the free 
streaming limit. 

 A very different approach that has many of the same features is the 
lattice Boltzmann equation (LBE) \cite{Romatschke:2011hm,Romatschke:2011qp}.
The LBE is a kinetic equation that acts on a very simple, discrete, 
velocity space. The LBE provides a very robust and efficient implementation 
of the Navier-Stokes equation in the limit of a short mean free path, 
and reduces to free streaming in the limit $l_{\it mfp}\to\infty$.

\begin{figure}[t!]
\begin{center}
\includegraphics*[width=7.5cm]{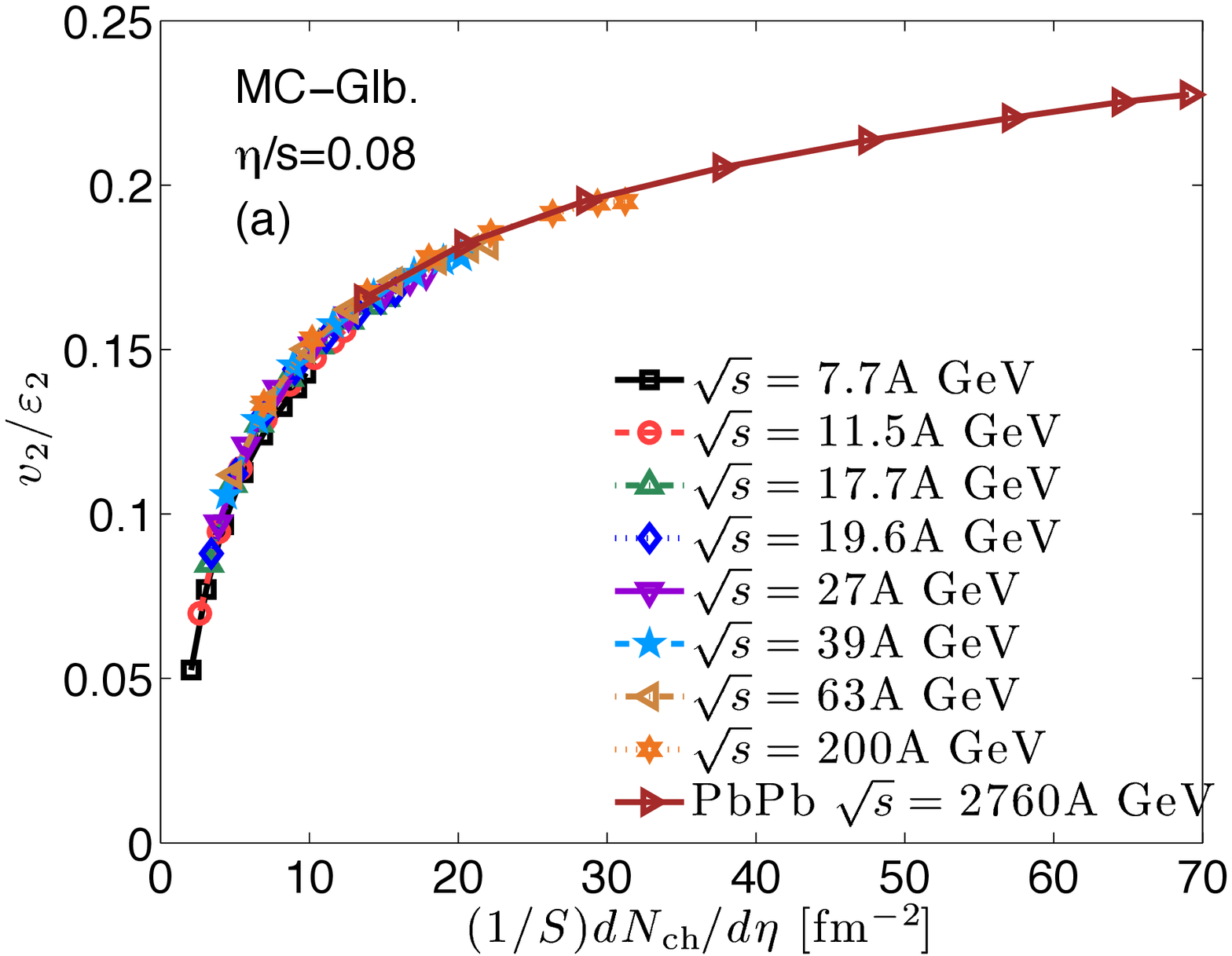}
\includegraphics*[width=7.5cm]{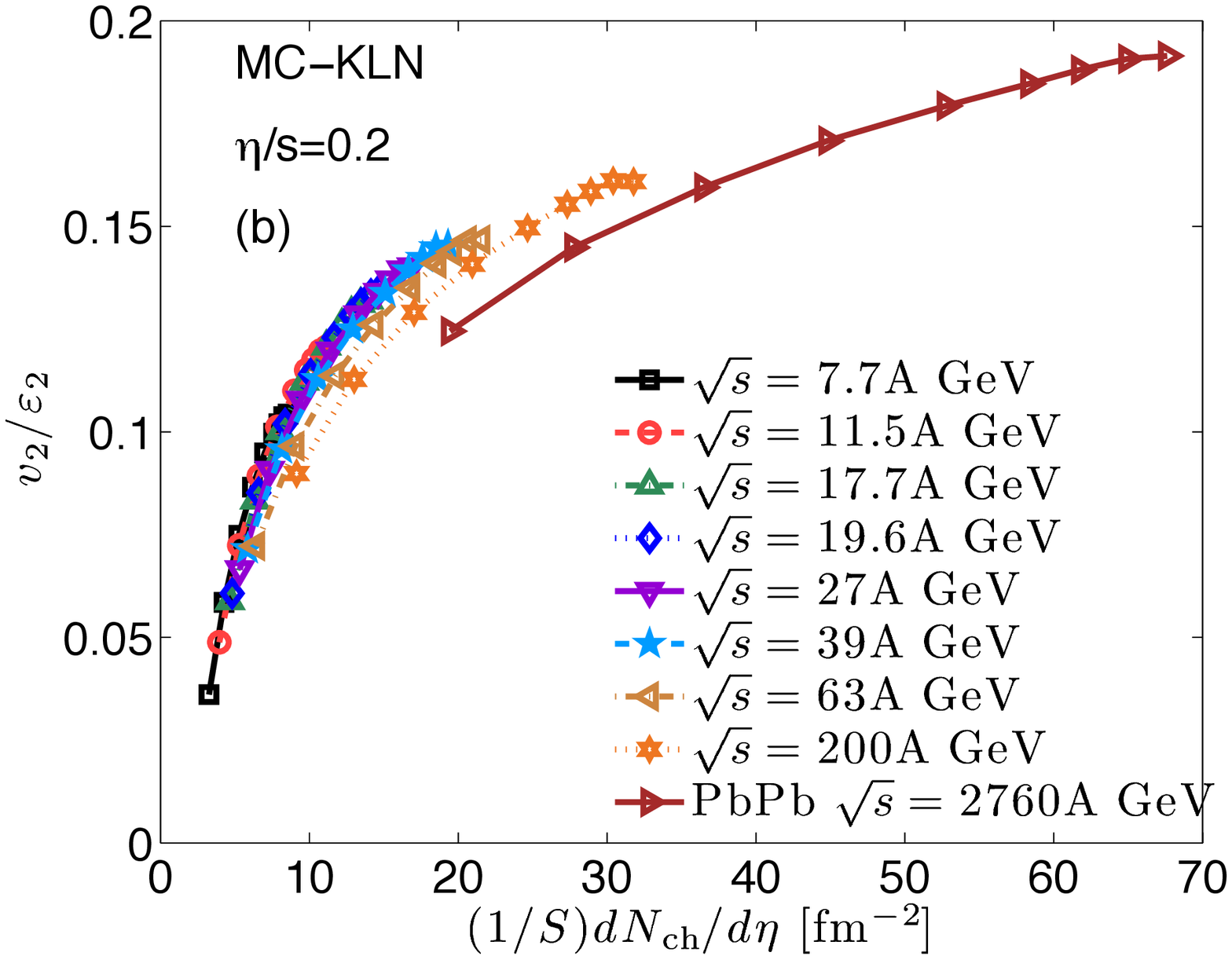}
\end{center}
\caption{\label{fig_dNdy_scal}
Eccentricity scaled elliptic flow $v_2$ plotted as a function of
the charged hadron multiplicity density $dN/dy$ divided by 
the overlap area $S$ for different collision energies, from
\cite{Shen:2012vn}. The left panel shows results from a hydrodynamic 
simulation for MC-Glauber initial conditions with $\eta/s= 0.08$, 
and the right panel shows a calculation with MC-KLN initial 
conditions and $\eta/s= 0.2$.}
\end{figure}

\vspace*{0.4cm} 
\addcontentsline{toc}{subsection}{Knudsen number scaling}
\noindent
{\it 23. Knudsen number scaling:} In Sect.~\ref{sec_flow} we argued 
that the local expansion parameter for the hydrodynamic gradient 
expansion in a heavy ion collision is given by $\eta/(s\tau T)$. 
In order to compare experimental data from collisions at different
beam energies, impact parameters, and nuclear mass numbers it is
also important to identify global variables that control the 
validity of hydrodynamics. An important step in this direction
was taken by Heiselberg and Levy, who studied elliptic flow
in the dilute limit \cite{Heiselberg:1998es}. The contribution
from single elastic scattering events is 
\be 
\label{v2_bal}
v_2(p_{1,T}) = \frac{\delta}{16S}\, \frac{dN}{dy}\, 
      \frac{v_{1,T}^2}{\langle v_{12}^2\rangle}\, 
      \langle v_{12,T}\sigma_{\it tr}\rangle\, , 
\hspace*{0.5cm}
\delta=\frac{\langle R_y^2-R_x^2\rangle}{\langle R_y^2+R_x^2\rangle}\, , 
\ee
where $dN/dy$ is the multiplicity per unit of rapidity, $S$ is the 
transverse overlap area, and $v_{1,T}$ is the transverse velocity. The 
symbol $\langle .\rangle$ denotes an average over the distribution
of particle 2, and $v_{12}$ is the relative velocity of particles 1 and 2. 
We have also defined the transport cross section $\sigma_{\it tr}$, the 
cross section weighted by $(1-\cos\theta)$, where $\cos\theta$ is the 
scattering angle. Finally, $R_x$ and $R_y$ are the radii of the overlap
region, and $\delta$ is the elliptic deformation. It is standard to 
characterize the elliptic deformation not in terms of $\delta$, but 
using the quantity $\epsilon_2$ defined by 
\be
\epsilon_2 = \frac{\langle y^2 -x^2\rangle}{\langle y^2+x^2 \rangle}\, ,
\ee
where the average is carried out using the energy density as a weight
function. Based on equ.~(\ref{v2_bal}) we predict
\be 
 \frac{v_2}{\epsilon_2} \sim \frac{1}{S} \frac{dN}{dy} 
   \langle \sigma \rangle \, . 
\ee
Following the arguments in Sect.~\ref{sec_kin} we expect that the parameter
$(1/S)(dN/dy) \langle \sigma \rangle$ also appears in fluid dynamics. This
is indeed the case, as we can see using the following argument 
\cite{Bhalerao:2005mm}. Consider a fireball of size $\bar{R}\simeq 
\sqrt{R_x^2+R_y^2}$ which is undergoing Bjorken expansion in the 
longitudinal direction. The time scale for transverse expansion is 
$\tau=\bar{R}/c_s$, and the density at this time is $n\sim 1/(c\tau S)
(dN/dy)$. This implies that the inverse Knudsen number is 
\be 
\frac{1}{{\it Kn}} = \frac{\bar{R}}{l_{\it mfp}} 
  = \bar{R}n\langle\sigma\rangle 
  = \frac{c_s}{c} \frac{1}{S}\frac{dN}{dy}\langle\sigma\rangle \, . 
\ee
Knudsen number scaling of $v_2/\epsilon_2$ was first studied by Voloshin 
and Poskanzer, see \cite{Voloshin:1999gs,Alt:2003ab}. The results compiled
in \cite{Alt:2003ab} demonstrate nice data collapse if different systems, 
centralities, and beam energies are plotted as a function of $(1/S)(dN/dy)$. 
The compilation also shows that $v_2/\epsilon_2$ rises almost linearly
with $(1/S)(dN/dy)$, and that the RHIC data at 200 GeV per nucleon
saturate the flow predicted by ideal hydrodynamics. A more recent 
analysis of data from the RHIC beam energy scan and $Pb+Pb$ collisions
at the LHC is shown in Fig.~\ref{fig_dNdy_scal} \cite{Shen:2012vn}.
There is some uncertainty related to different models for $\epsilon_2$,
which is reflected in the difference between the left and right panels. 
The main result is that there is data collapse, which is excellent 
for the Monte Carlo Glauber model, and not quite as good in the case 
of the KLN model. We also observe some curvature in $v_2/\epsilon_2$.
This means that there are viscous effects at high energy, and that 
there is no saturation of flow even at the LHC. 

 An important assumption in Fig.~\ref{fig_dNdy_scal} is that the 
effective cross section is not a function of the collision parameters.
At high temperature the quark gluon plasma is scale invariant and
we expect $\langle\sigma\rangle \sim s^{-2/3}$. Then 
\be 
 \frac{1}{{\it Kn}} \sim 
   \left(\frac{c_s}{c}\frac{dN}{dy}\right)^{1/3} \, , 
\ee
and the overlap area does not appear in the estimate for the Knudsen
number. This makes a significant difference when comparing $pA$ and 
$AA$ collisions, and there is some evidence that $dN/dy$ scaling is 
preferred by the data \cite{Basar:2013hea}.

\vspace*{0.4cm} 
\addcontentsline{toc}{subsection}{Sensitivity to the relaxation time}
\noindent
{\it 24. Sensitivity to the relaxation time:} We have seen that an 
important consistency check for the hydrodynamic description is to
show that the dependence on second order coefficients, like the 
relaxation time, is weak. On the other hand, we have also argued 
that second order hydrodynamics can be used to regularize instabilities
and acausal behavior of the Navier-Stokes equation. How can both
of these statements be correct? In particular, if second order 
terms serve as regulators then the $\tau_R\to 0$ limit cannot be
smooth.

 It is straightforward to compute the limiting speed of a shear 
wave in transient fluid dynamics \cite{Romatschke:2009im}
\be 
\label{v_max}
 v_{\it max} = \lim_{q\to\infty} \frac{\partial \omega}{\partial q} 
 = \left[ \frac{\eta}{({\cal E}+P)\tau_R}\right]^{1/2}\, .  
\ee
We observe that $v_{\it max}\leq 1$ is satisfied if $(\eta/s)<(\tau_RT)$. 
For $\eta/s\ll 1$ this implies that there is indeed a large window 
for $\tau_R T$ in which both causality and the constraint from the 
validity of the gradient expansion, $\tau_R T<1$, are satisfied. 
We also note that if acausal modes are excluded by incorporating an
explicit cutoff then the limit $\tau_R\to 0$ is smooth. 

\vspace*{0.6cm}
\addcontentsline{toc}{section}{Endnotes: Frontiers}
\noindent
{\bf\large 4 Frontiers}

\vspace*{0.4cm} 
\addcontentsline{toc}{subsection}{The role of the AdS/CFT correspondence}
\noindent
{\it 25. The role of the AdS/CFT correspondence:} The successful hydrodynamic 
description of heavy ion collisions outlined in Sect.~\ref{sec_flow} does
not rely on the AdS/CFT correspondence. Indeed, as we have emphasized, 
hydrodynamics is an effective theory of the long distance behavior of
non-equilibrium systems that does not depend on specific features of
the underlying microscopic theory. Nevertheless, the success of nearly
perfect fluid dynamics in describing heavy ion collisions at RHIC and
the LHC is frequently mentioned as one of the principal success stories of 
string theory and the AdS/CFT correspondence. This is indeed justified
for several reasons: 

\begin{enumerate}
\item {\it The possibility of nearly perfect fluidity:} The idea that
$\eta/s$ could be as small as $0.1$ was first discussed in the 
important work of Danielewicz and Gyulassy \cite{Danielewicz:1984ww}.
It should be noted, however, that this work can be interpreted as 
showing that i) the applicability of hydrodynamics in relativistic
heavy ion collisions requires $\eta/s\ll 1$, ii) a value of $\eta/s$ 
this small requires rather extreme assumptions about kinetic theory.
It is then reasonable to conclude that these assumptions are not likely 
to be realized in practice, and the focus in the years following the
publication of \cite{Danielewicz:1984ww} shifted from dissipative 
fluid dynamics to parton cascades \cite{Wang:1991,Geiger:1992} (see 
\cite{Rischke:1998fq,Muronga:2001zk} for rare exceptions). Interest in 
nearly perfect fluid dynamics was revived because of the experimental 
discoveries at RHIC, combined with the almost contemporaneous 
result that the strongly coupled fluid described by AdS/CFT satisfies
$\eta/s=1/(4\pi)$.

\item {\it Second order conformal fluid dynamics:} The general 
structure of the equations of relativistic conformal fluid dynamics
can be established purely based on symmetry arguments, but in practice
the equations were first found with the help of the AdS/CFT correspondence
\cite{Baier:2007ix}. In principle the Israel-Stewart equations form a 
consistent subset of the most general second order equations. However, in 
practice some studies employed truncations of the Israel-Stewart equations 
that are not consistent with conformal symmetry \cite{Heinz:2005bw}. The 
truncated equations exhibit significant dependence on the relaxation time 
\cite{Song:2007fn}, which becomes much weaker once the full Israel-Stewart 
equations are considered \cite{Song:2008si}. These differences could have 
been resolved without AdS/CFT, but historically the holographic flows 
found by Heller and Janik \cite{Heller:2007qt}, and the subsequent matching 
to second hydrodynamics provided by Baier et al.~\cite{Baier:2007ix} played 
a central role in explaining the relation between different approaches. 
More recently, anomalous transport coefficients were discovered using
the AdS/CFT correspondence \cite{Erdmenger:2008rm}. These transport
coefficients can be understood based on the general properties of fluid 
dynamics \cite{Son:2009tf}, but in the literature the presence of 
these terms had been missed. 

\item {\it Rapid hydrodynamization:} The hydrodynamic description of
elliptic flow in relativistic heavy ion collisions requires a very
short equilibration time $\tau_{\it eq}\sim 1$ fm/c. Equilibration 
can be understood in kinetic theories based on $2\to 2$ and $2\to 3$ 
scattering, and equilibration times are known to be further reduced 
by collective plasma effects. However, quantitative estimates give 
$\tau_{\it eq}\gsim (2-3)$ fm/c, see for example \cite{Baier:2011}. On 
the other hand, fast equilibration is natural in holographic theories 
\cite{Chesler:2010bi}. In addition to that, AdS/CFT shows that the 
Navier-Stokes description can be reliable even if non-equilibrium 
contributions to the pressure are large \cite{Chesler:2010bi,Heller:2011ju}, 
as is the case in the early stages of a heavy ion collision. 

\item {\it Absence of quasi-particles:} AdS/CFT provides an explicit
example of a fluid in which hydrodynamic behavior does not emerge
from an underlying kinetic theory. While we still do not know 
whether this is the correct picture for the quark gluon plasma
produced at RHIC and the LHC, the existence of an alternative to 
the quasi-particle paradigm has been very useful for studying 
the role of various assumptions in analyzing the data. 

\end{enumerate}

 Despite this impressive list we should emphasize that a lot of important 
work on relativistic fluid dynamics has little or no relation to AdS/CFT. 
A variety of schemes for transient higher order fluid dynamics were 
developed \cite{Geroch:1990bw,Ottinger:1998,Denicol:2012cn}, and these
schemes provided the tools to test the sensitivity of the analysis
of the RHIC data to poorly constrained high order transport coefficients, 
see \cite{Dusling:2007gi,Song:2007ux,Luzum:2008cw}. We note, in particular,
that even those implementations of higher order fluid dynamics that are 
based on the conformal, AdS/CFT inspired, second order equations make use 
of the idea of transient fluid dynamics, see equ.~(\ref{pi_relax}).
This approach emerges naturally in kinetic theory, but it is not a 
systematic approximation to fluid dynamics in AdS/CFT.

\vspace*{0.4cm} 
\addcontentsline{toc}{subsection}{Puzzles and challenges}
\noindent
{\it 26. Puzzles and challenges:} As noted in Sect.~\ref{sec_out}
there are number of puzzles related to the hydrodynamic description
of heavy ion collisions at RHIC and LHC. 

\begin{enumerate}
\item Approximate beam energy independence of the charged particle
elliptic flow $v_2(p_T)$: The elliptic flow of charged particles
has been measured over a large range of beam energies, from the 
low end of the RHIC beam energy scan, $\sqrt{s_{NN}}=7.7$ GeV, 
to the current LHC energy $\sqrt{s_{NN}}=2.76$ TeV
\cite{Aamodt:2010pa,Chatrchyan:2012ta,Adamczyk:2012ku}. In a given 
centrality class the results are essentially beam energy independent. 
Within hydrodynamics this is somewhat surprising because many 
variables, such as the lifetime of the system and $\eta/s$ are 
obviously changing. The result may be somewhat of an accident, 
because the $v_2$ of identified particles, and the $p_T$ integrated
$v_2$ do show beam energy dependence. 

\item Large photon elliptic flow: The photon $v_2(p_T)$ has been
measured at RHIC and LHC \cite{Adare:2011zr,Lohner:2012ct}, and the 
result is comparable (within sizable errors) to the elliptic flow 
of light hadrons. This is surprising, because photon emission is 
expected to be dominated by the early stages of the quark gluon plasma 
evolution before a significant collective flow can develop 
\cite{Chatterjee:2005de}.

\item Hydrodynamic flow in p+Pb collisions: Significant elliptic and
triangular flow has been observed in high multiplicity p+Pb collisions
at the LHC \cite{Chatrchyan:2013nka,Aad:2013fja,Abelev:2013wsa}. 
A particularly striking discovery is the mass ordering of $v_2(p_T)$
\cite{Abelev:2013wsa}, which is usually regarded as strong evidence
for collective expansion \cite{Heinz:2009xj}. The result is surprising, 
because the proton nucleus collisions have generally been regarded
as a control experiment in which dissipative corrections are too 
large for collective flow to develop. We should note, however, that 
the collective response to initial state fluctuations in nucleus-nucleus
collisions already indicates that the mean free path is very short, and
that hydrodynamic response can be seen on small scales. A simple
scaling analysis of hydrodynamic behavior in p+Pb collisions was 
recently presented in \cite{Basar:2013hea}, but we should note that initial 
state effects may well be important \cite{Dusling:2013oia,McLerran:2013oju}.

\end{enumerate}




\vspace*{1cm}
\begin{thebibliography}{199}

\bibitem{Martin:1963}
L.~P.~Kadanoff, P.~C.~Martin,
``Hydrodynamic equations and correlation functions,''
Ann.\ Phys.\ {\bf 24}, 419 (1963).

\bibitem{Forster}
D.~Forster, 
``Hydrodynamic Fluctuations, Broken Symmetry, and Correlation Functions'',
Addison Wesley (1995).

\bibitem{Reiner:1964}
M. Reiner,
``The Deborah Number,''
Phys. Today 17(1), 62 (1964).

\bibitem{Purcell:1977}
E.~M.~Purcell,
``Life at Low Reynolds Number,''
Am.\ J.\ of Phys. {\bf 45}, 3 (1977).

\bibitem{Danielewicz:1984ww}
P.~Danielewicz and M.~Gyulassy,
``Dissipative Phenomena in Quark Gluon Plasmas,''
Phys.\ Rev.\  D {\bf 31}, 53 (1985).

\bibitem{Kovtun:2004de}
P.~Kovtun, D.~T.~Son and A.~O.~Starinets,
``Viscosity in strongly interacting quantum field theories from black hole
physics,''
Phys.\ Rev.\ Lett.\  {\bf 94}, 111601 (2005)
[arXiv:hep-th/0405231].

\bibitem{Schafer:2009dj} 
T.~Sch\"afer and D.~Teaney,
``Nearly Perfect Fluidity: From Cold Atomic Gases to Hot Quark Gluon Plasmas,''
Rept.\ Prog.\ Phys.\  {\bf 72}, 126001 (2009)
[arXiv:0904.3107 [hep-ph]].

\bibitem{Adams:2012th} 
A.~Adams, L.~D.~Carr, T.~Sch\"afer, P.~Steinberg and J.~E.~Thomas,
``Strongly Correlated Quantum Fluids: Ultracold Quantum Gases, Quantum 
Chromodynamic Plasmas, and Holographic Duality,''
New J.\ Phys.\  {\bf 14}, 115009 (2012)
[arXiv:1205.5180 [hep-th]].

\bibitem{Burnett:1935}
D. Burnett, 
``The distribution of velocities in a slightly non-uniform gas, 
Proc.\ Lond.\ Math.\ Soc.\ {\bf 39} 385 (1935).

\bibitem{Garcia:2008}
L.~S.~Garcia-Colina, R.~M.~Velascoa, F.~J.~Uribea, 
``Beyond the Navier-Stokes equations: Burnett hydrodynamics''
Phys.\ Rep.\ {\bf 465} 149 (2008).

\bibitem{Grad:1949}
H.~Grad,
``On the kinetic theory of rarefied gases,''
Comm.\ Pure and Appl.\ Math.\ {\bf 2} 331 (1949).

\bibitem{Israel:1979wp}
W.~Israel and J.~M.~Stewart,
``Transient relativistic thermodynamics and kinetic theory,''
Annals Phys.\  {\bf 118}, 341 (1979).

\bibitem{Romatschke:2009im}
P.~Romatschke,
``New Developments in Relativistic Viscous Hydrodynamics,''
Int.\ J.\ Mod.\ Phys.\ E {\bf 19}, 1 (2010)
[arXiv:0902.3663 [hep-ph]].

\bibitem{Landau:elast}
L.~D.~Landau, E.~M.~Lifshitz,
``Theory of Elasticity'', 
Course of Theoretical Physics, Vol.VII, 
Pergamon Press (1959).

\bibitem{Cattaneo:1948}
C. Cattaneo, 
``Sulla conduzione del calore,''
Atti Sem.\ Mat.\ Fis.\ Univ.\ Modena {\bf 3} (1948) 3.

\bibitem{Muller:2006}
I.~M\"uller,
``A History of Thermodynamics,''
Springer, Heidelberg (2006).  

\bibitem{Muller:1967}
I.~M\"uller, 
``Zum Paradoxon der W\"armeleitungstheorie,''
Z.\ Phys.\ {\bf 198} (1967) 329.

\bibitem{Israel:1976}
W.~Israel, J.~M.~Stewart, 
``Thermodynamics of nonstationary and transient effects in 
a relativistic gas,''
Phys.\ Lett.\ A {\bf 58}, 213 (1967).

\bibitem{Baier:2007ix}
R.~Baier, P.~Romatschke, D.~T.~Son, A.~O.~Starinets and M.~A.~Stephanov,
``Relativistic viscous hydrodynamics, conformal invariance, and holography,''
JHEP {\bf 0804}, 100 (2008)
[arXiv:0712.2451 [hep-th]].

\bibitem{Chao:2011cy} 
J.~Chao and T.~Sch\"afer,
``Conformal symmetry and non-relativistic second order fluid dynamics,''
Annals Phys.\  {\bf 327}, 1852 (2012)
[arXiv:1108.4979 [hep-th]].

\bibitem{Kovtun:2011np} 
P.~Kovtun, G.~D.~Moore and P.~Romatschke,
``The stickiness of sound: An absolute lower limit on viscosity and 
the breakdown of second order relativistic hydrodynamics,''
Phys.\ Rev.\ D {\bf 84}, 025006 (2011)
[arXiv:1104.1586 [hep-ph]].

\bibitem{Chafin:2012eq} 
C.~Chafin and T.~Sch\"afer,
``Hydrodynamic fluctuations and the minimum shear viscosity 
of the dilute Fermi gas at unitarity,''
Phys.\ Rev.\ A {\bf 87}, 023629 (2013)
[arXiv:1209.1006 [cond-mat.quant-gas]].

\bibitem{Ahikari:2005}
R.~Adhikari, M.~E.~Cates, K.~Stratford, A.~J.~Wagner,
``Fluctuating lattice Boltzmann,''
Europhys.\ Lett.\ {\bf 71}, 473 (2205)
[arXiv:cond-mat/0402598 [cond-mat.stat-mech]].

\bibitem{Murase:2013tma} 
K.~Murase and T.~Hirano,
``Relativistic fluctuating hydrodynamics with memory functions and 
colored noises,''
arXiv:1304.3243 [nucl-th].

\bibitem{Bhatnagar:1954}
P.~L.~Bhatnagar, E.~P.~Gross, M.~Krook,
``A Model for Collision Processes in Gases,''
Phys.\ Rev.\ {\bf 94}, 511 (1954).

\bibitem{Chapman:1970}
S. Chapman and T. G. Cowling, 
``The Mathematical Theory of Non-Uniform Gases'',
Cambridge University Press, 3rd ed. (1970).

\bibitem{Baym:1962}
L.~P. Kadanoff, G.~Baym,
``Quantum statistical mechanics: Green's function methods 
in equilibrium and nonequilibrium problems,''
W.~A.~Benjamin, New York (1962).

\bibitem{Maxwell:1986}
J.~C.~Maxwell,
E.~Garber, S.~G.~Brush, C.~W.~F.~Everitt,
``Maxwell on Molecules and Gases,''
MIT Press (1986). 

\bibitem{Brush:1986}
S.~G.~Brush,
``The kind of motion we call heat,''
North Holland (1986).

\bibitem{Karsch:1986cq}
F.~Karsch and H.~W.~Wyld,
``Thermal Green's Functions And Transport Coefficients On The Lattice,''
Phys.\ Rev.\  D {\bf 35}, 2518 (1987).

\bibitem{Meyer:2007ic}
H.~B.~Meyer,
``A calculation of the shear viscosity in SU(3) gluodynamics,''
Phys.\ Rev.\  D {\bf 76}, 101701 (2007)
[arXiv:0704.1801 [hep-lat]].

\bibitem{Sakai:2007cm}
S.~Sakai and A.~Nakamura,
``Lattice calculation of the QGP viscosities - Present results and next
project,''
PoS {\bf LAT2007}, 221 (2007)
[arXiv:0710.3625 [hep-lat]].

\bibitem{Aarts:2007wj}
G.~Aarts, C.~Allton, J.~Foley, S.~Hands and S.~Kim,
``Spectral functions at small energies and the electrical conductivity in
hot, quenched lattice QCD,''
Phys.\ Rev.\ Lett.\  {\bf 99}, 022002 (2007)
[arXiv:hep-lat/0703008].

\bibitem{Meyer:2011gj} 
H.~B.~Meyer,
``Transport Properties of the Quark-Gluon Plasma: A Lattice QCD Perspective,''
Eur.\ Phys.\ J.\ A {\bf 47}, 86 (2011)
[arXiv:1104.3708 [hep-lat]].

\bibitem{Wlazlowski:2013owa} 
G.~Wlazlowski, P.~Magierski, A.~Bulgac and K.~J.~Roche,
``The temperature evolution of the shear viscosity in a unitary Fermi gas,''
Phys.\  Rev.\  A 88, {\bf 013639} (2013)
[arXiv:1304.2283 [cond-mat.quant-gas]].

\bibitem{Thorne:1986}
K.~S.~Thorne, R.~H.~Price, D.~A.~MacDonald,
``Black Holes: The Membrane Paradigm'',
Yale University Press, 1986.

\bibitem{Maldacena:1997re}
J.~M.~Maldacena,
``The large N limit of superconformal field theories and supergravity,''
Adv.\ Theor.\ Math.\ Phys.\  {\bf 2}, 231 (1998)
[Int.\ J.\ Theor.\ Phys.\  {\bf 38}, 1113 (1999)]
[arXiv:hep-th/9711200].

\bibitem{Son:2007vk}
D.~T.~Son and A.~O.~Starinets,
``Viscosity, Black Holes, and Quantum Field Theory,''
Ann.\ Rev.\ Nucl.\ Part.\ Sci.\  {\bf 57}, 95 (2007)
[arXiv:0704.0240 [hep-th]].

\bibitem{Gubser:2009md} 
S.~S.~Gubser and A.~Karch,
``From gauge string duality to strong interactions: A Pedestrian's Guide,''
Ann.\ Rev.\ Nucl.\ Part.\ Sci.\  {\bf 59}, 145 (2009)
[arXiv:0901.0935 [hep-th]].

\bibitem{CasalderreySolana:2011us} 
J.~Casalderrey-Solana, H.~Liu, D.~Mateos, K.~Rajagopal and U.~A.~Wiedemann,
``Gauge/String Duality, Hot QCD and Heavy Ion Collisions,''
arXiv:1101.0618 [hep-th].

\bibitem{DeWolfe:2013cua} 
O.~DeWolfe, S.~S.~Gubser, C.~Rosen and D.~Teaney,
``Heavy ions and string theory,''
arXiv:1304.7794 [hep-th].

\bibitem{Policastro:2002se} 
G.~Policastro, D.~T.~Son and A.~O.~Starinets,
``From AdS / CFT correspondence to hydrodynamics,''
JHEP {\bf 0209}, 043 (2002)
[hep-th/0205052].

\bibitem{Teaney:2006nc}
D.~Teaney,
``Finite temperature spectral densities of momentum and R-charge  
correlators in $N=4$ Yang Mills theory,''
Phys.\ Rev.\  D {\bf 74}, 045025 (2006)
[arXiv:hep-ph/0602044].

\bibitem{Policastro:2001yc}
G.~Policastro, D.~T.~Son and A.~O.~Starinets,
``The shear viscosity of strongly coupled $N = 4$ supersymmetric Yang-Mills
plasma,''
Phys.\ Rev.\ Lett.\  {\bf 87}, 081601 (2001)
[arXiv:hep-th/0104066].

\bibitem{Kovtun:2006pf}
P.~Kovtun and A.~Starinets,
``Thermal spectral functions of strongly coupled N = 4 supersymmetric
Yang-Mills theory,''
Phys.\ Rev.\ Lett.\  {\bf 96}, 131601 (2006)
[arXiv:hep-th/0602059].

\bibitem{Starinets:2002br} 
A.~O.~Starinets,
``Quasinormal modes of near extremal black branes,''
Phys.\ Rev.\ D {\bf 66}, 124013 (2002)
[hep-th/0207133].

\bibitem{Chesler:2010bi} 
P.~M.~Chesler and L.~G.~Yaffe,
``Holography and colliding gravitational shock waves in asymptotically 
AdS$_5$ spacetime,''
Phys.\ Rev.\ Lett.\  {\bf 106}, 021601 (2011)
[arXiv:1011.3562 [hep-th]].

\bibitem{Heller:2011ju} 
M.~P.~Heller, R.~A.~Janik and P.~Witaszczyk,
``The characteristics of thermalization of boost-invariant plasma from 
holography,''
Phys.\ Rev.\ Lett.\  {\bf 108}, 201602 (2012)
[arXiv:1103.3452 [hep-th]].

\bibitem{Heller:2013fn} 
M.~P.~Heller, R.~A.~Janik and P.~Witaszczyk,
``On the character of hydrodynamic gradient expansion in gauge theory plasma,''
Phys.\ Rev.\ Lett.\  {\bf 110}, 211602 (2013)
[arXiv:1302.0697 [hep-th]].

\bibitem{Kovtun:2003vj} 
P.~Kovtun and L.~G.~Yaffe,
``Hydrodynamic fluctuations, long time tails, and supersymmetry,''
Phys.\ Rev.\ D {\bf 68}, 025007 (2003)
[hep-th/0303010].

\bibitem{Bhattacharyya:2008jc}
S.~Bhattacharyya, V.~E.~Hubeny, S.~Minwalla and M.~Rangamani,
``Nonlinear Fluid Dynamics from Gravity,''
JHEP {\bf 0802}, 045 (2008)
[arXiv:0712.2456 [hep-th]].

\bibitem{Rangamani:2009xk}
M.~Rangamani,
``Gravity and Hydrodynamics: Lectures on the fluid-gravity correspondence,''
Class.\ Quant.\ Grav.\  {\bf 26}, 224003 (2009)
[arXiv:0905.4352 [hep-th]].

\bibitem{Cohen:2007qr}
T.~D.~Cohen,
``Is there a most perfect fluid consistent with quantum field theory?,''
Phys.\ Rev.\ Lett.\  {\bf 99}, 021602 (2007)
[arXiv:hep-th/0702136].

\bibitem{Mott:1972}
N.~F.~Mott,
``Conduction in non-crystalline systems IX: The minimum metallic 
conductivity,''
Phil.\ Magazine {\bf 26}, 1015 (1972).

\bibitem{Paalanen:1982}
M.~A.~Paalanen, T.~F.~Rosenbaum, G.~A.~Thomas, and R.~N.~Bhatt,
``Stress Tuning of the Metal-Insulator Transition at Millikelvin 
Temperatures,''
Phys.\ Rev.\ Lett.\ {\bf 48}, 1284 (1982).

\bibitem{Buchel:2003tz}
A.~Buchel and J.~T.~Liu,
``Universality of the shear viscosity in supergravity,''
Phys.\ Rev.\ Lett.\  {\bf 93}, 090602 (2004)
[arXiv:hep-th/0311175].

\bibitem{Iqbal:2008by}
N.~Iqbal and H.~Liu,
``Universality of the hydrodynamic limit in AdS/CFT and the membrane
paradigm,''
Phys.\ Rev.\  D {\bf 79}, 025023 (2009)
[arXiv:0809.3808 [hep-th]].

\bibitem{Herzog:2006gh}
C.~P.~Herzog, A.~Karch, P.~Kovtun, C.~Kozcaz and L.~G.~Yaffe,
``Energy loss of a heavy quark moving through N = 4 supersymmetric
Yang-Mills plasma,''
JHEP {\bf 0607}, 013 (2006)
[arXiv:hep-th/0605158].

\bibitem{CasalderreySolana:2006rq}
J.~Casalderrey-Solana and D.~Teaney,
``Heavy quark diffusion in strongly coupled N = 4 Yang Mills,''
Phys.\ Rev.\  D {\bf 74}, 085012 (2006)
[arXiv:hep-ph/0605199].

\bibitem{Gubser:2006bz}
S.~S.~Gubser,
``Drag force in AdS/CFT,''
Phys.\ Rev.\  D {\bf 74}, 126005 (2006)
[arXiv:hep-th/0605182].

\bibitem{Kats:2007mq}
Y.~Kats and P.~Petrov,
``Effect of curvature squared corrections in AdS on the viscosity of 
the dual gauge theory,''
JHEP {\bf 0901}, 044 (2009)
[arXiv:0712.0743 [hep-th]].

\bibitem{Cremonini:2011iq} 
S.~Cremonini,
``The Shear Viscosity to Entropy Ratio: A Status Report,''
Mod.\ Phys.\ Lett.\ B {\bf 25}, 1867 (2011)
[arXiv:1108.0677 [hep-th]].

\bibitem{Brigante:2007nu}
M.~Brigante, H.~Liu, R.~C.~Myers, S.~Shenker and S.~Yaida,
``Viscosity Bound Violation in Higher Derivative Gravity,''
Phys.\ Rev.\  D {\bf 77}, 126006 (2008)
[arXiv:0712.0805 [hep-th]].

\bibitem{Buchel:2009tt} 
A.~Buchel and R.~C.~Myers,
``Causality of Holographic Hydrodynamics,''
JHEP {\bf 0908}, 016 (2009)
[arXiv:0906.2922 [hep-th]].

\bibitem{Camanho:2010ru} 
X.~O.~Camanho, J.~D.~Edelstein and M.~F.~Paulos,
``Lovelock theories, holography and the fate of the viscosity bound,''
JHEP {\bf 1105}, 127 (2011)
[arXiv:1010.1682 [hep-th]].

\bibitem{Castin:2011}
Y.~Castin and F.~Werner,
``The Unitary Gas and its Symmetry Properties,''
in: Springer Lecture Notes in Physics ``BEC-BCS Crossover and
the Unitary Fermi gas. Wilhelm Zwerger (editor)
[arXiv:1103.2851 [cond-mat.quant-gas]].

\bibitem{Bloch:2007}
I.~Bloch, J.~Dalibard, W.~Zwerger,
``Many-Body Physics with Ultracold Gases''
Rev.\ Mod.\ Phys.\ {\bf 80}, 885 (2008)
[arXiv:0704.3011].

\bibitem{Giorgini:2008}
S.~Giorgini, L.~P.~Pitaevskii, S.~Stringari, 
``Theory of ultracold atomic Fermi gases''
Rev.\ Mod.\ Phys.\ {\bf 80} 1215 (2008)
[arXiv:0706.3360].

\bibitem{Schaefer:2013oba} 
T.~Sch\"afer and K.~Dusling,
``Bulk viscosity and conformal symmetry breaking in the dilute Fermi gas 
near unitarity,''
Phys.\ Rev.\ Lett.\ {\bf 111}, 120603 (2013)
[arXiv:1305.4688 [cond-mat.quant-gas]].

\bibitem{Massignan:2004}
P.~Massignan, G.~M.~Bruun, H.~Smith,
``Viscous relaxation and collective oscillations in a trapped Fermi 
gas near the unitarity limit'',
Phys.\ Rev.\ A {\bf 71}, 033607 (2005) 
[cond-mat/0409660].

\bibitem{Bruun:2005}
G.~M.~Bruun, H.~Smith,
``Viscosity and thermal relaxation for a resonantly interacting 
Fermi gas'',
Phys.\ Rev.\ A {\bf 72}, 043605 (2005) 
[cond-mat/0504734].

\bibitem{Ku:2011}
M.~J.~H.~Ku, A.~T. ~Sommer, L.~W.~Cheuk, M.~W.~Zwierlein,
``Revealing the Superfluid Lambda Transition in the Universal
Thermodynamics of a Unitary Fermi Gas,''
Science {\bf 335}, 563 (2012)
[arXiv:1110.3309 [cond-mat.quant-gas]].

\bibitem{Rupak:2007vp}
G.~Rupak and T.~Sch\"afer,
``Shear viscosity of a superfluid Fermi gas in the unitarity limit,''
Phys.\ Rev.\  A {\bf 76}, 053607 (2007)
[arXiv:0707.1520 [cond-mat.other]].

\bibitem{Enss:2010qh}
T.~Enss, R.~Haussmann, and W.~Zwerger,
``Viscosity and scale invariance in the unitary Fermi gas,''
Annals Phys.\  {326}, 770-796 (2011).
[arXiv:1008.0007 [cond-mat.quant-gas]].

\bibitem{Braby:2010tk} 
M.~Braby, J.~Chao and T.~Sch\"afer,
``Viscosity spectral functions of the dilute Fermi gas in kinetic theory,''
New J.\ Phys.\  {\bf 13}, 035014 (2011)
[arXiv:1012.0219 [cond-mat.quant-gas]].

\bibitem{Taylor:2010ju}
E.~Taylor and M.~Randeria,
``Viscosity of strongly interacting quantum fluids: spectral functions and
sum rules,''
Phys.\ Rev.\  {\bf A81}, 053610 (2010).
[arXiv:1002.0869 [cond-mat.quant-gas]].

\bibitem{Tan:2005}
S.~Tan, 
``Large momentum part of fermions with large scattering length,''
Ann.\ Phys.\ {\bf 323}, 2971 (2008).

\bibitem{Hofmann:2011qs}
J.~Hofmann,
``Current response, structure factor and hydrodynamic quantities 
of a two- and three-dimensional Fermi gas from the operator product 
expansion,''
 Phys.\ Rev.\ A {\bf 84}, 043603 (2011)
[arXiv:1106.6035 [cond-mat.quant-gas]].

\bibitem{oHara:2002}
K.~M.~O'Hara, S.~L.~Hemmer, M.~E.~Gehm, S.~R.~Granade, J.~E.~Thomas,
``Observation of a Strongly-Interacting Degenerate Fermi Gas of Atoms,''
Science {\bf 298}, 2179 (2002)
[cond-mat/0212463].

\bibitem{Schaefer:2009px}
T.~Sch\"afer and C.~Chafin,
``Scaling Flows and Dissipation in the Dilute Fermi Gas at Unitarity,''
Lect.\ Notes Phys.\  {836}, 375 (2012)
[arXiv:0912.4236 [cond-mat.quant-gas]].

\bibitem{Cao:2010wa}
C.~Cao, E.~Elliott, J.~Joseph, H.~Wu, J.~Petricka, T.~Sch\"afer
and J.~E.~Thomas,
``Universal Quantum Viscosity in a Unitary Fermi Gas,''
Science {331}, 58 (2011)
[arXiv:1007.2625 [cond-mat.quant-gas]].

\bibitem{Stringari:2004}
``Collective oscillations of a trapped superfluid Fermi gas 
near a Feshbach resonance''
S.~Stringari,
Europhys.\ Lett.\ {\bf 65}, 749 (2004)
[cond-mat/0312614].

\bibitem{Bulgac:2004}
A.~Bulgac and G.~F.~Bertsch,
``Collective Oscillations of a Trapped Fermi Gas near the Unitary Limit,''
Phys.\ Rev.\ Lett.\ {\bf 94}, 070401 (2005) 
[cond-mat/0404687].

\bibitem{Kinast:2004b}
J.~Kinast, A.~Turlapov, J.~E.~Thomas,
``Breakdown of Hydrodynamics in the Radial Breathing Mode of a 
Strongly-Interacting Fermi Gas,''
Phys.\ Rev.\ A {\bf 70}, 051401(R) (2004)
[arXiv:cond-mat/0408634 [cond-mat.soft]].

\bibitem{Bartenstein:2004}
M.~Bartenstein, A.~Altmeyer, S.~Riedl, S.~Jochim, C.~Chin, 
J.~Hecker Denschlag, and R.~Grimm,
``Collective Excitiations of a Degenerate Gas at the BEC-BCS Crossover,''
Phys.\ Rev.\ Lett.\ {\bf 92}, 203201 (2004) 
[cond-mat/0412712];

\bibitem{Son:2005tj}
D.~T.~Son,
``Vanishing bulk viscosities and conformal invariance of unitary Fermi gas,''
Phys.\ Rev.\ Lett.\  {\bf 98}, 020604 (2007)
[arXiv:cond-mat/0511721].

\bibitem{Elliott:2013}
E.~Elliott, J.~A.~Joseph, J.~E.~Thomas,
``Observation of conformal symmetry breaking and scale invariance in 
expanding Fermi gases,''
arXiv:1308.3162 [cond-mat.quant-gas].

\bibitem{Schafer:2007pr}
T.~Sch\"afer,
``The Shear Viscosity to Entropy Density Ratio of Trapped Fermions in the
Unitarity Limit,''
Phys.\ Rev.\  A {\bf 76}, 063618 (2007)
[arXiv:cond-mat/0701251].

\bibitem{Turlapov:2007}
A.~Turlapov, J.~Kinast, B.~Clancy, L.~Luo, J.~Joseph, J.~E.~Thomas,
``Is a Gas of Strongly Interacting Atomic Fermions a Nearly Perfect Fluid''
J.\ Low Temp.\ Phys.\ {\bf 150}, 567 (2008)
[arXiv:0707.2574].
 
\bibitem{Bruun:2007}
G.~M.~Bruun, H.~Smith,
``Frequency and damping of the Scissors Mode of a Fermi gas'',
Phys. Rev. A {\bf 76}, 045602 (2007) 
[arXiv:0709.1617].

\bibitem{Kajantie:2002wa} 
K.~Kajantie, M.~Laine, K.~Rummukainen and Y.~Schroder,
``The Pressure of hot QCD up to $g^6 \ln (1/g)$,''
Phys.\ Rev.\ D {\bf 67}, 105008 (2003)
[hep-ph/0211321].

\bibitem{Blaizot:2003tw}
J.~P.~Blaizot, E.~Iancu and A.~Rebhan,
``Thermodynamics of the high-temperature quark gluon plasma'',
in Quark Gluon Plasma 3, R.~Hwa, X.-N.~Wang, eds., (2003)
[hep-ph/0303185].

\bibitem{Braaten:1990it}
E.~Braaten and R.~D.~Pisarski,
``Calculation of the gluon damping rate in hot QCD,''
Phys.\ Rev.\  D {\bf 42}, 2156 (1990).

\bibitem{Blaizot:1999fq} 
J.~P.~Blaizot and E.~Iancu,
``Ultrasoft amplitudes in hot QCD,''
Nucl.\ Phys.\ B {\bf 570}, 326 (2000)
[hep-ph/9906485].

\bibitem{Baym:1990uj}
G.~Baym, H.~Monien, C.~J.~Pethick and D.~G.~Ravenhall,
``Transverse interactions and transport in relativistic quark - gluon and
electromagnetic plasmas,''
Phys.\ Rev.\ Lett.\  {\bf 64}, 1867 (1990).

\bibitem{Arnold:2000dr}
P.~Arnold, G.~D.~Moore and L.~G.~Yaffe,
``Transport coefficients in high temperature gauge theories. I:  Leading-log
results,''
JHEP {\bf 0011}, 001 (2000)
[arXiv:hep-ph/0010177].

\bibitem{Arnold:2003zc}
P.~Arnold, G.~D.~Moore and L.~G.~Yaffe,
``Transport coefficients in high temperature gauge theories. II: Beyond
leading log,''
JHEP {\bf 0305}, 051 (2003)
[arXiv:hep-ph/0302165].

\bibitem{Aoki:2006br}
Y.~Aoki, Z.~Fodor, S.~D.~Katz and K.~K.~Szabo,
``The QCD transition temperature: Results with physical masses in the
continuum limit,''
Phys.\ Lett.\  B {\bf 643}, 46 (2006)
[arXiv:hep-lat/0609068].

\bibitem{Bazavov:2011nk} 
A.~Bazavov, T.~Bhattacharya, M.~Cheng, C.~DeTar, H.~T.~Ding, S.~Gottlieb, 
R.~Gupta and P.~Hegde {\it et al.},
``The chiral and deconfinement aspects of the QCD transition,''
Phys.\ Rev.\ D {\bf 85}, 054503 (2012)
[arXiv:1111.1710 [hep-lat]].

\bibitem{Prakash:1993bt}
M.~Prakash, M.~Prakash, R.~Venugopalan and G.~Welke,
``Nonequilibrium properties of hadronic mixtures,''
Phys.\ Rept.\  {\bf 227}, 321 (1993).

\bibitem{Romatschke:2009ng} 
P.~Romatschke and D.~T.~Son,
``Spectral sum rules for the quark-gluon plasma,''
Phys.\ Rev.\ D {\bf 80}, 065021 (2009)
[arXiv:0903.3946 [hep-ph]].

\bibitem{Aarts:2002cc}
G.~Aarts and J.~M.~Martinez Resco,
``Transport coefficients, spectral functions and the lattice,''
JHEP {\bf 0204}, 053 (2002)
[arXiv:hep-ph/0203177].

\bibitem{Zhu:2012be} 
Y.~Zhu and A.~Vuorinen,
``The shear channel spectral function in hot Yang-Mills theory,''
JHEP {\bf 1303}, 002 (2013)
[arXiv:1212.3818 [hep-ph]].

\bibitem{Ding:2010ga} 
 H.~T.~Ding, A.~Francis, O.~Kaczmarek, F.~Karsch, E.~Laermann and W.~Soeldner,
 ``Thermal dilepton rate and electrical conductivity: An analysis of vector 
current correlation functions in quenched lattice QCD,''
Phys.\ Rev.\ D {\bf 83}, 034504 (2011)
[arXiv:1012.4963 [hep-lat]].

\bibitem{Heinz:2009xj}
U.~W.~Heinz,
``Early collective expansion: Relativistic hydrodynamics and the transport
properties of QCD matter,''
arXiv:0901.4355 [nucl-th].

\bibitem{Teaney:2009qa}
D.~A.~Teaney,
``Viscous Hydrodynamics and the Quark Gluon Plasma,''
arXiv:0905.2433 [nucl-th].

\bibitem{Cleymans:1992zc} 
J.~Cleymans and H.~Satz,
``Thermal hadron production in high-energy heavy ion collisions,''
Z.\ Phys.\ C {\bf 57}, 135 (1993)
[hep-ph/9207204].

\bibitem{BraunMunzinger:2009zz}
P.~Braun-Munzinger, J.~Wambach,
``Colloquium: Phase diagram of strongly interacting matter,''
Rev.\ Mod.\ Phys.\  {81}, 1031-1050 (2009).

\bibitem{Schnedermann:1993ws}
E.~Schnedermann, J.~Sollfrank and U.~W.~Heinz,
``Thermal phenomenology of hadrons from 200-A/GeV S+S collisions,''
Phys.\ Rev.\  C {\bf 48}, 2462 (1993)
[arXiv:nucl-th/9307020].

\bibitem{Ollitrault:1992} 
J.~-Y.~Ollitrault,
``Anisotropy as a signature of transverse collective flow,''
Phys.\ Rev.\ D {\bf 46}, 229 (1992).

\bibitem{Schenke:2012wb} 
B.~Schenke, P.~Tribedy and R.~Venugopalan,
``Fluctuating Glasma initial conditions and flow in heavy ion collisions,''
Phys.\ Rev.\ Lett.\  {\bf 108}, 252301 (2012)
[arXiv:1202.6646 [nucl-th]].

\bibitem{Kharzeev:2000ph} 
D.~Kharzeev and M.~Nardi,
``Hadron production in nuclear collisions at RHIC and high density QCD,''
Phys.\ Lett.\ B {\bf 507}, 121 (2001)
[nucl-th/0012025].

\bibitem{Adler:2003kt}
S.~S.~Adler {\it et al.}  [PHENIX Collaboration],
``Elliptic flow of identified hadrons in Au + Au collisions at  
$s_{NN}^{1/2}$ = 200 GeV,''
Phys.\ Rev.\ Lett.\  {\bf 91}, 182301 (2003).
[arXiv:nucl-ex/0305013].

\bibitem{Adams:2004bi}
J.~Adams {\it et al.}  [STAR Collaboration],
``Azimuthal anisotropy in Au + Au collisions at $s_{NN}^{1/2}$ = 200-GeV,''
Phys.\ Rev.\  C {\bf 72}, 014904 (2005).
[arXiv:nucl-ex/0409033].

\bibitem{Aamodt:2010pa} 
K. Aamodt {\it et al.}  [ALICE Collaboration],
``Elliptic flow of charged particles in Pb-Pb collisions at 2.76 TeV,''
Phys.\ Rev.\ Lett.\  {\bf 105}, 252302 (2010).
[arXiv:1011.3914 [nucl-ex]].

\bibitem{Alver:2010gr}
B. Alver, G. Roland,
``Collision geometry fluctuations and triangular flow in heavy-ion
collisions,''
Phys.\ Rev.\ {C81}, 054905 (2010)
[arXiv:1003.0194 [nucl-th]].

\bibitem{Alver:2008zza} 
B.~Alver {\it et al.},
``Importance of correlations and fluctuations on the initial source 
eccentricity in high-energy nucleus-nucleus collisions,''
Phys.\ Rev.\ C {\bf 77}, 014906 (2008)
[arXiv:0711.3724 [nucl-ex]].

\bibitem{Bjorken:1983}
J.~D.~Bjorken,
``Highly relativistic nucleus-nucleus collisions: 
The central rapidity region
Phys.\ Rev.\  D {\bf 27}, 140 (1983).

\bibitem{Gubser:2010ui} 
S.~S.~Gubser and A.~Yarom,
``Conformal hydrodynamics in Minkowski and de Sitter spacetimes,''
Nucl.\ Phys.\ B {\bf 846}, 469 (2011)
[arXiv:1012.1314 [hep-th]].

\bibitem{ATLAS:2012at} 
G.~Aad {\it et al.}  [ATLAS Collaboration],
``Measurement of the azimuthal anisotropy for charged particle production 
in $\sqrt{s_{NN}}=2.76$ TeV lead-lead collisions with the ATLAS detector,''
Phys.\ Rev.\ C {\bf 86}, 014907 (2012)
[arXiv:1203.3087 [hep-ex]].

\bibitem{Gale:2012rq} 
C.~Gale, S.~Jeon, B.~Schenke, P.~Tribedy and R.~Venugopalan,
``Event-by-event anisotropic flow in heavy-ion collisions from combined 
Yang-Mills and viscous fluid dynamics,''
Phys.\ Rev.\ Lett.\  {\bf 110}, 012302 (2013)
[arXiv:1209.6330 [nucl-th]].

\bibitem{Cooper:1974mv}
F.~Cooper and G.~Frye,
``Comment on the Single Particle Distribution in the Hydrodynamic
and Statistical Thermodynamic Models of Multiparticle Production,''
Phys.\ Rev.\  {D10}, 186 (1974).

\bibitem{Teaney:2003kp}
D.~Teaney,
``Effect of shear viscosity on spectra, elliptic flow, and Hanbury
Brown-Twiss radii,''
Phys.\ Rev.\ C {\bf 68}, 034913 (2003)
[nucl-th/0301099].

\bibitem{Heinz:2013th} 
U.~Heinz and R.~Snellings,
``Collective flow and viscosity in relativistic heavy-ion collisions,''
Ann.\ Rev.\ Nucl.\ Part.\ Sci.\  {\bf 63}, 123 (2013)
[arXiv:1301.2826 [nucl-th]].

\bibitem{Gale:2013da} 
C.~Gale, S.~Jeon and B.~Schenke,
``Hydrodynamic Modeling of Heavy-Ion Collisions,''
Int.\ J.\ Mod.\ Phys.\ A {\bf 28}, 1340011 (2013)
[arXiv:1301.5893 [nucl-th]].

\bibitem{Miller:2007ri}
M.~L.~Miller, K.~Reygers, S.~J.~Sanders and P.~Steinberg,
``Glauber modeling in high energy nuclear collisions,''
Ann.\ Rev.\ Nucl.\ Part.\ Sci.\  {\bf 57}, 205 (2007)
[arXiv:nucl-ex/0701025].

\bibitem{Martinez:2010sc} 
M.~Martinez and M.~Strickland,
``Dissipative Dynamics of Highly Anisotropic Systems,''
Nucl.\ Phys.\ A {\bf 848}, 183 (2010)
[arXiv:1007.0889 [nucl-th]].

\bibitem{Petersen:2008dd} 
H.~Petersen, J.~Steinheimer, G.~Burau, M.~Bleicher and H.~Stocker,
``A Fully Integrated Transport Approach to Heavy Ion Reactions with 
an Intermediate Hydrodynamic Stage,''
Phys.\ Rev.\ C {\bf 78}, 044901 (2008)
[arXiv:0806.1695 [nucl-th]].

\bibitem{vanderSchee:2013pia} 
W.~van der Schee, P.~Romatschke and S.~Pratt,
``A fully dynamical simulation of central nuclear collisions,''
Phys.\ Rev.\ Lett.\  {\bf 111}, 222302 (2013)
[arXiv:1307.2539 [nucl-th]].

\bibitem{Huovinen:2009yb} 
P.~Huovinen and P.~Petreczky,
``QCD Equation of State and Hadron Resonance Gas,''
Nucl.\ Phys.\ A {\bf 837}, 26 (2010)
[arXiv:0912.2541 [hep-ph]].

\bibitem{Bass:2000ib} 
S.~A.~Bass and A.~Dumitru,
``Dynamics of hot bulk QCD matter: From the quark gluon plasma to 
hadronic freezeout,''
Phys.\ Rev.\ C {\bf 61}, 064909 (2000)
[nucl-th/0001033].

\bibitem{Hirano:2005xf}
T.~Hirano, U.~W.~Heinz, D.~Kharzeev, R.~Lacey and Y.~Nara,
``Hadronic dissipative effects on elliptic flow in ultrarelativistic
heavy-ion collisions,''
Phys.\ Lett.\  B {\bf 636}, 299 (2006)
[arXiv:nucl-th/0511046].

\bibitem{Romatschke:2007mq}
P.~Romatschke and U.~Romatschke,
``Viscosity Information from Relativistic Nuclear Collisions: How Perfect is
the Fluid Observed at RHIC?,''
Phys.\ Rev.\ Lett.\  {\bf 99}, 172301 (2007)
[arXiv:0706.1522 [nucl-th]].

\bibitem{Dusling:2007gi}
K.~Dusling and D.~Teaney,
``Simulating elliptic flow with viscous hydrodynamics,''
Phys.\ Rev.\  C {\bf 77}, 034905 (2008)
[arXiv:0710.5932 [nucl-th]].

\bibitem{Song:2007ux}
H.~Song and U.~W.~Heinz,
``Causal viscous hydrodynamics in 2+1 dimensions for relativistic heavy-ion
collisions,''
Phys.\ Rev.\  C {\bf 77}, 064901 (2008)
[arXiv:0712.3715 [nucl-th]].

\bibitem{Song:2011qa} 
H.~Song, S.~A.~Bass and U.~Heinz,
``Elliptic flow in 200 A GeV Au+Au collisions and 2.76 A TeV Pb+Pb 
collisions: insights from viscous hydrodynamics + hadron cascade hybrid 
model,''
Phys.\ Rev.\ C {\bf 83}, 054912 (2011)
[Erratum-ibid.\ C {\bf 87}, 019902 (2013)]
[arXiv:1103.2380 [nucl-th]].

\bibitem{Luzum:2012wu} 
M.~Luzum and J.~-Y.~Ollitrault,
``Extracting the shear viscosity of the quark-gluon plasma from flow in 
ultra-central heavy-ion collisions,''
Nucl.\ Phys.\ A {\bf 904-905}, 377c (2013)
[arXiv:1210.6010 [nucl-th]].

\bibitem{Song:2008hj}
H.~Song and U.~W.~Heinz,
``Extracting the QGP viscosity from RHIC data -- a status report from viscous
hydrodynamics,''
J.\ Phys.\ G {\bf 36}, 064033 (2009)
[arXiv:0812.4274 [nucl-th]].

\bibitem{Huovinen:2013wma} 
P.~Huovinen,
``Hydrodynamics at RHIC and LHC: What have we learned?,''
Int.\ J.\ Mod.\ Phys.\ E {\bf 22}, 1330029 (2013)
[arXiv:1311.1849 [nucl-th]].

\bibitem{Elliott:2013b}
E.~Elliott, J.~A.~Joseph, J.~E.~Thomas,
``Anomalous minimum in the shear viscosity of a Fermi gas,''
arXiv:1311.2049 [cond-mat.quant-gas].

\bibitem{Mueller:2009}
M.~Mueller, J.~Schmalian, L.~Fritz,
``Graphene - a nearly perfect fluid,''
Phys.\ Rev.\ Lett.\ {\bf 103}, 025301 (2009)
[arXiv:0903.4178 [cond-mat.mes-hall]].

\bibitem{Guo:2010}
H.~Guo, D.~Wulin, C.-C.~Chien, K.~Levin,
``Perfect Fluids and Bad Metals: Transport Analogies Between Ultracold 
Fermi Gases and High $T_c$ Superconductors,''
New J.\ Phys.\ {\bf 13}, 075011 (2011)
[arXiv:1009.4678 [cond-mat.supr-con]].

\bibitem{Kurkela:2011ti} 
A.~Kurkela and G.~D.~Moore,
``Thermalization in Weakly Coupled Nonabelian Plasmas,''
JHEP {\bf 1112}, 044 (2011)
[arXiv:1107.5050 [hep-ph]].

\bibitem{Kharzeev:2007jp}
D.~E.~Kharzeev, L.~D.~McLerran and H.~J.~Warringa,
``The Effects of topological charge change in heavy ion collisions:
Event by event P and CP violation,''
Nucl.\ Phys.\ A {803}, 227 (2008)
[arXiv:0711.0950 [hep-ph]].

\bibitem{Kharzeev:2010gr}
D.~E.~Kharzeev and D.~T.~Son,
``Testing the chiral magnetic and chiral vortical effects in heavy ion
collisions,''
Phys.\ Rev.\ Lett.\  {106}, 062301 (2011)
[arXiv:1010.0038 [hep-ph]].

\bibitem{Erdmenger:2008rm}
J.~Erdmenger, M.~Haack, M.~Kaminski and A.~Yarom,
``Fluid dynamics of R-charged black holes,''
JHEP {0901}, 055 (2009)
[arXiv:0809.2488 [hep-th]].

\bibitem{Son:2009tf}
D.~T.~Son and P.~Surowka.
Hydrodynamics with Triangle Anomalies.
Phys.\ Rev.\ Lett.\  {103}, 191601 (2009)
[arXiv:0906.5044 [hep-th]].

\bibitem{Khalatnikov:1965}
I.~M.~Khalatnikov,
``Introduction to the Theory of Superfluidity'',
W.~A.~Benjamin, Inc. (1965). 

\bibitem{Dzyaloshinski:1980}
I.~E.~Dzyaloshinskii, G.~E.~Volovik, 
``Poisson brackets in condensed matter physics,''
Ann.\ Phys.\ {\bf 125}, 67 (1980).

\bibitem{Son:1999pa} 
D.~T.~Son,
``Hydrodynamics of nuclear matter in the chiral limit,''
Phys.\ Rev.\ Lett.\  {\bf 84}, 3771 (2000)
[hep-ph/9912267].

\bibitem{Son:2002ci} 
D.~T.~Son and M.~A.~Stephanov,
``Real time pion propagation in finite temperature QCD,''
Phys.\ Rev.\ D {\bf 66}, 076011 (2002)
[hep-ph/0204226].

\bibitem{Hohenberg:1977ym}
P.~C.~Hohenberg and B.~I.~Halperin,
``Theory Of Dynamic Critical Phenomena,''
Rev.\ Mod.\ Phys.\  {\bf 49}, 435 (1977).

\bibitem{Son:2004iv}
D.~T.~Son and M.~A.~Stephanov,
``Dynamic universality class of the QCD critical point,''
Phys.\ Rev.\  D {\bf 70}, 056001 (2004)
[arXiv:hep-ph/0401052].

\bibitem{Landau:fluid}
L.~D.~Landau, E.~M.~Lifshitz,
``Fluid Dynamics'', 
Course of Theoretical Physics, Vol.VI, 
Pergamon Press (1959).

\bibitem{Leutwyler:1993gf} 
H.~Leutwyler,
``Nonrelativistic effective Lagrangians,''
Phys.\ Rev.\ D {\bf 49}, 3033 (1994)
[hep-ph/9311264].

\bibitem{Greiter:1989qb} 
M.~Greiter, F.~Wilczek and E.~Witten,
``Hydrodynamic Relations in Superconductivity,''
Mod.\ Phys.\ Lett.\ B {\bf 3}, 903 (1989).

\bibitem{Son:2002zn} 
D.~T.~Son,
``Low-energy quantum effective action for relativistic superfluids,''
hep-ph/0204199.

\bibitem{Son:2005rv}
D.~T.~Son and M.~Wingate,
``General coordinate invariance and conformal invariance in nonrelativistic
physics: Unitary Fermi gas,''
Annals Phys.\  {\bf 321}, 197 (2006)
[cond-mat/0509786].

\bibitem{Rupak:2008xq} 
G.~Rupak and T.~Sch\"afer,
``Density Functional Theory for non-relativistic Fermions in the Unitarity 
Limit,''
Nucl.\ Phys.\ A {\bf 816}, 52 (2009)
[arXiv:0804.2678 [nucl-th]].

\bibitem{Martin:1973zz} 
P.~C.~Martin, E.~D.~Siggia and H.~A.~Rose,
``Statistical Dynamics of Classical Systems,''
Phys.\ Rev.\ A {\bf 8}, 423 (1973).

\bibitem{DeDominicis:1977fw} 
C.~De Dominicis and L.~Peliti,
``Field Theory Renormalization and Critical Dynamics Above $T_c$: 
Helium, Antiferromagnets and Liquid Gas Systems,''
Phys.\ Rev.\ B {\bf 18}, 353 (1978).

\bibitem{Khalatnikov:1983ak} 
I.~M.~Khalatnikov, V.~V.~Lebedev and A.~I.~Sukhorukov,
``Diagram Technique For Calculating Long Wave Fluctuation Effects,''
Phys.\ Lett.\ A {\bf 94}, 271 (1983).

\bibitem{Kovtun:2012rj} 
P.~Kovtun,
``Lectures on hydrodynamic fluctuations in relativistic theories,''
J.\ Phys.\ A {\bf 45}, 473001 (2012)
[arXiv:1205.5040 [hep-th]].

\bibitem{Kovtun:2014hpa} 
P.~Kovtun, G.~D.~Moore and P.~Romatschke,
``Towards an effective action for relativistic dissipative hydrodynamics,''
arXiv:1405.3967 [hep-ph].

\bibitem{Jensen:2012jh} 
K.~Jensen, M.~Kaminski, P.~Kovtun, R.~Meyer, A.~Ritz and A.~Yarom,
``Towards hydrodynamics without an entropy current,''
Phys.\ Rev.\ Lett.\  {\bf 109}, 101601 (2012)
[arXiv:1203.3556 [hep-th]].

\bibitem{Dubovsky:2011sj} 
S.~Dubovsky, L.~Hui, A.~Nicolis and D.~T.~Son,
``Effective field theory for hydrodynamics: thermodynamics, and the 
derivative expansion,''
Phys.\ Rev.\ D {\bf 85}, 085029 (2012)
[arXiv:1107.0731 [hep-th]].

\bibitem{Torrieri:2011ne} 
G.~Torrieri,
``Viscosity of An Ideal Relativistic Quantum Fluid: A Perturbative study,''
Phys.\ Rev.\ D {\bf 85}, 065006 (2012)
[arXiv:1112.4086 [hep-th]].

\bibitem{Levin:2013}
Y.~He, K.~Levin,
``Establishing Conservation Laws in Pair Correlated Many Body theories: 
T matrix Approaches,''
arXiv:1308.6793 [cond-mat.quant-gas].

\bibitem{Jeon:1995zm} 
S.~Jeon and L.~G.~Yaffe,
``From quantum field theory to hydrodynamics: Transport coefficients 
and effective kinetic theory,''
Phys.\ Rev.\ D {\bf 53}, 5799 (1996)
[hep-ph/9512263].

\bibitem{Landau:kin}
L.~D.~Landau, E.~M.~Lifshitz,
``Physical Kinetics'', 
Course of Theoretical Physics, Vol.X, 
Pergamon Press (1981).

\bibitem{Bruun:2006}
G.~M.~Bruun, H.~Smith,
``Shear viscosity and damping for a Fermi gas in the unitarity limit''
Phys.\ Rev.\ A {\bf 75}, 043612 (2007)
[cond-mat/0612460].

\bibitem{Ernst:1997}
M.~H.~Ernst, 
``Bogoliubov Choh Uhlenbeck theory: Cradle of modern kinetic theory,'' 
in: Progress in Statistical Physics, Proc. Int. Conf. on
Statistical Physics in Memory of Prof. Soon-Tahk Choh, Seoul, Korea, 
World Scientific Publ. Co., Singapore, 1998. Eds. W. Sung et al.
[arXiv:cond-mat/9707146 [cond-mat.stat-mech]].

\bibitem{CaronHuot:2008uh}
S.~Caron-Huot and G.~D.~Moore,
``Heavy quark diffusion in QCD and N=4 SYM at next-to-leading order,''
JHEP {\bf 0802}, 081 (2008)
[arXiv:0801.2173 [hep-ph]].

\bibitem{Denicol:2012cn} 
G.~S.~Denicol, H.~Niemi, E.~Molnar and D.~H.~Rischke,
``Derivation of transient relativistic fluid dynamics from the 
Boltzmann equation,''
Phys.\ Rev.\ D {\bf 85}, 114047 (2012)
[arXiv:1202.4551 [nucl-th]].

\bibitem{Weinberg:1972}
S.~Weinberg, 
``Gravitation and Cosmology'',
Wiley \& Sons (1972).

\bibitem{Moore:2010bu} 
G.~D.~Moore and K.~A.~Sohrabi,
``Kubo Formulae for Second-Order Hydrodynamic Coefficients,''
Phys.\ Rev.\ Lett.\  {\bf 106}, 122302 (2011)
[arXiv:1007.5333 [hep-ph]].

\bibitem{Cercignani:2002}
C.~Cercignani, G.~M.~Kremer, 
``The Relativistic Boltzmann Equation:
Theory and Applications'', 
Birkh\"auser Verlag (2002). 

\bibitem{Hong:2010at} 
J.~Hong and D.~Teaney,
``Spectral densities for hot QCD plasmas in a leading log approximation,''
Phys.\ Rev.\ C {\bf 82}, 044908 (2010)
[arXiv:1003.0699 [nucl-th]].

\bibitem{Landau:smII}
L.~D.~Landau, E.~M.~Lifshitz,
``Statistical Mechanics, Part II'', 
Course of Theoretical Physics, Vol.IX, 
Pergamon Press (1981).

\bibitem{Buchel:2004di}
A.~Buchel, J.~T.~Liu and A.~O.~Starinets,
``Coupling constant dependence of the shear viscosity in N=4 supersymmetric
Yang-Mills theory,''
Nucl.\ Phys.\  B {\bf 707}, 56 (2005)
[arXiv:hep-th/0406264].

\bibitem{Buchel:2008ac}
A.~Buchel,
``Shear viscosity of boost invariant plasma at finite coupling,''
Nucl.\ Phys.\  B {\bf 802}, 281 (2008)
[arXiv:0801.4421 [hep-th]].

\bibitem{Buchel:2008sh}
A.~Buchel,
``Resolving disagreement for eta/s in a CFT plasma at finite coupling,''
Nucl.\ Phys.\  B {\bf 803}, 166 (2008)
[arXiv:0805.2683 [hep-th]].

\bibitem{Myers:2008yi}
R.~C.~Myers, M.~F.~Paulos and A.~Sinha,
``Quantum corrections to $\eta/s$,''
Phys.\ Rev.\  D {\bf 79}, 041901 (2009)
[arXiv:0806.2156 [hep-th]].

\bibitem{Huot:2006ys}
S.~C.~Huot, S.~Jeon and G.~D.~Moore,
``Shear viscosity in weakly coupled N = 4 super Yang-Mills theory  compared
to QCD,''
Phys.\ Rev.\ Lett.\  {\bf 98}, 172303 (2007)
[arXiv:hep-ph/0608062].

\bibitem{Arnold:2011ja} 
P.~Arnold, D.~Vaman, C.~Wu and W.~Xiao,
``Second order hydrodynamic coefficients from 3-point stress tensor 
correlators via AdS/CFT,''
JHEP {\bf 1110}, 033 (2011)
[arXiv:1105.4645 [hep-th]].

\bibitem{York:2008rr}
M.~A.~York and G.~D.~Moore,
``Second order hydrodynamic coefficients from kinetic theory,''
Phys.\ Rev.\ D {\bf 79}, 054011 (2009)
[arXiv:0811.0729 [hep-ph]].

\bibitem{Son:2006em}
D.~T.~Son and A.~O.~Starinets,
``Hydrodynamics of R-charged black holes,''
JHEP {\bf 0603}, 052 (2006)
[arXiv:hep-th/0601157].

\bibitem{Buchel:2007mf}
A.~Buchel,
``Bulk viscosity of gauge theory plasma at strong coupling,''
Phys.\ Lett.\  B {\bf 663}, 286 (2008)
[arXiv:0708.3459 [hep-th]].

\bibitem{Gubser:2008sz}
S.~S.~Gubser, S.~S.~Pufu and F.~D.~Rocha,
``Bulk viscosity of strongly coupled plasmas with holographic duals,''
JHEP {\bf 0808}, 085 (2008)
[arXiv:0806.0407 [hep-th]].

\bibitem{DeWolfe:2011ts} 
O.~DeWolfe, S.~S.~Gubser and C.~Rosen,
``Dynamic critical phenomena at a holographic critical point,''
Phys.\ Rev.\ D {\bf 84}, 126014 (2011)
[arXiv:1108.2029 [hep-th]].

\bibitem{Kovtun:2005ev} 
P.~K.~Kovtun and A.~O.~Starinets,
``Quasinormal modes and holography,''
Phys.\ Rev.\ D {\bf 72}, 086009 (2005)
[hep-th/0506184].

\bibitem{Nunez:2003eq} 
A.~Nunez and A.~O.~Starinets,
``AdS / CFT correspondence, quasinormal modes, and thermal correlators 
in N=4 SYM,''
Phys.\ Rev.\ D {\bf 67}, 124013 (2003)
[hep-th/0302026].

\bibitem{Liao:2009gb} 
J.~Liao and V.~Koch,
``On the Fluidity and Super-Criticality of the QCD matter at RHIC,''
Phys.\ Rev.\ C {\bf 81}, 014902 (2010)
[arXiv:0909.3105 [hep-ph]].

\bibitem{Adams:2008wt}
A.~Adams, K.~Balasubramanian and J.~McGreevy,
``Hot Spacetimes for Cold Atoms,''
JHEP {\bf 0811}, 059 (2008)
[arXiv:0807.1111 [hep-th]].

\bibitem{Herzog:2008wg}
C.~P.~Herzog, M.~Rangamani and S.~F.~Ross,
``Heating up Galilean holography,''
JHEP {\bf 0811}, 080 (2008)
[arXiv:0807.1099 [hep-th]].

\bibitem{Cherman:2007fj} 
A.~Cherman, T.~D.~Cohen and P.~M.~Hohler,
``A Sticky business: The Status of the conjectured viscosity/entropy 
density bound,''
JHEP {\bf 0802}, 026 (2008)
[arXiv:0708.4201 [hep-th]].

\bibitem{Dobado:2007tm} 
A.~Dobado and F.~J.~Llanes-Estrada,
``On the violation of the holographic viscosity versus entropy KSS bound 
in non relativistic systems,''
Eur.\ Phys.\ J.\ C {\bf 51}, 913 (2007)
[hep-th/0703132].

\bibitem{Son:2007xw}
D.~T.~Son,
``Comment on 'Is There a 'Most Perfect Fluid' Consistent with Quantum Field
Theory?',''
Phys.\ Rev.\ Lett.\  {\bf 100}, 029101 (2008)
[arXiv:0709.4651 [hep-th]].

\bibitem{Braby:2010ec} 
M.~Braby, J.~Chao and T.~Sch\"afer,
``Thermal Conductivity and Sound Attenuation in Dilute Atomic Fermi Gases,''
Phys.\ Rev.\ A {\bf 82}, 033619 (2010)
[arXiv:1003.2601 [cond-mat.quant-gas]].

\bibitem{Joseph:2006}
J.~Joseph, B.~Clancy, L.~Luo, J.~Kinast, A.~Turlapov, J.~E.~Thomas,
``Sound propagation in a Fermi gas near a Feshbach resonance,''
Phys.\ Rev.\ Lett.\ {\bf 98}, 170401 (2007) 
[cond-mat/0612567].

\bibitem{Bruun:2011}
G.~M.~Bruun,
``Spin diffusion in Fermi gases,''
New J.\ Phys.\  {\bf 13}, 035005, (2011)
[arXiv:1012.1607 [cond-mat.quant-gas]].

\bibitem{Sommer:2011}
A.~Sommer, M.~Ku, G.~Roati, and M.~W.~Zwierlein.
``Universal spin transport in a strongly interacting Fermi gas,''
Nature {\bf 472}, 201 (2011)
[arXiv:1103.2337v1 [cond-mat.quant-gas]]

\bibitem{Bruun:2011b}
G.~M.~Bruun, C.~J.~Pethick,
``Spin diffusion in trapped clouds of strongly interacting cold atoms,''
Phys.\ Rev.\ Lett.\ {\bf 107}, 255302 (2011)
[arXiv:1109.5709 [cond-mat.quant-gas]].

\bibitem{Schaefer:2014xma} 
T.~Sch\"afer,
``Second order fluid dynamics for the unitary Fermi gas from kinetic theory,''
arXiv:1404.6843 [cond-mat.quant-gas].

\bibitem{Braaten:2008uh} 
E.~Braaten and L.~Platter,
``Exact Relations for a Strongly Interacting Fermi Gas from the Operator 
Product Expansion,''
Phys.\ Rev.\ Lett.\  {\bf 100}, 205301 (2008)
[arXiv:0803.1125 [cond-mat.other]].

\bibitem{Sagi:2012}
Y.~Sagi, T.~E.~Drake, R.~Paudel, D.~S.~Jin,
``Measurement of the Homogeneous Contact of a Unitary Fermi Gas,''
Phys.\ Rev.\ Lett.\ {\bf 109}, 220402 (2012)
[arXiv:1208.2067 [cond-mat.quant-gas]].

\bibitem{Drut:2010yn} 
J.~E.~Drut, T.~A.~Lahde and T.~Ten,
``Momentum Distribution and Contact of the Unitary Fermi gas,''
Phys.\ Rev.\ Lett.\  {\bf 106}, 205302 (2011)
[arXiv:1012.5474 [cond-mat.stat-mech]].

\bibitem{Yu:2009}
Z.~Yu, G.~M.~Bruun, G.~Baym,
``Short-range correlations and entropy in ultracold atomic Fermi gases,''
Phys.\ Rev.\ A {\bf 80}, 023615 (2009)
[arXiv:0905.1836].

\bibitem{Braaten:2010if} 
E.~Braaten,
``Universal Relations for Fermions with Large Scattering Length,''
Lect.\ Notes Phys.\  {\bf 836}, 193 (2012)
[arXiv:1008.2922 [cond-mat.quant-gas]].

\bibitem{Son:2008ye}
D.~T.~Son,
``Toward an AdS/cold atoms correspondence: a geometric realization of the
Schroedinger symmetry,''
Phys.\ Rev.\  D {\bf 78}, 046003 (2008)
[arXiv:0804.3972 [hep-th]].

\bibitem{Balasubramanian:2008dm}
K.~Balasubramanian and J.~McGreevy,
``Gravity duals for non-relativistic CFTs,''
Phys.\ Rev.\ Lett.\  {\bf 101}, 061601 (2008)
[arXiv:0804.4053 [hep-th]].
 
\bibitem{Hagen:1972pd}
C.~R.~Hagen,
``Scale and conformal transformations in galilean-covariant field theory,''
Phys.\ Rev.\  D {\bf 5}, 377 (1972).

\bibitem{Nishida:2007pj} 
Y.~Nishida and D.~T.~Son,
``Nonrelativistic conformal field theories,''
Phys.\ Rev.\ D {\bf 76}, 086004 (2007)
[arXiv:0706.3746 [hep-th]].

\bibitem{Maldacena:2008wh}
J.~Maldacena, D.~Martelli and Y.~Tachikawa,
``Comments on string theory backgrounds with non-relativistic conformal
symmetry,''
JHEP {\bf 0810}, 072 (2008)
[arXiv:0807.1100 [hep-th]].

\bibitem{Rangamani:2008gi}
M.~Rangamani, S.~F.~Ross, D.~T.~Son and E.~G.~Thompson,
``Conformal non-relativistic hydrodynamics from gravity,''
JHEP {\bf 0901}, 075 (2009)
[arXiv:0811.2049 [hep-th]].

\bibitem{Janiszewski:2012nf} 
S.~Janiszewski and A.~Karch,
``String Theory Embeddings of Nonrelativistic Field Theories and Their 
Holographic Horava Gravity Duals,''
Phys.\ Rev.\ Lett.\  {\bf 110}, 081601 (2013)
[arXiv:1211.0010 [hep-th]].

\bibitem{Janiszewski:2014ewa} 
S.~Janiszewski, A.~Karch, B.~Robinson and D.~Sommer,
``Charged black holes in Horava gravity,''
arXiv:1401.6479 [hep-th].

\bibitem{Bekaert:2011cu} 
X.~Bekaert, E.~Meunier and S.~Moroz,
``Towards a gravity dual of the unitary Fermi gas,''
Phys.\ Rev.\ D {\bf 85}, 106001 (2012)
[arXiv:1111.1082 [hep-th]].

\bibitem{Nikolic:2007zz} 
P.~Nikolic and S.~Sachdev,
``Renormalization-group fixed points, universal phase diagram, 
and 1/N expansion for quantum liquids with interactions near the 
unitarity limit,''
Phys.\ Rev.\ A {\bf 75}, 033608 (2007)
[cond-mat/0609106 [cond-mat.supr-con]].

\bibitem{Veilette:2007}
M.~Y.~Veillette, D.~E.~Sheehy, L.~Radzihovsky,
``Large-N expansion for unitary superfluid Fermi gases,''
Phys.\ Rev.\ A {\bf 75}, 043614 (2007)
[arXiv:cond-mat/0610798 [cond-mat.other]].

\bibitem{Nishida:2007mr} 
Y.~Nishida, D.~T.~Son and S.~Tan,
``Universal Fermi Gas with Two- and Three-Body Resonances,''
Phys.\ Rev.\ Lett.\  {\bf 100}, 090405 (2008)
[arXiv:0711.1562 [cond-mat.other]].

\bibitem{Schafer:2010dv} 
T.~Sch\"afer,
``Dissipative fluid dynamics for the dilute Fermi gas at unitarity: 
Free expansion and rotation,''
Phys.\ Rev.\ A {\bf 82}, 063629 (2010)
[arXiv:1008.3876 [cond-mat.quant-gas]].

\bibitem{Guery:1999}
D.~Guery-Odelin, F.~Zambelli, J.~Dalibard, and S.~Stringari,  
``Collective oscillations of a classical gas confined in harmonic traps''
Phys.\ Rev.\ A {\bf 60} 4851 (1999). 

\bibitem{Pedri:2002}
P.~Pedri, D.~Guery-Odelin and S.~Stringari,  
``Dynamics of a classical gas including dissipative and mean-field effects''
Phys.\ Rev.\ A {\bf 68} 043608 (2003)
[cond-mat/0305624].

\bibitem{Menotti:2002}
C.~Menotti, P.~Pedri, S.~Stringari,
``Expansion of an interacting Fermi gas,''
Phys.\ Rev.\ Lett.\ {\bf 89}, 250402 (2002)
[cond-mat/0208150].

\bibitem{Dusling:2011dq} 
K.~Dusling and T.~Sch\"afer,
``Elliptic flow of the dilute Fermi gas: From kinetics to hydrodynamics,''
Phys.\ Rev.\ A {\bf 84}, 013622 (2011)
[arXiv:1103.4869 [cond-mat.stat-mech]].

\bibitem{Kinast:2005}
J.~Kinast, A.~Turlapov, J.~E.~Thomas,
``Two Transitions in the Damping of a Unitary Fermi Gas,''
Phys.\ Rev.\ Lett.\  {\bf 94}, 170404 (2005)  
[cond-mat/0502507].

\bibitem{Cao:2011fh}
C.~Cao, E.~Elliott, H.~Wu and J.~E.~Thomas,
``Searching for Perfect Fluids: Quantum Viscosity in a Universal Fermi Gas,''
New J.\ Phys.\  {\bf 13} (2011) 075007
[arXiv:1105.2496 [cond-mat.quant-gas]].

\bibitem{Arnold:2006fz}
P.~Arnold, C.~Dogan and G.~D.~Moore,
``The bulk viscosity of high-temperature QCD,''
Phys.\ Rev.\  D {\bf 74}, 085021 (2006)
[arXiv:hep-ph/0608012].

\bibitem{Alford:2007xm} 
M.~G.~Alford, A.~Schmitt, K.~Rajagopal and T.~Sch\"afer,
``Color superconductivity in dense quark matter,''
Rev.\ Mod.\ Phys.\  {\bf 80}, 1455 (2008)
[arXiv:0709.4635 [hep-ph]].

\bibitem{Svetitsky:1987gq}
B.~Svetitsky,
``Diffusion of charmed quark in the quark - gluon plasma,''
Phys.\ Rev.\  D {\bf 37}, 2484 (1988).

\bibitem{Moore:2004tg}
G.~D.~Moore and D.~Teaney,
``How much do heavy quarks thermalize in a heavy ion collision?,''
Phys.\ Rev.\  C {\bf 71}, 064904 (2005)
[arXiv:hep-ph/0412346].

\bibitem{Gubser:2010ze} 
S.~S.~Gubser,
``Symmetry constraints on generalizations of Bjorken flow,''
Phys.\ Rev.\ D {\bf 82}, 085027 (2010)
[arXiv:1006.0006 [hep-th]].

\bibitem{Shuryak:2009} 
E.~Shuryak,
``The Cone, the Ridge and the Fate of the Initial State Fluctuations 
in Heavy Ion Collisions,''
Phys.\ Rev.\ C {\bf 80}, 054908 (2009)
[Erratum-ibid.\ C {\bf 80}, 069902 (2009)]
arXiv:0903.3734 [nucl-th]].

\bibitem{Staig:2010} 
P.~Staig, E.~Shuryak,
``The Fate of the Initial State Fluctuations in Heavy Ion Collisions: 
II. The Fluctuations and Sounds,''
Phys.\ Rev.\ C {\bf 84}, 034908 (2011)
[arXiv:1008.3139 [nucl-th]].

\bibitem{Staig:2011} 
P.~Staig and E.~Shuryak,
``The Fate of the Initial State Fluctuations in Heavy Ion Collisions: 
III. The Second Act of Hydrodynamics,''
Phys.\ Rev.\ C {\bf 84}, 044912 (2011)
[arXiv:1105.0676 [nucl-th]].

\bibitem{Kapusta:2011} 
J.~I.~Kapusta, B.~Muller and M.~Stephanov,
 ``Relativistic Theory of Hydrodynamic Fluctuations with Applications 
to Heavy Ion Collisions,''
Phys.\ Rev.\ C {\bf 85}, 054906 (2012)
[arXiv:1112.6405 [nucl-th]].

\bibitem{Mocsy:2011} 
A.~Mocsy and P.~Sorensen,
``Analyzing the Power Spectrum of the Little Bangs,''
Nucl.\ Phys.\ A {\bf 855}, 241 (2011)
[arXiv:1101.1926 [hep-ph]].

\bibitem{Lacey:2013qua} 
R.~A.~Lacey, A.~Taranenko, J.~Jia, D.~Reynolds, N.~N.~Ajitanand, 
J.~M.~Alexander, Y.~Gu and A.~Mwai,
``Beam energy dependence of the viscous damping of anisotropic flow,''
arXiv:1305.3341 [nucl-ex].

\bibitem{Gardim:2011} 
F.~G.~Gardim, F.~Grassi, M.~Luzum and J.~-Y.~Ollitrault,
``Mapping the hydrodynamic response to the initial geometry in heavy-ion 
collisions,''
Phys.\ Rev.\ C {\bf 85}, 024908 (2012)
[arXiv:1111.6538 [nucl-th]].

\bibitem{Teaney:2012} 
D.~Teaney and L.~Yan,
``Non linearities in the harmonic spectrum of heavy ion collisions with 
ideal and viscous hydrodynamics,''
Phys.\ Rev.\ C {\bf 86}, 044908 (2012)
[arXiv:1206.1905 [nucl-th]].

\bibitem{Floerchinger:2013tya} 
S.~Floerchinger, U.~A.~Wiedemann, A.~Beraudo, L.~Del Zanna, G.~Inghirami 
and V.~Rolando,
``How (non-) linear is the hydrodynamics of heavy ion collisions?,''
arXiv:1312.5482 [hep-ph].

\bibitem{Blaizot:1987nc} 
J.~P.~Blaizot and A.~H.~Mueller,
``The Early Stage of Ultrarelativistic Heavy Ion Collisions,''
Nucl.\ Phys.\ B {\bf 289}, 847 (1987).

\bibitem{Baier:2000sb} 
R.~Baier, A.~H.~Mueller, D.~Schiff and D.~T.~Son,
``'Bottom up' thermalization in heavy ion collisions,''
Phys.\ Lett.\ B {\bf 502}, 51 (2001)
[hep-ph/0009237].

\bibitem{Casalderrey-Solana:2013aba} 
J.~Casalderrey-Solana, M.~P.~Heller, D.~Mateos and W.~van der Schee,
``From full stopping to transparency in a holographic model of heavy ion 
collisions,''
Phys.\ Rev.\ Lett.\  {\bf 111}, 181601 (2013)
[arXiv:1305.4919 [hep-th]].

\bibitem{Gubser:2012gy} 
S.~S.~Gubser,
``Complex deformations of Bjorken flow,''
Phys.\ Rev.\ C {\bf 87}, 014909 (2013)
[arXiv:1210.4181 [hep-th]].

\bibitem{Kolb:2000fh}
P.~F.~Kolb, P.~Huovinen, U.~W.~Heinz and H.~Heiselberg,
``Elliptic flow at SPS and RHIC: From kinetic transport to  hydrodynamics,''
Phys.\ Lett.\ B {\bf 500}, 232 (2001)
[arXiv:hep-ph/0012137].

\bibitem{Molnar:2007} 
D.~Molnar and M.~Gyulassy,
``Saturation of elliptic flow and the transport opacity of the gluon 
plasma at RHIC,''
Nucl.\ Phys.\ A {\bf 697}, 495 (2002)
[Erratum-ibid.\ A {\bf 703}, 893 (2002)]
[nucl-th/0104073].

\bibitem{Xu:2004mz} 
Z.~Xu and C.~Greiner,
``Thermalization of gluons in ultrarelativistic heavy ion collisions 
by including three-body interactions in a parton cascade,''
Phys.\ Rev.\ C {\bf 71}, 064901 (2005)
[hep-ph/0406278].

\bibitem{Xu:2007jv} 
Z.~Xu, C.~Greiner and H.~Stocker,
``PQCD calculations of elliptic flow and shear viscosity at RHIC,''
Phys.\ Rev.\ Lett.\  {\bf 101}, 082302 (2008)
[arXiv:0711.0961 [nucl-th]].

\bibitem{Chen:2013} 
J.~-W.~Chen, J.~Deng, H.~Dong and Q.~Wang,
``Shear and Bulk Viscosities of a Gluon Plasma in Perturbative QCD: 
Comparison of Different Treatments for the $gg\leftrightarrow ggg$ 
Process,''
Phys.\ Rev.\ C {\bf 87}, 024910 (2013)
[arXiv:1107.0522 [hep-ph]].

\bibitem{Fochler:2013epa} 
O.~Fochler, J.~Uphoff, Z.~Xu and C.~Greiner,
``Radiative parton processes in perturbative QCD: An improved version 
of the Gunion and Bertsch cross section from comparisons to the exact result,''
Phys.\ Rev.\ D {\bf 88}, 014018 (2013)
[arXiv:1302.5250 [hep-ph]].

\bibitem{Florkowski:2010} 
W.~Florkowski and R.~Ryblewski,
``Highly-anisotropic and strongly-dissipative hydrodynamics for 
early stages of relativistic heavy-ion collisions,''
Phys.\ Rev.\ C {\bf 83}, 034907 (2011)
[arXiv:1007.0130 [nucl-th]].

\bibitem{Martinez:2010} 
M.~Martinez and M.~Strickland,
 ``Dissipative Dynamics of Highly Anisotropic Systems,''
Nucl.\ Phys.\ A {\bf 848}, 183 (2010)
[arXiv:1007.0889 [nucl-th]].

\bibitem{Romatschke:2011hm} 
P.~Romatschke, M.~Mendoza and S.~Succi,
``A fully relativistic lattice Boltzmann algorithm,''
Phys.\ Rev.\ C {\bf 84}, 034903 (2011)
[arXiv:1106.1093 [nucl-th]].

\bibitem{Romatschke:2011qp} 
P.~Romatschke,
``Relativistic (Lattice) Boltzmann Equation with Non-Ideal Equation of State,''
Phys.\ Rev.\ D {\bf 85}, 065012 (2012)
[arXiv:1108.5561 [gr-qc]].

\bibitem{Shen:2012vn} 
C.~Shen and U.~Heinz,
``Collision Energy Dependence of Viscous Hydrodynamic Flow in Relativistic 
Heavy-Ion Collisions,''
Phys.\ Rev.\ C {\bf 85}, 054902 (2012)
[Erratum-ibid.\ C {\bf 86}, 049903 (2012)]
[arXiv:1202.6620 [nucl-th]].

\bibitem{Heiselberg:1998es} 
H.~Heiselberg and A.-M.~Levy,
``Elliptic flow and HBT in noncentral nuclear collisions,''
Phys.\ Rev.\ C {\bf 59}, 2716 (1999)
[nucl-th/9812034].

\bibitem{Bhalerao:2005mm} 
R.~S.~Bhalerao, J.-P.~Blaizot, N.~Borghini and J.-Y.~Ollitrault,
``Elliptic flow and incomplete equilibration at RHIC,''
Phys.\ Lett.\ B {\bf 627}, 49 (2005)
[nucl-th/0508009].

\bibitem{Voloshin:1999gs} 
S.~A.~Voloshin and A.~M.~Poskanzer,
``The Physics of the centrality dependence of elliptic flow,''
Phys.\ Lett.\ B {\bf 474}, 27 (2000)
[nucl-th/9906075].

\bibitem{Alt:2003ab} 
C.~Alt {\it et al.}  [NA49 Collaboration],
``Directed and elliptic flow of charged pions and protons in Pb + Pb 
collisions at 40-A-GeV and 158-A-GeV,''
Phys.\ Rev.\ C {\bf 68}, 034903 (2003)
[nucl-ex/0303001].

\bibitem{Basar:2013hea} 
G.~Basar and D.~Teaney,
``A scaling relation between pA and AA collisions,''
arXiv:1312.6770 [nucl-th].

\bibitem{Wang:1991} 
X.-N.~Wang and M.~Gyulassy,
``HIJING: A Monte Carlo model for multiple jet production in pp, pA and 
AA collisions,''
Phys.\ Rev.\ D {\bf 44}, 3501 (1991).

\bibitem{Geiger:1992} 
K.~Geiger and B.~Muller,
``Dynamics of parton cascades in highly relativistic nuclear collisions,''
Nucl.\ Phys.\ B {\bf 369}, 600 (1992).

\bibitem{Rischke:1998fq} 
D.~H.~Rischke,
``Fluid dynamics for relativistic nuclear collisions,''
In ``Cape Town 1998, Hadrons in dense matter and hadrosynthesis'' 21-70
[nucl-th/9809044].

\bibitem{Muronga:2001zk} 
A.~Muronga,
``Second order dissipative fluid dynamics for ultrarelativistic nuclear 
collisions,''
Phys.\ Rev.\ Lett.\  {\bf 88}, 062302 (2002)
[Erratum-ibid.\  {\bf 89}, 159901 (2002)]
[nucl-th/0104064].

\bibitem{Heinz:2005bw} 
U.~W.~Heinz, H.~Song and A.~K.~Chaudhuri,
``Dissipative hydrodynamics for viscous relativistic fluids,''
Phys.\ Rev.\ C {\bf 73}, 034904 (2006)
[nucl-th/0510014].

\bibitem{Song:2007fn} 
H.~Song and U.~W.~Heinz,
``Suppression of elliptic flow in a minimally viscous quark-gluon plasma,''
Phys.\ Lett.\ B {\bf 658}, 279 (2008)
[arXiv:0709.0742 [nucl-th]].

\bibitem{Song:2008si} 
H.~Song and U.~W.~Heinz,
``Multiplicity scaling in ideal and viscous hydrodynamics,''
Phys.\ Rev.\ C {\bf 78}, 024902 (2008)
[arXiv:0805.1756 [nucl-th]].

\bibitem{Heller:2007qt} 
M.~P.~Heller and R.~A.~Janik,
``Viscous hydrodynamics relaxation time from AdS/CFT,''
Phys.\ Rev.\ D {\bf 76}, 025027 (2007)
[hep-th/0703243 [HEP-TH]].

\bibitem{Baier:2011} 
R.~Baier, A.~H.~Mueller, D.~Schiff and D.~T.~Son,
``Does parton saturation at high density explain hadron multiplicities 
at LHC?,''
arXiv:1103.1259 [nucl-th].

\bibitem{Geroch:1990bw}
R.~P.~Geroch and L.~Lindblom,
``Dissipative relativistic fluid theories of divergence type,''
Phys.\ Rev.\  D {\bf 41}, 1855 (1990).

\bibitem{Ottinger:1998}
H.~C.~{\"O}ttinger, 
``Relativistic and nonrelativistic description of fluids with 
anisotropic heat conduction,''
Physica A {\bf 254} 433 (1998).

\bibitem{Luzum:2008cw}
M.~Luzum and P.~Romatschke,
``Conformal Relativistic Viscous Hydrodynamics:  Applications to RHIC results
at $s_{NN}^{1/2}$ = 200 GeV,''
Phys.\ Rev.\  C {\bf 78}, 034915 (2008)
[arXiv:0804.4015 [nucl-th]].

\bibitem{Chatrchyan:2012ta} 
S.~Chatrchyan {\it et al.}  [CMS Collaboration],
``Measurement of the elliptic anisotropy of charged particles produced 
in PbPb collisions at nucleon-nucleon center-of-mass energy = 2.76 TeV,''
Phys.\ Rev.\ C {\bf 87}, 014902 (2013)
[arXiv:1204.1409 [nucl-ex]].

\bibitem{Adamczyk:2012ku} 
L.~Adamczyk {\it et al.}  [STAR Collaboration],
``Inclusive charged hadron elliptic flow in Au + Au collisions at 
$\sqrt{s_{NN}}$ = 7.7 - 39 GeV,''
Phys.\ Rev.\ C {\bf 86}, 054908 (2012)
[arXiv:1206.5528 [nucl-ex]].

\bibitem{Adare:2011zr} 
A.~Adare {\it et al.}  [PHENIX Collaboration],
``Observation of direct-photon collective flow in $\sqrt{s_{NN}}=200$ 
GeV Au+Au collisions,''
Phys.\ Rev.\ Lett.\  {\bf 109}, 122302 (2012)
[arXiv:1105.4126 [nucl-ex]].

\bibitem{Lohner:2012ct} 
D.~Lohner [ALICE Collaboration],
``Measurement of Direct-Photon Elliptic Flow in Pb-Pb Collisions at 
$\sqrt{s_{NN}} = 2.76$ TeV,''
J.\ Phys.\ Conf.\ Ser.\  {\bf 446}, 012028 (2013)
[arXiv:1212.3995 [hep-ex]].

\bibitem{Chatterjee:2005de} 
R.~Chatterjee, E.~S.~Frodermann, U.~W.~Heinz and D.~K.~Srivastava,
``Elliptic flow of thermal photons in relativistic nuclear collisions,''
Phys.\ Rev.\ Lett.\  {\bf 96}, 202302 (2006)
[nucl-th/0511079].

\bibitem{Chatrchyan:2013nka} 
S.~Chatrchyan {\it et al.}  [CMS Collaboration],
``Multiplicity and transverse momentum dependence of two- and four-particle 
correlations in pPb and PbPb collisions,''
Phys.\ Lett.\ B {\bf 724}, 213 (2013)
[arXiv:1305.0609 [nucl-ex]].

\bibitem{Aad:2013fja} 
G.~Aad {\it et al.}  [ATLAS Collaboration],
``Measurement with the ATLAS detector of multi-particle azimuthal 
correlations in p+Pb collisions at $\sqrt{s_{NN}}$=5.02 TeV,''
Phys.\ Lett.\ B {\bf 725}, 60 (2013)
[arXiv:1303.2084 [hep-ex]].

\bibitem{Abelev:2013wsa} 
B.~B.~Abelev {\it et al.}  [ALICE Collaboration],
``Long-range angular correlations of pi, K and p in p-Pb collisions 
at $\sqrt{s_{NN}}$ = 5.02 TeV,''
Phys.\ Lett.\ B {\bf 726}, 164 (2013)
[arXiv:1307.3237 [nucl-ex]].

\bibitem{Dusling:2013oia} 
K.~Dusling and R.~Venugopalan,
``Comparison of the color glass condensate to dihadron correlations in 
proton-proton and proton-nucleus collisions,''
Phys.\ Rev.\ D {\bf 87}, 094034 (2013)
[arXiv:1302.7018 [hep-ph]].

\bibitem{McLerran:2013oju} 
L.~McLerran, M.~Praszalowicz and B.~Schenke,
``Transverse Momentum of Protons, Pions and Kaons in High Multiplicity 
pp and pA Collisions: Evidence for the Color Glass Condensate?,''
Nucl.\ Phys.\ A {\bf 916}, 210 (2013)
[arXiv:1306.2350 [hep-ph]].





\end{thebibliography}
\end{document}